\documentclass[11pt]{article}
\usepackage[utf8]{inputenc}
\usepackage{setspace}
\usepackage{float}
\usepackage{multirow}
\usepackage{graphicx}
\usepackage{amsmath}
\usepackage{accents}
\usepackage{amssymb}
\usepackage{algorithm}
\usepackage{amsthm}
\usepackage{natbib}
\usepackage{subcaption}
\usepackage[usenames,dvipsnames]{xcolor}
\usepackage{bbm}
\usepackage{algpseudocode}
\usepackage{booktabs}
\usepackage[colorlinks,linkcolor=blue,citecolor=blue,urlcolor=blue]{hyperref}
 
\usepackage{macros}
\usepackage[left=1.1in,top=1in,right=1.1in,bottom=1in]{geometry}

\def\title#1{{\Large\bf  \begin{center} #1 \vspace{0pt} \end{center}  } }
\def\authors#1{{\large\bf \begin{center} #1 \vspace{0pt} \end{center} } }
\def\university#1{{\sl \begin{center} #1 \vspace{0pt} \end{center} } }
\def\inst#1{\unskip$^{#1}$}

\begin{document}

\title{Bayesian Variable Selection and Sparse Estimation for
High-Dimensional Graphical Models}

\bigskip

\authors{Anwesha Chakravarti\inst{1}, Naveen N. Narisetty, Feng Liang\inst{2}
     }

\smallskip

%
%

 \university{
 \inst{1} University of Illinois at Urbana Champaign, Department of Statistics, United States, anwesha5@illinois.edu \\
 \inst{2} University of Illinois at Urbana Champaign, Department of Statistics, United States, liangf@illinois.edu }

\bigskip

\noindent {\large\bf Abstract}

\medskip

We introduce a novel Bayesian approach for both covariate selection and
sparse precision matrix estimation in the context of high-dimensional Gaussian graphical models involving multiple responses. Our approach provides a sparse estimation of the three distinct sparsity structures: the regression coefficient matrix, the conditional dependency structure among responses, and between responses and covariates. 
This contrasts with existing methods, which typically focus on any two of these structures but seldom achieve simultaneous sparse estimation for all three. A key aspect of our method is that it leverages the structural sparsity information gained from the presence of irrelevant covariates in the dataset to introduce covariate-level sparsity in the precision and regression coefficient matrices. This is achieved through a Bayesian conditional random field model using a hierarchical spike and slab prior setup. Despite the non-convex nature of the problem, we establish statistical accuracy for points in the high posterior density region, including the maximum-a-posteriori (MAP) estimator. We also present an efficient Expectation-Maximization (EM) algorithm for computing the estimators. Through simulation experiments, we demonstrate the competitive performance of our method, particularly in scenarios with weak signal strength in the precision matrices. Finally, we apply our method to a bike-share dataset, showcasing its predictive performance.
\bigskip

\noindent {\large\bf Keywords}

\medskip

\noindent Bayesian variable selection, Gaussian conditional random field, Bayesian regularization, Spike and slab Lasso prior, Graphical models.

%
%





\section{Introduction}

In high-dimensional data settings, where the number of responses and covariates can be substantial, using graphical models to statistically estimate dependence structures among responses and covariates enhances our understanding of the relationship between them and improves our prediction ability. Mathematically, this setup can be formulated as follows: Consider $p$ response variables $(Y_1,Y_2,\dots,Y_p)$ and $q$ covariates $(X_1,X_2,\dots,X_q)$, many of which do not affect the responses. Our key goals are:
\begin{enumerate}
    \item Estimating the sparse conditional dependency structure between the multiple responses $(Y_i,Y_j)$ through a precision matrix estimation, denoted as $\Lambda$,
    \item Estimating the sparse conditional dependency structure between the responses-covariates duo $(Y_i,X_j)$ through a precision matrix estimation, denoted as $\Theta$,
    \item Estimating the regression coefficient matrix to predict the multivariate responses $Y$ based on the covariates $X$, denoted as $B$.
\end{enumerate}

This problem commonly occurs in various applications. For example, consider a study that aims to uncover the relationships between genes in response to a particular drug treatment. In this case, the gene expression levels of multiple genes are the response variables, and the covariates may include information about the drug dosage, treatment duration, or patient-specific factors. By incorporating these covariates into a graphical model and recovering the sparsity structures in the data, researchers can examine how the relationships between the expression levels of the genes change in the presence of the drug, enabling them to identify genes that are directly or indirectly affected by the treatment. The regression coefficient matrix can then be used for prediction tasks. Other examples where the sparse estimation of precision matrices and regression coefficient matrix is found useful include cell signaling data \citep{friedman2008sparse}, brain fMRI data \citep{honorio2012variable}, finance \citep{sohn2012joint}, energy demand prediction, wind power forecasting \citep{wytock2013sparse}, global weather predictions \citep{radosavljevic2014neural} and various other health and environmental domains. 

When dealing with a large number of covariates, especially with limited observations, many covariates may not significantly contribute to explaining a particular response. It is also natural to have totally irrelevant covariates that do not explain any of the responses. For instance, in the context of our example above, it is realistic for health
centers to collect vast amounts of information from patients, most of which is nonessential. Hence, effective covariate selection becomes vital, particularly in identifying these totally irrelevant variables. It is worth emphasizing that when a row in $\Theta$ (a $q\times p$ matrix) or a column in $B$ (a $p\times q$ matrix) consists entirely of zeros, it implies that the corresponding covariate does not influence any of the responses. Thus, one of our primary objectives is to develop a model capable of incorporating this structural sparsity by providing a row-wise group sparse estimate of $\Theta$ and a column-wise group sparse estimate of $B$. Due to the relation $B = -\Lambda^{-1}\Theta^T$, we simplify this to a single problem, as further elaborated in Section 2.

Many methods in existing literature study individual aspects of our goal of estimating $\Lambda, \Theta$, and $B$. One common approach for estimating the conditional dependency matrices $(\Theta,\Lambda)$ is within a joint Gaussian graphical model framework \citep{friedman2008sparse,ravikumar2011high,li2019expectation,gan2019bayesian,li2022transfer, mohammadi2023accelerating} assuming a joint multivariate normal distribution on $(X, Y)$. However, this method entails estimating the dependence structure between the $X$'s along with $(\Theta, \Lambda)$, leading to computational inefficiency, mainly when the dimension of $X$ is large relative to $Y$. Alternatively, for the simultaneous sparse estimation of $(\Theta,\Lambda)$, Gaussian conditional random fields (GCRF) with an $\ell_1$-penalization  \citep{wytock2013sparse,yuan2014partial} and with penalization induced by spike and slab priors \citep{gan2022bayesian} have been proposed, and their estimation accuracy have been widely studied.  Furthermore, within the context of multiple regression, various approaches have been employed to simultaneously estimate $(B,\Lambda)$ while imposing sparsity assumptions on these parameters \citep{rothman2010sparse,yin2011sparse,sohn2012joint,cai2013covariate, deshpande2019simultaneous, Osborne2020LatentNE,zhang2022high}. 

One limitation of previous studies is their failure to address the structural sparsity stemming from the presence of totally irrelevant covariates. In other words, these approaches lack a straightforward mechanism for discarding variables that do not impact any of the responses.
The aforementioned prior works also treat the three goals as two separate problems, estimating $(\Theta,\Lambda)$ or estimating $(B,\Lambda)$. While the relation $B = -\Lambda^{-1}\Theta^T$, which is a consequence of the multiple regression model, inherently links the two estimation procedures, methods designed for sparse estimation of $(\Theta,\Lambda)$ may not necessarily yield sparse estimates for $B$ and vice-versa. This causes a gap in our understanding of the dependence between the $X$'s and $Y$'s since $\Theta$ and $B$ give us different aspects of this dependency. While a sparse $\Theta$ offers insights into the conditional independence of $X_i$ and $Y_j$, a sparse $B$ explains the linear dependence of the response means on the covariates, making it imperative to bridge this gap. 

Our primary contribution lies in developing a methodology that generates row-wise group sparse estimates for $\Theta$ alongside sparse estimates for $\Lambda$, resulting in column-wise group sparse estimates for $B$. The approach merges concepts from Gaussian conditional random fields with multiple regression to simultaneously estimate the sparsity structures for all three crucial parameters, $(\Lambda, \Theta, B)$. Our proposed methodology adopts a hierarchical Bayesian prior setup with spike and slab Lasso (SSL) priors \citep{rovckova2014emvs,rovckova2018bayesian,rovckova2018spike, Bai2020SpikeandSlabML} and estimates the parameters through a maximum a posteriori (MAP) estimation via a non-convex regularization problem. Our computational contribution involves developing an efficient EM algorithm to compute the MAP estimators. This EM algorithm draws inspiration from prior works in this field, such as \citet{wytock2013sparse} and \citet{gan2022bayesian}. 

Our theoretical findings establish an optimal point in the high probability density (HPD) region for $\Theta$ and $\Lambda$, which ensures support recovery consistency and, most importantly, column recovery for the corresponding $B$, guaranteeing accurate variable selection. This optimal ($\Theta$,$\Lambda$) also achieves optimal convergence in the infinity norm. Furthermore, for all $(\Theta,\Lambda) \in HPD$ and their corresponding estimates of $B$, we establish an optimal convergence rate in the Frobenius norm.

\comment{Our theoretical findings establish that within the high probability density (HPD) region for $\Theta$ and $\Lambda$, as well as for the associated $B$ obtained using these estimates, an optimal convergence rate in the Frobenius norm exists. Furthermore, we demonstrate at least one optimal point within this region that achieves optimal convergence in the infinity norm. This optimal point also ensures support recovery consistency for $\Theta$ and $\Lambda$ and, most importantly, column recovery for $B$, guaranteeing accurate variable selection. The convergence rates for all three parameters are on par with recent works \citep{yin2011sparse,cai2013covariate,yuan2014partial,gan2022bayesian}.  Additionally, through our simulation studies, we observe that in high dimensional situations with a large number of responses and when the strength of the conditional dependencies is small, our method showcases enhanced performance in correctly estimating the sparse conditional dependency structures by leveraging the information gained from the group sparsity.}

Our paper is structured as follows. Section 2 explains the model formulation, outlining the incorporation of group sparsity and the connection between Gaussian conditional random fields (GCRF) and the multiple regression framework. Section 3 provides an extensive description of the Bayesian methodology discussed above and additional information on estimation and structure recovery. Section 4 contains detailed theoretical results on estimation accuracy and support recovery consistency. In Section 5, we present experimental results that illustrate the competitive performance of our method in both simulated scenarios and when applied to real-world data. We conclude with some final remarks in Section 6.

\section{Model formulation}

Our work explores high-dimensional settings where most covariates are completely irrelevant to the outputs, while some key covariates may influence multiple output variables. 
Our objective is to gain insights into the conditional dependencies within the $Y$'s through a sparse $\Lambda$, conditional dependency structure among the $Y$'s and the $X$'s through a row-wise group sparse $\Theta$ and to estimate the column sparse regression coefficient matrix $B$. In mathematical terms, a sparse element in each matrix implies:
\begin{align*}
    \Theta_{ij} = 0 &\iff X_{i} \ind Y_{j} \text{ given } X_{-(i)}, Y_{-(j)};\\
    B_{ij} = 0 &\iff E(Y_{i}|X) \text{ does not depend on }X_{j} ;\\
    \Lambda_{ij} = 0 &\iff Y_{i} \ind Y_{j} \text{ given } X,Y_{-(i,j)},
\end{align*}
where $X_{-(k)}$ denotes all the covariates except $X_k$ and $Y_{-(k)}$ denotes all the responses except $Y_k$. Thus, a row-sparse $\Theta$ and column-sparse $B$ help identify the key variables, as a row in $\Theta$ or a column in $B$ filled entirely with zeros indicates that the corresponding covariate does not affect any of the responses.

Two frameworks can be employed to estimate the above parameters. The first uses Gaussian conditional random fields (GCRFs), initially introduced by \citet{lafferty2001conditional},  to jointly estimate $(\Theta,\Lambda$). The GCRF model assumes the conditional density of $Y$ given $X$ as

\begin{equation}
\label{eq:GCRF}
    p(Y|X,\Lambda,\Theta) \propto \sqrt{\det(\Lambda)}\exp\left\{-\frac12Y^T\Lambda Y - X^T\Theta Y\right\},
\end{equation}
and has gained widespread use in the context of estimation of $(\Theta, \Lambda)$, as evident in subsequent works such as \citet{yin2011sparse,wytock2013sparse,yuan2014partial,gan2022bayesian}.
The second framework uses a multivariate regression approach, also known as the covariate-adjusted graphical model, for jointly estimating $(B, \Lambda)$ \citep{rothman2010sparse, cai2013covariate, consonni2017objective, deshpande2019simultaneous,zhang2022high}. This model assumes normality for the conditional distribution of the responses given the covariates and can be expressed as
\begin{equation}
\label{eq:covadjgraphmodel}
    Y|X \sim N(BX,\Lambda^{-1}).
\end{equation}
By considering $B = -\Lambda^{-1}\Theta^T$, the distribution in (\ref{eq:covadjgraphmodel}) can be written as 
$$p(Y|X,\Lambda,\Theta) \propto det(\Lambda)^{1/2}\exp\left(-\frac{1}{2}\big((Y+\Lambda^{-1}\Theta^TX)^T\Lambda(Y+\Lambda^{-1}\Theta^TX)\big) \right),$$
which simplifies to be equivalent to (\ref{eq:GCRF}). Therefore, reparameterizing the multiple regression model in terms of $\Theta$ and $\Lambda$ yields the GCRF model with $B = -\Lambda^{-1}\Theta^T$. Consequently, we can utilize either framework to obtain estimates for the three parameters $(\Lambda,\Theta,B)$ by estimating either $(\Theta, \Lambda)$ or $(B,\Lambda)$ and then using $B = -\Lambda^{-1}\Theta^T$ for the third. The challenge with existing methods in the literature lies in achieving group sparse estimates for both $\Theta$ and $B$ simultaneously.

In our work, by focusing on obtaining either a row-sparse estimate for $\Theta$ or a column-sparse estimate for $B$, we inherently achieve the necessary structural sparsity for the other parameter. Due to the relation $B = -\Lambda^{-1}\Theta^T$, if $\Theta$ has rows consisting of all zeroes, even if $\Lambda^{-1}$ is dense, $B$ will be column-sparse because a $\textbf{0}$-row in $\Theta$ will cause $B$ to have a $\textbf{0}$-column. This is demonstrated in Figure \ref{fig:GS}. Similarly, a column-wise group sparsity on $B$ ensures that the sparse estimation of $(B,\Lambda)$ results in a row-sparse $\Theta$. Hence, obtaining group sparse estimates for both $\Theta$ and $B$ reduces to finding a group sparse estimate for any one of them. 
\begin{figure}
    \centering
    \includegraphics[scale = 0.3]{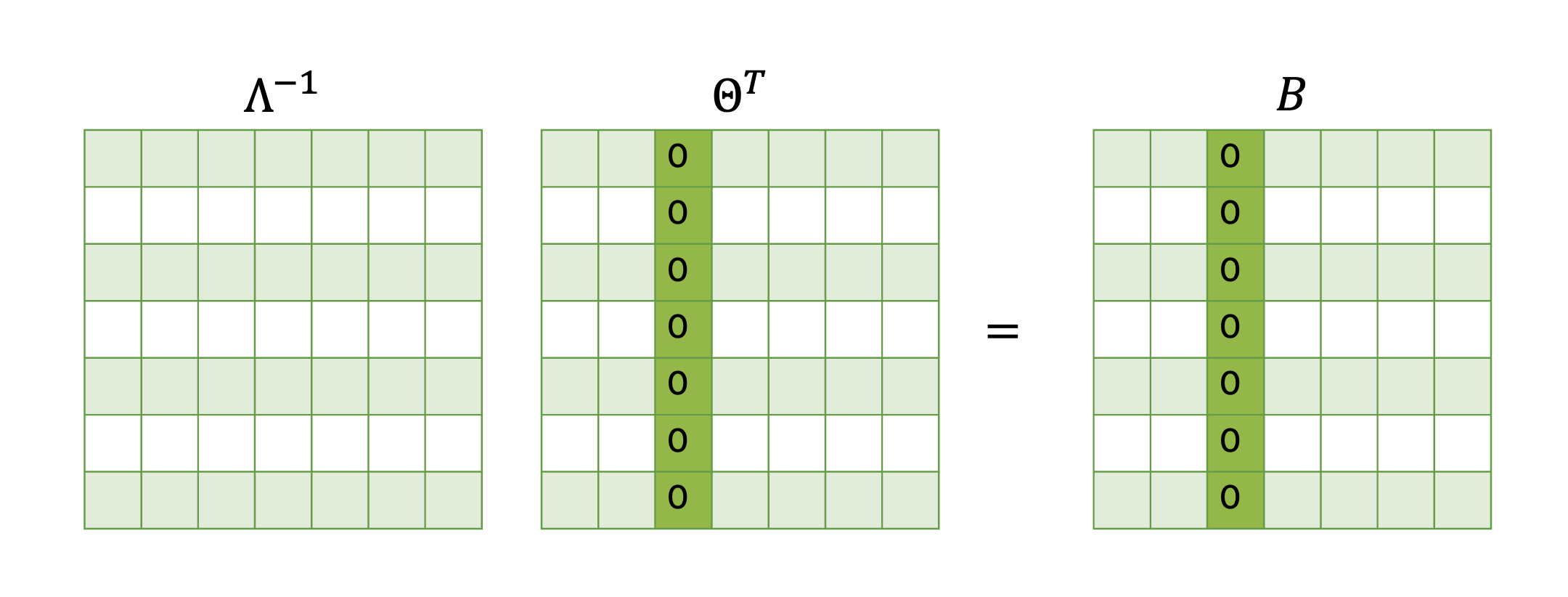}
    \caption{Demonstration of the use of group sparsity}
    \label{fig:GS}
\end{figure}

While both the approaches, estimating $(\Theta,\Lambda)$ to get a group sparse $B$, and estimating $(B,\Lambda)$ to get a group sparse $\Theta$ are valid methods, the log-likelihood function of the GCRF is convex, while the one from
the multivariate regression is not convex \citep{yuan2014partial}. Thus, we formulate a method for the simultaneous sparse estimation of $(\Lambda,\Theta,B)$ using the GCRF framework. We induce the structural sparsity in our estimates by choosing appropriate priors.

\comment{
\subsection{Conditional Gaussian random fields for the joint estimation of $(\Theta,\Lambda)$}
One common approach for estimating the conditional dependency matrices $(\Theta,\Lambda)$ is to model them within a joint Gaussian graphical model framework. Under the assumption that the covariates X are normally distributed, the joint Gaussian graphical model (GGM) states that $(X,Y)$ jointly follow a multivariate normal distribution $$\begin{pmatrix}X\\Y \end{pmatrix} \sim N\left(0,\Omega^{-1}\right), \quad \text{where } \Omega = \begin{pmatrix}
    \Omega_{XX} & \Theta\\\Theta^T & \Lambda
\end{pmatrix}.$$
Several GGM estimation techniques, including the widely-used graphical LASSO \citep{friedman2008sparse}, exist for the estimation of $\Omega$ and, consequently, the estimation of $\Theta$ and $\Lambda$.
However, a major drawback of this method is that it involves estimating $\Omega_{XX}$. In scenarios where the dimension of $X$ is considerably large, especially when compared to the dimension of $Y$, this can lead to a substantial computational burden. Such computational overhead is less than ideal, particularly when our primary focus lies in estimating $\Theta$ and $\Lambda$ and not $\Omega_{XX}$. 

In response to these challenges, we shift our focus to the Gaussian Conditional Random Field (GCRF) model, initially introduced by \citet{lafferty2001conditional}. The GCRF model assumes the conditional density function of $Y$ given $X$ as:

\begin{equation}
\label{eq:GCRF}
    p(Y|X,\Lambda,\Theta) \propto \sqrt{\det(\Lambda)}\exp\left\{-\frac12Y^T\Lambda Y - X^T\Theta Y\right\},
\end{equation}

\noindent where $\Theta$ is
a matrix of dimension $q \times p$ and $\Lambda$ is a $p \times p$ positive definite and symmetric matrix. This model has gained widespread use in the context of estimation of $(\Theta, \Lambda)$, as evident in subsequent works such as \citet{yin2011sparse,wytock2013sparse,yuan2014partial,gan2022bayesian}.

\subsection{Multivariate regression framework for the joint estimation of $(B,\Lambda)$}

The above conditional dependence structure from the GCRF model can be alternatively expressed as a multivariate regression model, also known as covariate-adjusted graphical model, which is commonly used for the joint estimation of $(B,\Lambda)$ \citep{rothman2010sparse, cai2013covariate, consonni2017objective, deshpande2019simultaneous,zhang2022high}. Under the covariate-adjusted graphical model, the conditional distribution of the responses given the covariates can be expressed as

\begin{equation}
\label{eq:covadjgraphmodel}
    Y|X \sim N(BX,\Lambda^{-1}); \quad B = -\Lambda^{-1}\Theta^T,
\end{equation}
 where $B$ is the regression coefficient matrix and $\Lambda$ is the error precision matrix. This model gives us a way to express $B$, as a function of the precision matrices $\Theta$ and $\Lambda$ which serve as our motivation to use the concept of group sparsity as explained in Section 2.2. While both the approaches, estimating $(\Theta,\Lambda)$ to get a sparse $B$, and estimating $(B,\Lambda)$ to get a sparse $\Theta$ are valid methods, the log-likelihood function of the GCRF is convex, while the one from
the multivariate regression is not convex \citep{yuan2014partial}. Thus, we formulate a method for the simultaneous sparse estimation of $(\Lambda,\Theta,B)$ using the GCRF framework.

\subsection{Dependence estimation using group sparsity}
Since we concentrate on

Given our focus on data with numerous irrelevant covariates, we want to obtain estimates for $\Theta$ and $B$ that can identify the relevant variables by setting the parameters associated with irrelevant variables to zero. Mathematically, this means attaining
row-sparse estimates for $\Theta$ and column-sparse estimates for $B$, 
As will be detailed in Section 2.4, the regression coefficient matrix $B$ can be expressed as $-\Lambda^{-1}\Theta^T$. This relation enables us to simultaneously estimate $\Theta$ and $B$ with the required structural sparsity alongside a sparse estimate for $\Lambda$. 
In particular, we can induce row-wise group sparsity on $\Theta$ to ensure that the sparse estimation of $(\Theta,\Lambda)$ results in a column-sparse $B$. Due to the relation $B = -\Lambda^{-1}\Theta^T$, if $\Theta$ has rows consisting of all zeroes, even if $\Lambda^{-1}$ is dense, $B$ will be sparse because a $\textbf{0}$-row in $\Theta$ will cause $B$ to have a $\textbf{0}$-column. This is demonstrated in Figure \ref{fig:GS}. Similarly, we can induce column-wise group sparsity on $B$ to ensure that the sparse estimation of $(B,\Lambda)$ results in a row-sparse $\Theta$. In this work, we induce the structural sparsity in our estimates by choosing appropriate priors. 

\begin{figure}[h]
    \centering
    \includegraphics[scale = 0.3]{Paper plots and sections/Plots/Group Sparsity.png}
    \caption{Demonstration of the use of group sparsity}
    \label{fig:GS}
\end{figure}

Using the concept of group sparsity is particularly useful in our context since a row filled with zeros in the matrix $\Theta$ ($\Theta_{ki} = 0$ for all $i$) implies that the corresponding $X$-variable ($X_{(k)}$) is independent to all the $Y$-variables conditioned on the other $X$'s. Thus, estimating $(B,\Lambda)$ with a column-wise group sparse estimate of $B$ leading to a row-wise group sparse estimate of $\Theta$, helps us identify the covariates that have dependency structures with the responses. Correspondingly, a column filled with zeros in the matrix $B$ ($B_{ik} = 0$ for all $i$) implies that the corresponding $X$-variable ($X_{(k)}$) has no effect on the means of any of the $Y$'s. In other words, this particular covariate ($X_{(k)}$) does not contribute to the conditional expectation of any of the response variables ($Y$) (i.e. $ E(Y_{(i)}|X_{(k)}) = 0$ for all $ i$). Thus, estimating $(\Theta,\Lambda)$ with a row-wise group sparse estimate of $\Theta$ leading to a column-wise group sparse estimate of $B$ facilitates variable selection. 

}

\section{Proposed adaptive Bayesian regularization framework}
We propose a Bayesian regularization framework to sparsely
estimate $(\Theta,\Lambda)$ for the GCRF model such that it leads to group-sparse estimates for $\Theta$ and $B$. In particular, we place appropriate priors on $\Theta$ and $\Lambda$ to induce the desired structural sparsity in the estimates. The prior likelihood is a product of the priors on $\Theta$ and $\Lambda$, $\Pi(\Theta,\Lambda) = \Pi(\Theta)\Pi(\Lambda).$ The negative posterior distribution $L(\Theta,\Lambda)$ corresponding to the prior distribution is then minimized to find the MAP (maximum a posteriori) estimators for $\Theta$ and $\Lambda$:
\begin{equation*}
    \min_{\Theta,\Lambda \succeq 0} L(\Theta,\Lambda).
\end{equation*}
We propose two different approaches for estimating the regression coefficient matrix $B$ from our estimates of $(\Theta, \Lambda)$:
\begin{enumerate}
    \item A natural approach is to use plug-in estimation in the equation $B = -\Lambda^{-1}\Theta^T$. However, when $p$ is large, the error from inverting $\Lambda$, a $p\times p$ matrix, accumulates. 
    
    \item An alternate approach involves dropping the completely irrelevant covariates, identified as $X_i$ if $\hat \Theta_{i.} = \textbf{0}$. Then the remaining covariates are used to train multiple regression models, with each model corresponding to a single response $Y_i$ regressed on the remaining $X$'s. This approach is better suited for scenarios with large $p$.
    
\end{enumerate}
The advantages and disadvantages of using each of these estimates are detailed in our theoretical results in Section 4. We now provide a detailed explanation of each aspect of the prior formulation and estimation process for the optimal $(\Theta,\Lambda)$.

\subsection{Prior formulation}
For our prior set-up, we place a spike and slab LASSO (SSL) prior \citep{rovckova2018bayesian} on $\Lambda$ and a mixture of the SSL prior and only a spike prior on $\Theta$

\begin{align}
\label{eq:prior_dist}
    \Pi(\Theta) &= \prod_i\bigg[\rho\prod_j\bigg( \et LP(\Theta_{ij},\vtone)+(1-\et)LP(\Theta_{ij},\vtzero)\bigg)\nonumber\\& \quad\quad\quad\quad\quad+(1-\rho)\prod_jLP(\Theta_{ij},\vtzero)\bigg],\nonumber\\
    \Pi(\Lambda) &= \prod_{i<j}\bigg(\el LP(\Lambda_{ij},\vlone )+ (1-\el)LP(\Lambda_{ij},\vlzero )\bigg).
\end{align}
To understand the motivation behind the choice of the priors, we present an alternative hierarchical form of our prior setup, which will also be used to formulate our EM algorithm. Consider the row-level binary indicator $r_i^\Theta$ and the element-level binary indicators $r_{ij}^\Theta$ and $r_{ij}^\Lambda$. These indicators serve as latent variables, which provide an alternative way of specifying our priors. The superscript ($\Theta$ or $\Lambda$) denotes the specific parameter to which each indicator corresponds. $r_i^\Theta$ indicates whether a complete row in $\Theta$ is zero ($r_i^\Theta = 0$) or not  ($r_i^\Theta = 1$), thus indicating the row-wise sparsity. $r_{ij}^\Theta$ and $r_{ij}^\Lambda$ indicate the sparsity of the corresponding entries $\Theta_{ij}$'s and $\Lambda_{ij}$'s. The row-level binary indicator ($r_i^\Theta$) induces a hierarchical sparsity structure on $\Theta$ since $r_{ij}^\Theta = 0$ when $r_i^\Theta = 0$, but can take both $0$ and $1$ as values when $r_i^\Theta = 1$.  

\begin{figure}[t]
    \centering
    \includegraphics[scale = 0.3]{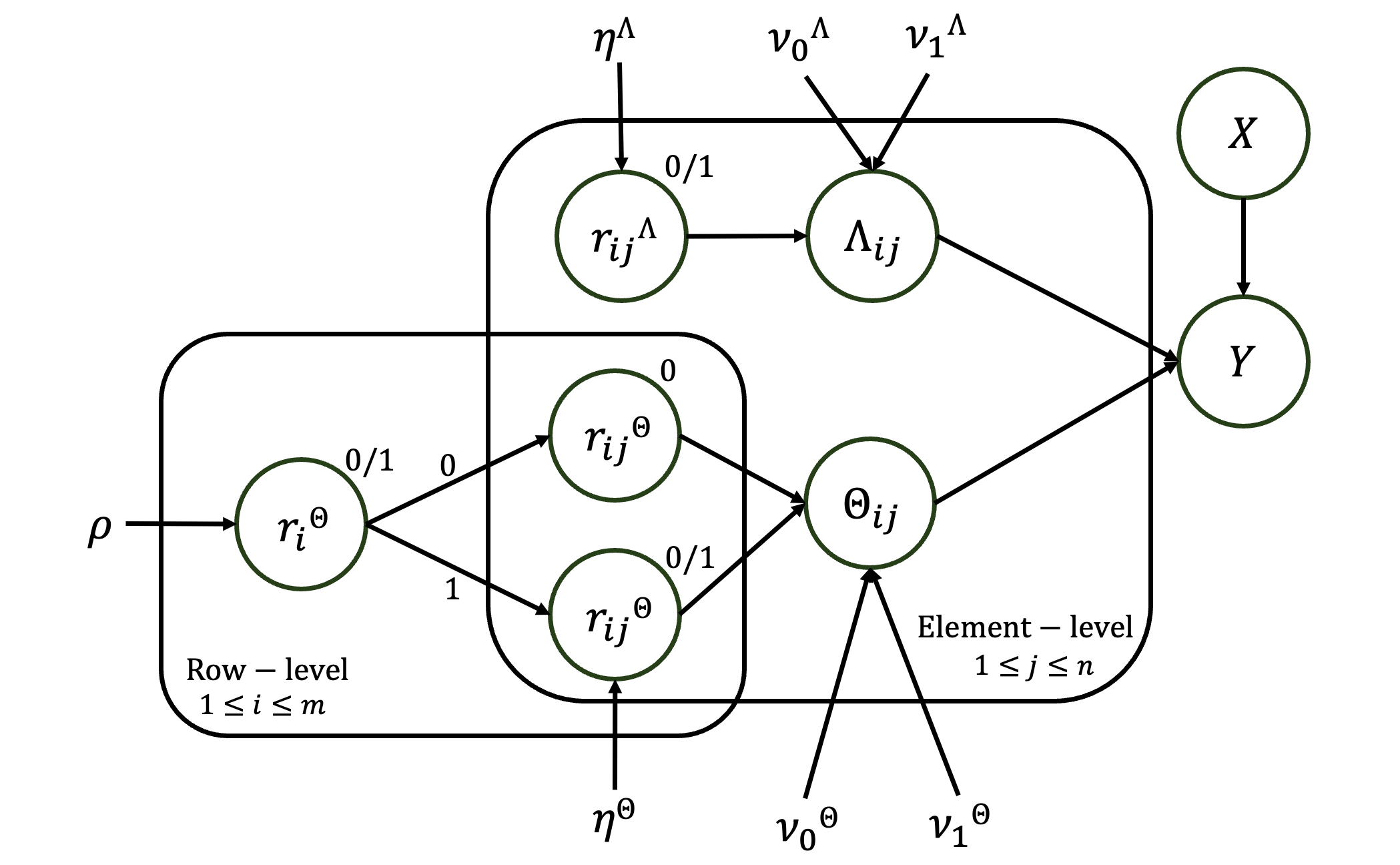}
    \caption{A graphical representation of our prior setup.}
    \label{fig:flowchart}
\end{figure}

We use spike and slab LASSO (SSL) priors, which have a well-established history within the Bayesian variable selection domain \citep{Bai2020SpikeandSlabML}, to achieve the desired sparsity patterns. The ``spike" part of the prior induces values to be close to zero, while the ``slab" part accounts for the possibility that some elements can take non-zero values. In particular, given the binary indicators $r_i^\Theta$, $r_{ij}^\Theta$ and $r_{ij}^\Lambda$, we place a spike prior when the indicator $r_{.}=0$ and a slab prior when $r_{.}=1$.   The diagonal entries of $\Lambda$ are generated from a uniform distribution. Since this leads to the corresponding prior likelihood for the diagonal entries to be independent of the parameters, they do not play a role in the posterior likelihood function, and thus, we avoid the diagonal entries in our notation. A similar hierarchical Bayesian framework has been applied to simultaneously estimate multiple graphical models in \citet{yang2021gembag}. Our prior specification is graphically represented for ease of understanding in Figure \ref{fig:flowchart} and can be summarized as follows: 
\comment{
For our prior set-up, we place a spike and slab LASSO (SSL) prior on $\Lambda$ and a mixture of the SSL prior and only a spike prior on $\Theta$ (Figure \ref{fig:flowchart}). SSL priors have a well-established history within the Bayesian variable selection domain \citep{Bai2020SpikeandSlabML}. Utilizing a spike prior on an entire row of $\Theta$ contributes to achieving group sparsity by effectively encouraging the entire row to converge towards zero. The element-wise sparisty in the remaining rows of $\Theta$ and in $\Lambda$ are achieved using the SSL prior.  Thus the corresponding prior likelihood is a constant in the parameters and does not play a role in the posterior likelihood function. 
}
\begin{align}
\label{eq:priorsetup}
\text{\textbf{For $\Theta$:}} &\quad r_{i}^\Theta \sim Bern(\rho); \quad r_{ij}^\Theta|r_{i}^\Theta =1 \sim Bern(\et); \quad r_{ij}^\Theta|r_{i}^\Theta =0 \sim \delta_0(r_{ij}^\Theta); \nonumber\\
&\quad \quad \quad \Theta_{ij}|r_{ij}^\Theta=1\sim LP(\Theta_{ij},\vtone); \quad \Theta_{ij}|r_{ij}^\Theta=0\sim LP(\Theta_{ij},\vtzero) \nonumber,\\
\text{\textbf{For $\Lambda$:}}&\quad  r_{ij}^\Lambda \sim Bern(\el); \quad \Lambda_{ij}|r_{ij}^\Lambda=1\sim LP(\Lambda_{ij},\vlone);\nonumber\\&\quad \quad \quad \Lambda_{ij}|r_{ij}^\Lambda=0\sim LP(\Lambda_{ij},\vlzero),
\end{align}
where $LP(\Phi,\nu) = \frac{1}{2\nu}\exp\left\{-\frac{|\Phi|}{\nu}\right\}$.  Using this formulation, we get estimates of $\Theta$ and $\Lambda$ by optimizing the posterior log-likelihood. 

\comment{
$$\gamma_i \sim \text{Bern}(\rho);$$
$$\Theta_{ij}|(\gamma_i = 1) \sim \et LP(\Theta_{ij},\vtone) + (1-\et)LP(\Theta_{ij},\vtzero);$$
$$\Theta_{ij}|(\gamma_i = 0) \sim LP(\Theta_{ij},\vtzero);$$
$$\Lambda_{ij} \sim \el LP(\Lambda_{ij},\vlone) + (1-\el)LP(\Lambda_{ij},\vlzero)\; \text{for }i<j;$$
$$\Lambda_{ii} \sim Unif(a,b),$$}

\subsection{Adaptive regularization and MAP estimation of $(\Theta, \Lambda)$}\label{sec3.2}
To get estimates for $(\Theta,\Lambda)$, we use the maximum a posteriori (MAP) estimator, which leads to a penalized likelihood optimization problem with an adaptive penalty. The adaptive nature of the penalty function helps reduce bias in the estimates. Due to the GCRF model (\ref{eq:GCRF}), given a random sample of $n$ observations, $(X_i,Y_i)_{i = 1}^n$, our corresponding log-likelihood function is
\begin{equation}
\label{eq:GCRFlik}
    l(\Theta,\Lambda) = \frac{n}{2}\bigg(\log\det(\Lambda) - tr(S_{yy}\Lambda + 2S_{xy}^T\Theta + \Lambda^{-1}\Theta^TS_{XX}\Theta)\bigg),
\end{equation}
where $S_{yy} = \frac 1n \sum_{i=1}^n Y_iY_i^T$, $S_{xy} = \frac 1n \sum_{i=1}^n X_iY_i^T$ and $S_{xx} = \frac 1n \sum_{i=1}^n X_iX_i^T$. Additionally, using the prior distribution functions defined in (\ref{eq:prior_dist}), we get the negative log-posterior $L(\Theta,\Lambda)$:
\begin{equation}
    \label{eq:postlik}
    L(\Theta,\Lambda) = -l(\Theta,\Lambda) + \sum_i Pen_{\text{MSS}}(\Theta_i) + \sum_{i<j}Pen_{\text{SS}}(\Lambda_{ij}) + const,
\end{equation}
where $\sum_iPen_{\text{MSS}}(\Theta_i) := -\log\Pi(\Theta)$ and $\sum_{i<j}Pen_{\text{SS}}(\Lambda_{ij}):= -\log\Pi(\Lambda)$.
\comment{
\begin{align*}
    \log\Pi(\Theta) &:= -\sum_iPen_{\text{MSS}}(\Theta_i)\\&= \sum_i\log\Bigg[\rho\prod_j\left(\frac{\et}{2\vtone}e^{-\frac{|\Theta_{ij}|}{\vtone}}+\frac{1-\et}{2\vtzero}e^{-\frac{|\Theta_{ij}|}{\vtzero}}\right)+\\&\quad \quad \quad \quad (1-\rho)\prod_j\frac{1}{2\vtzero}e^{-\frac{|\Theta_{ij}|}{\vtzero}}\Bigg].\\
    \log\Pi(\Lambda) &= \sum_{i<j}\underbrace{\log\left(\frac{\el}{2\vlone}\exp\left\{-\frac{|\Lambda_{ij}|}{\vlone}\right\} + \frac{(1-\el)}{2\vlzero}\exp\left\{-\frac{|\Lambda_{ij}|}{\vlzero}\right\}\right)}_{-Pen_{\text{SS}}(\Lambda_{ij})}.
\end{align*}
These expressions lead to the definition of the negative log-posterior $L(\Theta,\Lambda)$:
\begin{equation}
    \label{eq:postlik}
    L(\Theta,\Lambda) = -l(\Theta,\Lambda) + \sum_i Pen_{\text{MSS}}(\Theta_i) + \sum_{i<j}Pen_{\text{SS}}(\Lambda_{ij}).
\end{equation}Here, the likelihood term corresponding to the diagonal elements in $\Lambda$ which equals $\sum_i \log \Pi_{Unif}(\Lambda_{ii})$ does not depend on $\Lambda$ and $\Theta$, and thus it remains constant with respect to these variables. }
With the negative log-posterior defined as above, finding the MAP estimator for $(\Theta,\Lambda)$ is equivalent to solving the optimization problem:
\begin{equation}
    \argmin_{\Theta,\Lambda \succeq 0, \|\Lambda\|_2 <R} L(\Theta,\Lambda).
    \label{eq6}
\end{equation}
Here, we impose two key conditions: $\Lambda \succeq 0$ and $|\Lambda|_2 < R$, where the latter represents a bounded second norm of $\Lambda$. Such constraints have been previously used in \citet{gan2022bayesian}. The first condition is essential as it ensures that $\Lambda$ remains positive definite, a requirement stemming from its role as the inverse of a covariance matrix (\ref{eq:covadjgraphmodel}). 
While the second constraint introduces some limitations to the high-dimensional parameter space, it is important to note that the upper bound $R$ is flexible and can vary depending on the values of $n$, $p$, and $q$. Consequently, it can be set to relatively large values, making this restriction less severe and adaptable to the specific characteristics of the problem at hand.

The minimizer of (\ref{eq6}) has a natural interpretation as the penalized likelihood estimator using the penalty functions which are induced by
the Bayesian SSL priors. 
We will now introduce a proposition that demonstrates the concavity and the adaptive nature of the penalty functions $Pen_{\text{MSS}}(\Theta_i)$ and $Pen_{\text{SS}}(\Lambda_{ij})$ in 
(\ref{eq:postlik}). These findings extend the results regarding the derivatives of the spike and slab prior outlined in Lemma 1 of \cite{rovckova2018spike}. Before we delve into the statement of the proposition, we introduce some terms. Let 
$\eta_1(\phi)$ and $\eta_2(\phi)$ be defined as follows:
\begin{align}
\label{eq:etas}
    \eta_1(\phi) &= \frac{\eta^{\phi} LP\left(\phi,\nu_1\right)}{\eta^{\phi} LP\left(\eta^{\phi},\nu_1^{\phi}\right)+(1-\eta^{\phi}) LP\left(\eta^{\phi},\nu_0^{\phi}\right)},\nonumber\\   \eta_2(\phi) &= \frac{\rho S_1(\phi)}{\rho S_1(\phi) + (1-\rho)S_2(\phi)},
\end{align}
\comment{
Furthermore, the proposition provides explicit forms for the derivatives of the two penalty functions $Pen_{\text{MSS}}(\Theta_i)$ and $Pen_{\text{SS}}(\Lambda_{ij})$. Before we get into the proposition's statement, we introduce some definitions. Let $Z(\Lambda_{ij})$ and $W(\Theta_i)$ be binary random variables defined as 
\begin{equation}
    Z(\Lambda_{ij}) = \begin{cases}
    \frac{1}{\vtone} & \text{w.p.  } \eta_1(\Lambda_{ij})\\
    \frac{1}{\vtzero} & \text{w.p.  } 1-\eta_1(\Lambda_{ij})
\end{cases}, 
\label{eq:defZ}
\end{equation}
and
\begin{equation}
     W(\Theta_i) = \begin{cases}
    \frac{1}{\vtone} & \text{w.p.  } \eta_1(\Theta_{ij})\eta_2(\Theta_i) \\
    \frac{1}{\vtzero} & \text{w.p.  } 1-\eta_1(\Theta_{ij})\eta_2(\Theta_i) 
\end{cases},
\label{eq:defW}
\end{equation}

}
where  $S_1(\phi) = \prod_{j}\left(\eta^{\phi}LP(\phi_j,\nu_1^\phi) + (1-\eta^{\phi})LP(\phi_j,\nu_0^\phi)\right)$ and $S_2(\phi) = \prod_j LP(\phi_j,\nu_0^\phi)$. From the definition of the terms $\eta_1(.)$ and $\eta_2(.)$, it is worth noting that $\eta_1(.)$ gives the probability of an individual element belonging to the ``non-slab" component (or the ``spike" component) and $\eta_2(.)$ gives the probability of a complete row belonging to the ``non-slab" component. The proposition is as follows: 

\begin{prop} \label{prop1} With $\eta_1(\phi)$ and $\eta_2(\phi)$ as defined as in (\ref{eq:etas}), the following holds: 
\begin{itemize}
    \item[(i)] $Pen_{\text{SS}}(\Lambda_{ij})$ and $Pen_{\text{MSS}}(\Theta_i)$ are concave functions and 
    \item[(ii)] The first derivatives of $Pen_{\text{SS}}(\Lambda_{ij})$  and  $Pen_{\text{MSS}}(\Theta_i)$ are given by
    $$\frac{\partial}{\partial|\Lambda_{ij}|} Pen_{\text{SS}}(\Lambda_{ij}) =   \frac{\eta_1(\Lambda_{ij})}{\vlone} + \frac{1-\eta_1(\Lambda_{ij})}{\vlzero},$$
    $$\frac{\partial}{\partial|\Theta_{ij}|}Pen_{\text{MSS}}(\Theta_i) = \frac{\eta_1(\Theta_{ij})\eta_2(\Theta_i)}{\vtone} + \frac{1-\eta_1(\Theta_{ij})\eta_2(\Theta_i)}{\vtzero}.$$
\end{itemize}
\end{prop}
According to \cite{SCAD}, an effective adaptive penalty function should possess the properties of unbiasedness and sparsity. From the proposition above, we observe that the derivative of our penalty functions can be interpreted as a weighted mean of a small penalty $1/\nu_1$ and a large penalty $1/\nu_0$ with the weights $\eta$ and $1-\eta$ respectively representing the conditional probabilities of belonging to the ``spike" and ``slab" components. Consequently, when the parameter ($\Theta$ or $\Lambda$) is large, the derivative of the penalty is small. This means that significant fluctuations in the parameter values have little impact on the level of penalization. Essentially, the existence of the small penalty term reduces the bias due to over-shrinkage. Conversely, when the parameter is small, the derivative of the penalty functions is large. This effectively leads to a thresholding rule, driving certain entries towards zero and eliminating noisy contributions. \comment{This behaviour contributes to the sparsity property of our penalty function. Thus, our penalty function thresholds adaptively, resulting in estimators that are nearly unbiased.} 

\subsection{Optimization via an EM algorithm}
\begin{algorithm}[t]
\caption{EM Algorithm for Optimization of Posterior Log Likelihood}\label{alg:cap}
\begin{algorithmic}
\Require $X, Y$
\Ensure Optimized parameters $\Theta, \Lambda, B$
\State \textbf{Initialize:} $\Theta \gets 0$, $\Lambda \gets I$

\While{(not converged)}
\State \textbf{(E-Step:)}
\State Calculate $p^{\Theta} = \Expect{r^{\Theta}},\; p^{\Lambda} = \Expect{r^{\Lambda}}$ 

\State \textbf{(M-Step:)}
\While{(not converged)}
\State Determine active sets $A_\Theta, A_\Lambda$, using (\ref{thetaactiveset})
\State Compute the Newton updates $\Delta_\Theta, \Delta_\Lambda$ for the elements in the active sets by optimizing (\ref{eq:Mstepopt}). 
\State Set $\Delta_\Theta, \Delta_\Lambda$ = 0 for elements in the non-active sets ($A_{\Theta}^C,A_{\Lambda}^C$)
\State Determine step size $\alpha$ using backtracking line
search
\State Update $\Theta \gets \Theta + \alpha\Delta_\Theta,\;\; \Lambda \gets \Lambda + \alpha\Delta_\Lambda $

\EndWhile
\EndWhile

\If{$p$ is small}
    \State Compute $B = -\Lambda^{-1}\Theta^T$
\Else 
    \State Drop covariate $X_i$ if $\Theta_{i.} = \textbf{0}$. Create a new design matrix $\Tilde{X}$ with only the selected covariates. 
    \State Compute every row of $B$ as $B_{j.}^T = \hat\beta^{OLS}_{j}$, where $\hat\beta^{OLS}_{j}$ is the ordinary least square estimator for the model $$M_j: Y_{.j} = \Tilde{X}\beta_{j}+\varepsilon_j; \quad j = \{1,2,\dots, p\}$$
\EndIf
\end{algorithmic}
\end{algorithm}

To solve the penalized optimization problem defined in (\ref{eq6}), we develop an efficient EM algorithm (Algorithm \ref{alg:cap}), which uses a cyclic coordinate descent method. The proposed EM algorithm for the GCRF model is motivated by similar past works \citep{rovckova2014emvs,wytock2013sparse,rovckova2016fast,gan2022bayesian}. First, we reparametrize the objective function (\ref{eq6}) using the latent variables $r^\Theta_{ij}$ and $r^\Lambda_{ij}$ and the alternative prior setup defined in \ref{eq:priorsetup}. In the E-step of our algorithm, at every iteration of the cyclic coordinate descent, we find estimates $p^\Theta_{ij}$ and $p^\Lambda_{ij}$ for $r^\Theta_{ij}$ and $r^\Lambda_{ij}$ respectively. The expectation of the reparametrized objective function is then computed to derive $Q(\Phi|\Phi^{(t)})$ $(\Phi =( \Theta,\Lambda))$ which is minimized in the M-Step:
\begin{equation}
\label{eq:Mstepopt}
     Q(\Phi|\Phi^{(t)}) = -l(\Phi)+\sum_{i<j}\left(\frac{p_{ij}^{\Phi(t)}}{\nu_1^\Phi}+\frac{1-p_{ij}^{\Phi(t)}}{\nu_0^\Phi} \right)|\Phi_{ij}|+const. 
\end{equation}
The fundamental concept underlying our M-step involves an iterative process where we sequentially construct a second-order approximation to the M-step objective function $Q(\Phi|\Phi^{(t)})$. Subsequently, the cyclic coordinate descent algorithm transforms the problem into a simplified Lasso form. Solving this Lasso problem yields the Newton update step. Due to the sparse nature of our estimated parameters, we make our algorithm faster by only updating the elements in the active set during our coordinate descent instead of updating all the elements. We define the active set for both $\Theta$ and $\Lambda$ as follows:
\begin{equation}\left|\big(\nabla_\Phi l(\Theta,\Lambda)\big)_{ij}\right|> \left|\big(\nabla_\Phi Pen(\Phi)\big)_{ij}\right| \quad \text{ or } \quad \Phi_{ij} \neq 0 \label{thetaactiveset}, \end{equation}
\comment{
\begin{itemize}
    \item For $\Theta$:
    
    \item For $\Lambda$:
    \begin{equation}\left|\big(\nabla_\Lambda l(\Theta,\Lambda)\big)_{ij}\right|> \left|\big(\nabla_\Lambda Pen(\Lambda)\big)_{ij}\right| \quad \text{ or } \quad \Lambda_{ij} \neq 0, \label{lambdaactiveset}\end{equation}
\end{itemize}
}
where $\Phi$ = ($\Theta$,$\Lambda$), $Pen(\Theta) = \sum_i Pen_{\text{MSS}}(\Theta_i)$ and $Pen(\Lambda) = \sum_{i<j} Pen_{\text{SS}}(\Lambda_{ij})$. Using the estimates of $\Lambda$ and $\Theta$ obtained, $B$ is estimated using plug-in when $p$ is small or by training multiple linear regression models when p is large. Further details on the EM algorithm can be found in the Supplementary Materials.

\subsection{Structure Recovery based on Marginal Inclusion Probabilities}
An important aspect of our method is its ability to detect the sparsity patterns and recover the structure of the matrices. We quantify the uncertainty of these sparsity patterns by means of the marginal inclusion probabilities of the parameters. To achieve this, we use the binary indicators $r_{ij}^\Theta$ and $r_{ij}^\Lambda$, as defined in Section 3.1, which serve the purpose of indicating whether the elements within the corresponding matrices $(\Theta,\Lambda)$ are nonzero or not.
\comment{
introduce a reparametrization of the prior configuration using binary indicator matrices denoted by $R^\Lambda = ((r_{ij}^\Lambda))$ and $R^\Theta = ((r_{ij}^\Theta))$. These binary matrices serve the purpose of indicating whether the elements within the corresponding matrices $(\Theta,\Lambda)$ are nonzero or not. The prior reparametrization is as follows:
\begin{align*}
    \text{\textbf{For $\Lambda$:}}&\quad  r_{ij}^\Lambda \sim Bern(\el); \quad \Lambda_{ij}|r_{ij}^\Lambda=1\sim LP(\Lambda_{ij},\vlone);\\&\quad \quad \quad \Lambda_{ij}|r_{ij}^\Lambda=0\sim LP(\Lambda_{ij},\vlzero).\\
\text{\textbf{For $\Theta$:}} &\quad r_{i}^\Theta \sim Bern(\rho); \quad r_{ij}^\Theta|r_{i}^\Theta =1 \sim Bern(\et); \quad r_{ij}^\Theta|r_{i}^\Theta =0 \sim \delta_0(r_{ij}^\Theta);\\
&\quad \quad \quad \Theta_{ij}|r_{ij}^\Theta=1\sim LP(\Theta_{ij},\vtone); \quad \Theta_{ij}|r_{ij}^\Theta=0\sim LP(\Theta_{ij},\vtzero).
\end{align*}
}
The marginal inclusion probability for an element of $\Lambda$ is given by the probability of the corresponding $r_{ij}^\Lambda =1$ which can be computed to be
\begin{align*}
    P(r_{ij}^\Lambda=1|\Lambda) = \eta_1(\Lambda_{ij}).
\end{align*}
Similarly, the marginal inclusion probability for an element of $\Theta$ is given by the probability of the corresponding $r_{ij}^\Theta =1$ which can be computed to be  
\begin{equation*}
    P(r_{ij}^\Theta=1|\Theta) = P(r_i^\Theta =1|\Theta_i)P(r_{ij}^\Theta=1|r_i^\Theta=1,\Theta_{ij}) = \eta_2(\Theta_{i})\eta_1(\Theta_{ij}),
\end{equation*}
with $\eta_1(.)$ and $\eta_2(.)$ as defined in \ref{eq:etas}. From the above equation, we can understand how information is shared across rows due to the induced row-wise sparsity. The marginal inclusion probability for $\Theta_{ij}$ is factored into two parts. The first part given by $P(r_i^\Theta =1|\Theta_i) = \eta_2(\Theta_{i})$ gives information regarding the row-wise inclusion probability and the second part $P(r_{ij}^\Theta=1|r_i^\Theta=1,\Theta_{ij}) = \eta_1(\Theta_{ij})$ gives information regarding the within-row inclusion probability. Thus, the product of the two considers both the row-wise and individual-level sparsity patterns for $\Theta$ while performing sparse structure recovery. Thus, given the MAP estimator for $(\Theta,\Lambda)$ we get from (\ref{eq6}), we can estimate the sparsity pattern by thresholding the posterior inclusion probabilities by some $t \in [0,1]$ as follows:
\begin{equation*}
    r_{ij}^\Lambda =1\iff \eta_1(\Lambda_{ij}) > t; \quad r_{ij}^\Theta =1\iff \eta_1(\Theta_{ij})\eta_2(\Theta_i) > t.
\end{equation*}

This approach facilitates the identification of non-zero elements, aiding in recovering the structural sparsity for the matrices involved in our model.

\section{Theoretical Results}
In this Section, we provide theoretical results to support the effectiveness of our proposed method in recovering the structural sparsity of the parameters and achieving accurate variable selection. Additionally, we offer theoretical guarantees on the convergence of our estimates. Detailed proof of these results can be found in the Supplementary Materials. We begin by defining the notation used in the main results.

\subsubsection*{Notation:} 
To conduct the theoretical studies, we assume that our data $Y$ has been generated based on a fixed set of true parameters $(\Theta^0,\Lambda^0)$ and consider the true data-generating distribution to be sub-Gaussian with 
the random covariate vector $X$ having covariance $\Sigma^0_{xx}$. This is a frequentist data generation mechanism that is quite common in literature \citep{ishwaran2005spike, castillo2015bayesian,narisetty2014bayesian,  gan2019bayesian, gan2022bayesian}. We define the High Posterior Density (HPD) region:
\begin{align*}
HPD &= \{(\Theta,\Lambda): \Pi(\Theta,\Lambda|Data)\geq \Pi(\Theta^0,\Lambda^0|Data)\}\\
&= \{(\Theta,\Lambda): L(\Theta,\Lambda)\leq L(\Theta^0,\Lambda^0)\},
\end{align*}
where $L(\Theta,\Lambda)$ is the negative log posterior defined in (\ref{eq6}). Thus, the HPD region contains all the parameter values with a posterior probability as much as the true value given the data.  Let the signal set for $\Lambda$: $S^\Lambda = \{(i,j): \Lambda_{ij}^0 \neq 0\}$ denote the set of element-wise signals in $\Lambda$, and signal sets for $\Theta$: $S_{row}^\Theta = \{(i,j): \Theta_{i.}^0 \neq \vec{0}\}$ denote the row-wise signals and $S^\Theta = \{(i,j): \Theta_{ij}^0 \neq 0\}$ denote the element-wise signals in $\Theta$. 
This leads us to define the joint sparsity sets $S^0_{row} := S_{row}^\Theta\bigcup S^\Lambda$ and $S^0 := S^\Theta\bigcup S^\Lambda$. Additionally, let $\Tilde{q}$ denote the number of relevant variables in the data: $\Tilde{q} = card(\{i: \Theta^0_{i.} \neq 0\})$. For a matrix $\Phi$, we define $\lambda_{max}(\Phi),\lambda_{min}(\Phi)$ to be the largest and smallest eigenvalues of a matrix $\Phi$. We also define for $\Phi$ the Frobenius norm $\|\Phi\|_F := \sqrt{\sum_i\sum_j|\Phi_{ij}|^2}$, the elementwise max norm $\|\Phi\|_\infty := \max_{i,j}\Phi_{ij}$ and the absolute row sum matrix norm $\vertiii{\Phi}_\infty := \max_{i}\sum_j|\Phi_{ij}|$. Finally, we define the following constants
\begin{equation*}
    c_H = \frac n2\vertiii{H^{-1}_{S^0S^0}}_\infty, \quad c_{\Theta^0} = \vertiii{\Theta^{0^T}}_\infty, \quad c_{\Lambda^0} = \vertiii{(\Lambda^{0})^{-1}}_\infty,
\end{equation*}
where $H$ is the Hessian matrix $\nabla^2l(\Theta,\Lambda)$ and $H^{-1}_{S^0S^0}$ is the submatrix of $H^{-1}$ with the rows and columns indexed by $S_0$.

In addition to the notation above, we assume $\vtzero = \vlzero$, $\vtone = \vlone$, and $\et = \el$ referring to them as $\vone$, $\vzero$, and $\eta$ for simplicity.  The conditions required for the terms corresponding to $\Lambda$ and $\Theta$ are nearly identical, and we will specify any differences when they arise. Our analysis also accommodates the growth of the quantities $(\nu_0, \nu_1, R)$, as well as model sizes $p, q$, and $|S^0_{row}|$, with the sample size $n$. However, for the sake of convenience, we omit the dependence on $n$ in our notation. 

\subsection{Preliminary results on the likelihood function}
Before presenting our theoretical findings, we outline
the key assumptions and properties of our likelihood function necessary for our theoretical results. These assumptions are not uncommon in the Gaussian Conditional Random Fields literature, with similar assumptions being considered in \citet{yuan2014partial} and \citet{gan2022bayesian}.

\noindent \textbf{Assumptions:} Assume that \begin{itemize}

    \item[(a)]  $X$ satisfies the following $s_0$-sparse restricted isometry property condition:
$$\begin{cases}
\inf\left(\frac{u^TS_{xx}u}{u^T\Sigma^0_{xx}u}: \; u\neq 0,\|u\|_0\leq s_0\right) \geq 0.5,\\
\sup\left(\frac{u^TS_{xx}u}{u^T\Sigma^0_{xx}u}: \; u\neq 0,\|u\|_0\leq s_0\right) \leq 1.5,\\
\frac{\lambda_{max}[(\Theta^0)^TS_{xx}\Theta^0]}{\lambda_{max}[(\Theta^0)^T\Sigma^0_{xx}\Theta^0]} \leq 1.4.
\end{cases}$$ with $s_0 = |S^0_{row}| + \lceil4(\rho_2/\rho_1)\alpha^2|S^0_{row}|\rceil$, where $\rho_1 = 0.5\min(\lambda_{max}(\Lambda^0)^{-1},\lambda_{min}(\Sigma^0_{xx}))$ and $\rho_2 = 1.5\lambda_{max}(\Sigma^0_{xx})$
\item[(b)] The sample size $n$ satisfies: $n > (p+s_0)\log(p+q)/\varepsilon_0$ where $\varepsilon_0$ is a constant between $(0,1)$.
\end{itemize}
\citet{candes2007dantzig} shows that in our setting of a sub-Gaussian $X$, with the corresponding $\Sigma_{xx}^0$ having eigenvalues that meet specific regularity conditions, assumption (a) holds with a high probability when assumption (b) is satisfied. Thus, the assumptions are not very restrictive. 

Under the assumptions stated above, we can establish three important properties of the likelihood function, which are crucial for proving our theoretical results. While these properties have been established in prior works, we include them here for completeness and to keep our paper self-contained. The first property is that the derivative of the likelihood function is bounded. Specifically, as a direct consequence of Proposition 4 in \citet{yuan2014partial} we have $\|\nabla l(\Theta^0,\Lambda^0)\|_{\infty} \leq K\sqrt{\log(10(p+q)^2/\varepsilon_0)/n}$ with probability $1-\varepsilon_0$, where $\varepsilon_0$ is any constant in $(0,1)$ and $K$ is a constant that depends on $\Theta^0, \Lambda^0$ and $\Sigma^0_{xx}$.

Moreover, our likelihood function is strongly convex in the HPD region, a result supported by the following properties. Let the local restricted strong convexity (LRSC) constant, a quantity that measures the local curvature of $l(\Phi)$ at $\Phi^0$, be defined as:
\begin{align*}
    \beta(\Phi^0;r,\alpha) = \inf\bigg\{\left\langle\frac{\nabla l(\Phi^0) - \nabla l(\Phi^0 + \Delta),\Delta}{\|\Delta\|^2_2} \right\rangle; \quad \|\Delta\|_2 \leq r, \|\Delta_{(S^0_{row})^C}\|_1 \leq \alpha \|\Delta_{S^0_{row}}\|_1\bigg\}.
\end{align*}
and let \begin{align*}
    r_0 &= min[0.5\lambda_{min}(\Lambda^0),0.13\sqrt{\lambda_{max}[(\Theta^0)^T\Sigma^0_{xx}\Theta^0]/\rho_2}],\\
    \beta_0 &= \left\{\frac{\rho_1}{40\lambda_{max}(\Lambda^0)}\cdot min\left[1,\frac{\lambda_{min}(3\Lambda^0)}{16\lambda_{max}[(\Theta^0)^T\Sigma^0_{xx}\Theta^0]}\right] \right\}.
\end{align*}
Then, under the assumptions stated above, we have the LRSC constant $\beta(\Phi^0;r;\alpha) \geq n\beta_0$ for $r \leq r_0$.  This property
follows from Proposition 3 from \citet{yuan2014partial} and Lemma 2 in \citet{gan2022bayesian}, and gives a bound for the LRSC constant in the cone  $\|\Delta_{(S^0_{row})^C}\|_1 \leq \alpha \|\Delta_{S^0_{row}}\|_1$  showing that the likelihood function is strongly convex within this cone. Finally, using arguments similar to Lemma 3 in \cite{gan2022bayesian} it can be shown that 
if $1/\vone>2\|l(\Theta^0,\Lambda^0)\|_{\infty}$, then for any $(\Theta,\Lambda) = (\Theta^0,\Lambda^0) + \Delta$ so that $(\Theta,\Lambda) \in HPD$, we have 
\begin{equation} \label{eq:cone}
    \|\Delta_{(S^0_{row})^C}\|_1\leq  \alpha \|\Delta_{S^0_{row}}\|_1,
\end{equation} where $\alpha = 1 + 2\vone/\vzero.$
 This confirms that all the points in the HPD belong to the cone (\ref{eq:cone}), thus establishing the strong convexity of the likelihood function in the HPD region.

\subsection{Convergence and sparse structure recovery of the precision matrices $(\Theta, \Lambda)$ }

\subsubsection{Rate of convergence for all the points in the HPD}
 In our first theorem, we demonstrate that every $\Theta$ and $\Lambda$ within the HPD region closely approximates the true parameter with an optimal level of statistical precision in the Frobenius norm. 
\begin{theorem}\label{Thm1}

Let the following conditions hold along with the assumptions stated above:
\begin{enumerate}
    \item[(i)]The hyperparameters $\vone$ and $\vzero$ follow $$\frac{1}{n\vzero}\in\mathcal{O}\left(\sqrt{\frac{\log(p+q)}{n}}\right), \frac{1}{n\vone} \in \mathcal{O}\left(\sqrt{\frac{\log(p+q)}{n}}\right)$$
    and
    $$1/\vone>2\|\nabla l(\Theta^0,\Lambda^0)\|_{\infty};$$
    
    \item[(ii)] $R<\frac{2\lambda_{max}(\Lambda^0)\sqrt{r_0}}{\epsilon_n}$, where R is the matrix norm bound for $\Lambda$ and $\epsilon_n$ is as defined below;
\end{enumerate}
then for any $(\Theta,\Lambda) \in HPD$, with probability at least $1-\epsilon_0$, we have 

    $$\|(\Theta,\Lambda) - (\Theta^0,\Lambda^0)\|_F \leq\epsilon_n:= \frac{C}{\beta_0}\sqrt{\frac{|S^0_{row}|\log(p+q)}{n}} \rightarrow 0,$$ where $\epsilon_0$ is a constant in $(0,1)$. 
\end{theorem}

The above theorem shares notable similarities in its conditions with Theorem 1 in \citet{gan2022bayesian}. Our results also align with \citet{gan2022bayesian} in terms of the convergence rates for $(\Theta,\Lambda)$ that is, if the sample size follows assumption (b), then $\epsilon_n$ goes to zero. A similar result has been demonstrated for the Lasso penalty function by \citet{yuan2014partial} concerning the global optimum for $(\Theta,\Lambda)$. Turning our attention to methods that focus on the estimation of $(B,\Lambda)$, \citet{yin2011sparse} and \citet{cai2013covariate} also provide comparable convergence rates for the global optimum for $\Lambda$. However, our results are advantageous over the latter methods since we show our convergence rates for all points in the HPD, not only for the global optimum. 

\subsubsection{Faster convergence rates and sparsistency for a local optimum}

Our following theorem presents stronger estimation and selection accuracy results, specifically for at least one locally stationary point within the HPD. Additionally, we demonstrate that this optimal stationary point achieves sparsistency, meaning that the parameter estimates are correctly set to zero in places where the true parameter is indeed zero.

\begin{theorem}
\label{Thm2}
    Let the assumptions stated above hold. Then with probability $1-\epsilon_0$ there exists a stationary point $(\hat\Theta,\hat\Lambda) \in HPD$ such that $$\hat\Theta_{(S^\Theta)^C} = \hat\Lambda_{(S^\Lambda)^C} = 0,$$ and     $$\max\left\{\|\hat \Theta - \Theta^0\|_\infty,\|\hat \Lambda - \Lambda^0\|_\infty\right\}\leq r_n := 4C_3c_H\sqrt{\frac{\log(p+q)}{n}},$$where $\epsilon_0$ is a constant in $(0,1)$, if the following conditions hold:
    \begin{enumerate}
    \item[(i)] The hyper-parameters $\vone,\vzero$ and $\rho$ follow condition (i) stated in Theorem \ref{Thm1} along with the following:
    $$\left(\frac{\vone}{\vzero}\right)^{p+2}\frac{(1-\rho)}{\rho(\eta)^p}\leq \epsilon_1(p+q)^{\epsilon_2}; \quad \left(\frac{\vone}{\vzero}\right)^3\frac{(1-\eta)}{\eta}\leq \epsilon_1(p+q)^{\epsilon_2};\quad \frac{1}{n\vzero}>C_4\sqrt{\frac{\log(p+q)}{n}}$$
    where $\epsilon_1,\epsilon_2,C_1,C_3$ and $C_4$ are positive constants with $C_3>C_1$, $C_4>C_1$ and $\epsilon_2 = (C_3 - C_1)c_H(C_4 - C_1)$;
        \item[(ii)] The minimum signal strength for $\Theta$ and $\Lambda$ are both greater than $r_n$ i.e. $\min\{\Theta^0_{min},\Lambda^0_{min}\} > r_n$;
        \item[(iii)] $r_n < \min\bigg\{\frac{2C_4\sqrt{\log(p+q)/n}}{\left(1/2c_H + c_H\right)}\frac{(1-\eta)\vzero}{(1-\eta)\vzero+ \eta\vone},\frac{1}{3dc_{\Lambda^0}},\frac{c_{\Theta^0}}{2d},\frac{1}{7416d^2c_{\Theta^0}^2c_{\Lambda^0}^4\rho_2}\bigg\}$ and $\beta_0 > \frac{C_1^2\log(p+q)}{4(p+q)^{\epsilon_2}}$.
    \end{enumerate}
\end{theorem}
\textit{Remark 1:} If we consider $\vtone \neq \vlone$ and $\vtzero \neq \vlzero$, then the first condition in condition (i) is only required for the parameters corresponding to $\Theta$. 

\textit{Remark 2:}
The estimators for $\Theta,\Lambda$ and $B$ obtained in Theorem \ref{Thm2} are unique if there exists an $m>0$, such that $\beta_0 \geq \left(\frac 14 + \frac p8\right)\frac{C_1^2\log(p+q)}{(p+q)^{\epsilon_2}}+ \frac{m}{2n}$, where $\epsilon_2 = (C_3-C_1)c_H(C_4 - C_1)$ with $C_1,C_3,C_4$ as defined in Theorem \ref{Thm2}.

In the above theorem, condition (i) enables our estimates to achieve adaptive shrinkage by regulating the rate of $1/\nu_0$ and $1/\nu_1$, as explained in Section \ref{sec3.2}. Furthermore, contrasting condition (i) in Theorem \ref{Thm2} with that in Theorem \ref{Thm1}, we observe that Theorem \ref{Thm2} needs $\eta$ to fall strictly between $0$ and $1$, a requirement absent in Theorem \ref{Thm1}. Hence, the Lasso penalty, being a specific case of the spike and slab penalty where $\eta$ is constrained to be either $0$ or $1$, fails to meet the conditions. Condition (ii) is the beta-min condition, commonly used in sparse parameter estimation, requiring the minimum signal strength to be sufficiently large in the true model to help identify the relevant parameters. A construction-based proof for Theorem \ref{Thm2} motivated by similar proofs in \citet{ravikumar2011high,wytock2013sparse,loh2013regularized,gan2019bayesian,gan2022bayesian} is provided in the Supplementary materials. 

The convergence rates for $(\Theta,\Lambda)$ in Theorem 2 are comparable to the results in \citet{wytock2013sparse} and \cite{gan2022bayesian}, though the former method requires the mutual incoherence condition, which is more restrictive than our conditions (see Section 3.3 of \citet{gan2022bayesian} for a comprehensive discussion). While our results for $(\Theta,\Lambda)$ mirror the results from \cite{gan2022bayesian}, our group-sparse setup has the additional advantage of providing column-sparse estimates for $B$, which we elaborate on in Section 4.3.  
Additionally, \citet{yin2011sparse} and \citet{cai2013covariate} provide similar results for their estimates of $\Lambda$. 

\subsection{Convergence and structure recovery for the regression coefficient matrix $B$}
In Section 3, we introduce two different approaches for estimating the regression coefficient matrix from our estimates of $(\Theta, \Lambda)$.
We now present structure recovery and convergence results for both these estimates. Given that our framework does not directly optimize for the best estimate of $B$, achieving perfect convergence rates without any trade-offs is quite challenging. Consequently, we also provide a discussion on the pros and cons of using each of the estimates in terms of their convergence rates. 

\vspace{2mm}
\noindent \textbf{Results for the plug-in estimator:}

The plug-in estimator involves estimating the regression coefficient matrix as $\hat B = -\hat \Lambda^{-1}\hat\Theta^T$ where $(\hat \Theta,\hat \Lambda)$ are the estimates for $(\Theta,\Lambda)$. 
Recall, Theorem \ref{Thm2} proves that the optimal stationary point for $\Theta$ achieves row sparsistency i.e. $\hat \Theta_{S^\Theta_{row}} \neq 0$ and $\hat \Theta_{(S^\Theta_{row})^C} = 0$. As a direct consequence of Theorem \ref{Thm2}, we obtain column sparsistency for the plug-in estimate $\hat B$.  This result is presented in Corollary \ref{cor1}. Proving this corollary is straightforward using the insights on combined group sparsity of $\Theta$ and $B$ provided in Section 2. Note that column sparsistency implies that the totally irrelevant covariates in the true model are assigned entirely zero columns in the estimated regression coefficient matrix. Thus, this result showcases that our method can achieve effective variable selection. 

\begin{corollary} \label{cor1} Let $(\hat \Theta,\hat \Lambda)$ be as defined in Theorem \ref{Thm2}. If the conditions stated in Theorem \ref{Thm2} hold, the plug-in estimate $\hat B :=-\hat \Lambda^{-1}\hat\Theta^T$ has the following property, with probability $1-\epsilon_0$:
 $$\hat B_{(S^B_{col})^C} = 0,$$ where $S_{col}^B$ is the transpose of $S_{row}^\Theta$ denoting the set of column-wise signals in $B^0$ and $\epsilon_0$ is a constant in $(0,1)$.
\end{corollary}
In fact, the above result trivially generalizes for any $\hat B$ estimated using $\hat B = \Phi\hat \Theta^T$, where $\Phi$ is a $p \times p$ matrix and $\hat \Theta$ is an estimate of $\Theta$ that achieves row sparsistency. In addition to the sparsity structure recovery, we also establish some convergence results for the plug-in estimate. However, this estimator requires the inversion of a $p \times p$ matrix, the error of which adds up as $p$ grows with $n$ to infinity. Thus, we present the convergence results for this estimator only under a fixed $p$ scenario.

\begin{corollary}\label{cor2}
    Under a fixed $p$, for any $(\Theta,\Lambda) \in HPD$, if the conditions stated in Theorem \ref{Thm1} hold, the estimate $B := -\Lambda^{-1}\Theta^T$ has 
\begin{equation*}
    \|B - B^0\|_F \leq \frac{F_\Lambda\epsilon_n}{1-F_\Lambda\epsilon_n}(1+F_\Lambda F_\Theta),
\end{equation*}
with probability greater than $1-\epsilon_0$, where $\epsilon_0$ is a constant, $F_\Lambda := \|(\Lambda^0)^{-1}\|_F$ and $F_\Theta:= \|\Theta^0\|_F$.
\end{corollary}

In the above corollary, the Frobenius norm of $\Lambda^{-1}$ grows in the order $\sqrt{p}$ and the Frobenius norm of $\Theta^0 = \mathcal{O}(|S^\Theta|)$. Since we consider a sparse setup, the number of signals present in $\Theta^0$, denoted by $|S^\Theta|$, is expected to be small compared to $n$. Therefore, in the fixed $p$ scenario, both $F_\Lambda$ and $F_\Theta$ are controlled, leading to the bound of the Frobenius norm of the difference between the estimate and the true parameter approaching zero. An advantage of this estimate is that our result encompasses all points within the high probability density (HPD) region, including the global optimum. This contrasts with other methods such as \citet{cai2013covariate}, which focus solely on results for the optimum point. The limitation of this estimate is when $p$ grows with $n$, the bound is too loose, as $F_\Lambda$ is no longer controlled. It is worth mentioning here that while a theoretical convergence result cannot be established in this scenario, our simulations (in Section 5) indicate that the plug-in estimate performs well for reasonably large values of $p$.
Alternatively, where the direct plug-in is not useful, our result for $\hat\Theta$ gives us another effective way of estimating $B$.

\vspace{2mm}
\noindent \textbf{Results for estimating $B$ via multiple linear regression models:}

In this approach, we estimate the regression coefficient matrix by excluding any $X_i$ for which the optimal $\hat\Theta$ from Theorem \ref{Thm2} yields $\hat\Theta_{i.} = \textbf{0}$. Due to the row sparsistency of $\hat \Theta$, we effectively drop variables corresponding to $\textbf{0}$-columns in $B^0$, thus achieving perfect column-structure recovery for $B$. This is equivalent to the result for the plug-in estimate in Corollary \ref{cor1}.

Let $\Tilde{X}_{n\times \Tilde{q}} = (X_1,X_2,\dots,X_{\Tilde{q}})$ represent the design matrix comprising the variables selected through this process. Note that we have exactly $\Tilde{q}$ many variables remaining since we attain perfect variable selection. Subsequently, we estimate $\Tilde{B}$ using $p$ linear regression models, where each model $M_j$ is defined as: 
$$M_j: Y_{.j} = \Tilde{X}\Tilde{B}_{j.}^T+\varepsilon_j; \quad j = \{1,2,\dots, p\}.$$
In particular, every row of $\Tilde{B}$ (denoted as $\Tilde{B}_{j.}$) is estimated using the ordinary least squares estimator for model $M_j$. The estimate for $B$ can then be obtained by appending $q-\Tilde{q}$ many $\textbf{0}$-columns to $\hat{\Tilde{B}}$ corresponding to the irrelevant covariates. That is, $\hat B = \left[\hat{\Tilde{B}}_{\Tilde{q}\times p}: \textbf{0}_{(q-\Tilde{q})\times p}\right]$. The following corollary gives the convergence rate for the estimated $B$. 

\begin{corollary}
    If the conditions for Theorem \ref{Thm2} hold along with the condition $\lambda_{min}(\Tilde{X})>0$, then with probability greater than $1-\mathcal{O}\{(p+q)^{-1}\}$, the estimate $\hat B$ using the above approach satisfies
\begin{equation*}
   \frac{1}{\sqrt{p}} \|\hat B - B^0\|_F \leq \frac{K}{\lambda_{min}(\Tilde{X})}\sqrt{\frac{\Tilde{q}+\log\left(p+q\right)}{n}},
\end{equation*}
 where K is a constant.
\end{corollary}

The above corollary follows from standard results for ordinary least square estimators for regression coefficients \citep{Mit_regr}. Our estimation error for $B$ has the same rate as the bound described in \cite{cai2013covariate}, even though our method does not explicitly optimize for an estimate of $B$. Additionally, while \cite{cai2013covariate} provides convergence results for $B$, they lack similar results for $\Theta$. 
Comparing the two estimation approaches for $B$, note that our results for the plug-in estimator require the conditions for Theorem \ref{Thm1} to hold, whereas the multiple linear regressions estimator requires the conditions for Theorem \ref{Thm2}, which are stronger. Additionally, the results we provide here are solely for the optimal point and not all points in the HPD, unlike the plug-in estimator. Thus, when $p$ is reasonable, using the plug-in approach is preferable.

\comment{
\begin{theorem}\label{Thm5}
 Let $\hat\Theta$ and $\hat\Lambda$ be as defined in Theorem \ref{Thm2}. If the conditions stated in Theorem \ref{Thm2} hold, the diagonal estimate
 $\hat B =\hat D^{-1}\Theta^T$, where $D = diag(\Lambda)$ has with probability $1-\epsilon_0$
\begin{equation*}
    \frac{1}{\sqrt{p}}\|\hat B - B^0\|_F \leq K_1\epsilon_n + K_2\frac{1}{\sqrt{p}}(|S^\Lambda|-p)|S^\Theta|,
\end{equation*}
where $K_1,K_2$ are positive constants, $F_{off}:= \|\Lambda_{off}^0\|_F$ and $\Lambda_{off} := \Lambda - diag(\Lambda)$.
\end{theorem}
}

\comment{
Note that we get the diagonal estimator for $B$ by replacing the term $\hat\Lambda^{-1}$ with $\hat D^{-1}$ in the plug-in estimator, where $\hat D = diag(\hat \Lambda)$.  We approximate $\Lambda^{-1}$ with its diagonal to address the limitation of the plug-in estimator, effectively controlling the estimation error.
However, this introduces additional errors due to the approximation of $\Lambda$ with its diagonal. The bound for the Frobenius norm given in Theorem  \ref{Thm5} can be divided into two components: the estimation error, which is $\mathcal{O}(\epsilon_n)$ and converges to zero, and the approximation error, $K_2\frac{1}{\sqrt{p}}(|S^\Lambda|-p)|S^\Theta|$. Considering our sparse setup, both $|S^\Lambda|-p$, representing the number of signals in the off-diagonal part of $\Lambda^0$, and $|S^\Theta|$, are expected to be small. Hence, as $p$ increases, this approximation error diminishes.

Our estimation error for $B$ when $p$ grows with $n$ has the same rate as the bound described in \cite{cai2013covariate}, even though our method does not explicitly optimize for an estimate of $B$. Additionally, it is important to note that while \cite{cai2013covariate} provides convergence results for $B$, they lack similar results for $\Theta$. Furthermore, our results encompass all points within the high probability density (HPD) region, including the global optimum. In contrast, \cite{cai2013covariate} focuses solely on results for the optimum point. 

As corollaries to theorem \ref{Thm2}, we establish that the estimates of $B$ obtained from the optimal stationary point for $(\Theta,\Lambda)$ also exhibit fast convergence characteristics. Furthermore, they attain column sparsistency, which means that covariates having no impact on the responses in the true model are assigned entirely zero columns in the estimated regression coefficient matrix. While the convergence rates for $(\Theta,\Lambda)$ are the same as in \citet{gan2022bayesian}, the column sparsistency of $B$ is a unique feature of our approach and directly stems from the group sparse structure we employ.

\noindent \textbf{Results for estimation via multiple linear regression models:}}

\comment{

\begin{corollary} \label{cor1}Let $\hat\Theta$ and $\hat\Lambda$ be as defined in Theorem \ref{Thm2}. If the conditions stated in Theorem \ref{Thm2} hold and assuming $p$ to be fixed, with probability $1-\epsilon_0$, $\hat B := -\hat\Lambda^{-1}\hat\Theta^T$ has the following properties:

\begin{itemize}
    \item $\hat B_{(S^B_{col})^C} = 0$, where $S_{col}^B = \{(i,j): B_{.j}^0 \neq \vec{0}\}$ denote the set of column-wise signals in $B^0$;
    \item The max norm of the difference of the estimate and the true value of $B$ is bounded as follows $$\|\hat B - B^0\|_\infty \leq  \left(\frac{pc_{\Lambda}^2r_n}{1-c_{\Lambda}r_n}\right)\left(c_\Theta + r_n\right) + c_\Lambda r_n$$
\end{itemize}
\end{corollary}

\begin{corollary} \label{cor2}Let $\hat\Theta$ and $\hat\Lambda$ be as defined in Theorem \ref{Thm2}. If the conditions stated in Theorem \ref{Thm2} hold and assuming $p$ grows with $n$, with probability $1-\epsilon_0$, $\hat B := -\hat D^{-1}\hat\Theta^T$ where $\hat D = diag(\hat\Lambda)$ has the following properties:

\begin{itemize}
    \item $\hat B_{(S^B_{col})^C} = 0$, where $S_{col}^B = \{(i,j): B_{.j}^0 \neq \vec{0}\}$ denote the set of column-wise signals in $B^0$;
    \item The max norm of the difference of the estimate and the true value of $B$ is bounded as follows $$\|\hat B - B^0\|_\infty \leq K_3r_n + K_4\|\Lambda^0\|_\infty\|\Theta^0\|_\infty$$
\end{itemize}
\end{corollary}

}

\comment{
#---------------------------------------------------------------------------#

The proof of the above lemma can be seen in \citet{gan2022bayesian}

\subsubsection*{Local Strong Convexity}

\begin{proof}

\textbf{Case 1: $\forall \; i: \Theta_0 = 0$ we have $\Theta_i^0 = 0$ }

In this case we have from Lemma \ref{Lemma3}, $\|\Delta_{S_0^C}\|_1 \leq \alpha \|\Delta_{S_0}\|_1$ and thus the proof follows from \citet{gan2022bayesian}

\textbf{Case 2: $\exists \; i: \Theta_i = 0$ but $\Theta_i^0 \neq 0$}

Consider the negative log-likelihood function 

$$l(\Phi) = -\frac{n}{2}\bigg(\log\det(\Lambda) - tr(S_{yy}\Lambda + 2S_{xy}\Theta + \Lambda^{-1}\Theta^TS_{xx}\Theta)\bigg)$$

Note that $tr(S_{yy}\Lambda + 2S_{xy}\Theta + \Lambda^{-1}\Theta^TS_{XX}\Theta) = tr(S_{yy}\Lambda)+tr(2S_{xy}\Theta)+tr(\Lambda^{-1}\Theta^TS_{xx}\Theta)$

Let us define 
\begin{itemize}
    \item[] $S^*_{xy} := S_{xy}$ with column i removed if $i \in S_0^\Theta$
    \item[] $S^*_{xx} := S_{xx}$ with row and column i removed if $i \in S_0^\Theta$
    \item[] $\Theta^* := \Theta$ with the row i removed if $i \in S_0^\Theta$
\end{itemize}

Then it is easy to proof that $S_{xy}\Theta = S^*_{xy}\Theta^*$ and $\Lambda^{-1}\Theta^TS_{xx}\Theta = \Lambda^{-1}(\Theta^*)^TS^*_{xx}\Theta^*$

$$\therefore l(\Phi) = -\frac{n}{2}\bigg(\log\det(\Lambda) - tr(S_{yy}\Lambda + 2S^*_{xy}\Theta^* + \Lambda^{-1}(\Theta^*)^TS^*_{xx}\Theta^*)\bigg)$$

Note that $\Theta^*$ has no $\vec{0}$-rows. Thus we have $\Phi = \Phi^0 + \Delta$ where $\Delta$ lies in the curve $\|\Delta_{S_0^C}\|_1 \leq \alpha \|\Delta_{S_0}\|_1$ and therefore we the proof again follows as above. 

\end{proof}

\subsubsection*{Rate of Convergence for all points in HPD}
\begin{theorem}
Assume Assumption 1 holds with $s_0$ defined as above. If

\begin{enumerate}
    \item The prior hyper-parameters $\nu_0$, $\nu_1$ satisfy:
    $$\frac{2\nabla\|l(\Phi^0)\|_\infty}{n}\leq \frac{1}{n\nu_1} = C_1\sqrt{\frac{\log(p+q)}{n}}; \quad \frac{1}{n\nu_0} = C_0\sqrt{\frac{\log(p+q)}{n}}$$ for some constants $C_0 \geq C1$;
    \item The matrix norm bound R satisfies $R <\frac{2\lambda_{min}(\Lambda^0)\sqrt{r_0}}{\varepsilon_n}$
    \item the sample size $n$ satisfies $n \geq \log(10(p + q)^2/\varepsilon_0)$
\end{enumerate}

then for any $\Phi$ from the HPD region, we have
$$\|\Phi - \Phi^0\|_F \leq \varepsilon_n := \frac{C_0 + C_1}{\beta_0}\sqrt{\frac{|S_0|\log(p+q)}{n}} + \frac{1}{\sqrt{\beta_0}}\sqrt{\frac{2\log(\rho/1-\rho)S_0}{n\sqrt{n}C_1\sqrt{\log(p+q)}}}$$
with probability no less than $1-\varepsilon_0$ where $\varepsilon_0$ is a constant from $(0,1)$.
\end{theorem}

\begin{proof}
By Lemma \ref{Lemma1} we have $\gamma_n = \|\nabla l(\Phi^0)\|_\infty \leq K^*\sqrt{n\log(10(p+q)^2)/\eta)} \approx \sqrt{n\log(p+q)}$

By Lemma \ref{Lemma2} we have $\beta(\Phi^0;r,\alpha) \geq n\beta_0 \quad r \leq r_0$

Let $\Delta \Phi - \Phi^0$. We first assume that $\|\Delta\|^2_F \leq r_0$ which we prove at the end. 

Since $\Phi$ belongs to the HPD we have
$$0 \geq l(\Phi^0 + \Delta) - l(\Phi^0) + Pen(\Phi^0 + \Delta) - Pen(\Phi^0)$$

Now $l(\Phi^0 + \Delta) \geq l(\Phi^0) + (\Phi^0 +\Delta - \Phi^0)\nabla l(\Phi^0) + (\Phi^0 +\Delta - \Phi^0)^T\nabla^2 l(\Phi^0)(\Phi^0 +\Delta - \Phi^0)$ by second order expansion

Thus, $l(\Phi^0 + \Delta) - l(\Phi^0) \geq \langle \Delta,\nabla l(\Phi^0)\rangle + \langle \Delta^2,\nabla^2 l(\Phi^0)\rangle$
$$\implies l(\Phi^0 + \Delta) - l(\Phi^0) \geq \langle \Delta,\nabla l(\Phi^0)\rangle + n\beta_0\|\Delta\|^2_F$$
\begin{align*}
    \therefore 0 &\geq \langle \Delta,\nabla l(\Phi^0)\rangle + n\beta_0\|\Delta\|^2_F + Pen(\Phi^0 + \Delta) - Pen(\Phi^0)\\
    &\geq \langle \Delta,\nabla l(\Phi^0)\rangle + n\beta_0\|\Delta\|^2_F + Pen(\Lambda^0 + \Delta_\Lambda) - Pen(\Lambda^0) + Pen(\Theta^0 + \Delta_\Theta) - Pen(\Theta^0)
\end{align*}

\textbf{Case 1: $\forall \; i: \Theta_0 = 0$ we have $\Theta_i^0 = 0$ }

The proof follows from \citet{gan2022bayesian}

\textbf{Case 2: $\exists \; i: \Theta_i = 0$ but $\Theta_i^0 \neq 0$}

\begin{align*}
    0 &\geq \langle \Delta,\nabla l(\Phi^0)\rangle + n\beta_0\|\Delta\|^2_F + \left(\frac{1}{\nu_1}\right)\|\Delta_{S_0^C}\|_1 - \left(\frac{1}{\nu_0}\right)\|\Delta_{S_0}\|_1 - 2\nu_1k\\
    & \text{ where } k = card\bigg(\{i:\Theta_i = 0\}\bigcap S_0^\Theta\bigg)\log\left(\frac{\rho}{1-\rho}\right)\\
    & \geq -\gamma_n(\|\Delta_{S_0^C}\|_1 + \|\Delta_{S_0}\|_1) + n\beta_0\|\Delta\|^2_F + \left(\frac{1}{\nu_1}\right)\|\Delta_{S_0^C}\|_1 - \left(\frac{1}{\nu_0}\right)\|\Delta_{S_0}\|_1 - 2\nu_1k\\
    &\geq (n\beta_0\|\Delta\|^2_F - 2\nu_1k) + \underbrace{\left(\frac{1}{\nu_1} - \gamma_n\right)\|\Delta_{S_0^C}\|_1}_{>0} - \left(\frac{1}{\nu_0} + \gamma_n\right)\|\Delta_{S_0}\|_1
\end{align*}

$$\implies 0 \geq (n\beta_0\|\Delta\|^2_F - 2\nu_1k) - \left(\frac{1}{\nu_0} + \gamma_n\right)\|\Delta_{S_0}\|_1$$

$$\therefore 2\nu_1k+ \left(\frac{1}{\nu_0} + \gamma_n\right)\|\Delta_{S_0}\|_1 \geq n\beta_0\|\Delta\|^2_F$$

$$\implies \|\Delta\|^2_F \leq \frac{2\nu_1k}{n\beta_0}+ \frac{\left(\frac{1}{\nu_0} + \gamma_n\right)\|\Delta_{S_0}\|_1}{n\beta_0}$$

Note that $\gamma_n + \frac{1}{\nu_0} < \frac{1}{\nu_1} + \frac{1}{\nu_0}$

$$\implies \|\Delta\|^2_F \leq \frac{2\nu_1k}{n\beta_0} + \left(\frac{1}{n\nu_0}+ \frac{1}{n\nu_1}\right)\frac{\|\Delta_{S_0}\|_1}{\beta_0}$$

Using $\|\Delta_{S_0}\|_1 \leq \sqrt{|S_0|}\|\Delta_{S_0}\|_F \leq \sqrt{|S_0|}\|\Delta\|_F$

$$\implies \|\Delta\|_F^2 \leq \frac{2\nu_1}{n\beta_0} + \left(\frac{1}{n\nu_0}+ \frac{1}{n\nu_1}\right)\frac{\sqrt{|S_0|}\|\Delta\|_F}{\beta_0}$$

Solving the above quadratic form we get,

$$\|\Delta\|_F \leq \left(\frac{1}{n\nu_0}+ \frac{1}{n\nu_1}\right)\frac{\sqrt{|S_0|}}{2\beta_0} + \frac{1}{2}\sqrt{\left(\frac{1}{n\nu_0}+ \frac{1}{n\nu_1}\right)^2\frac{|S_0|}{\beta_0^2} + \frac{8\nu_1k}{n\beta_0}}$$

Now since $k = card\bigg(\{i:\Theta_i = 0\}\bigcap S_0^\Theta\bigg)\log\left(\frac{\rho}{1-\rho}\right) \leq |S_0|\log\left(\frac{\rho}{1-\rho}\right)$

\begin{align*}
    \therefore \|\Delta\|_F &\leq \left(\frac{1}{n\nu_0}+ \frac{1}{n\nu_1}\right)\frac{\sqrt{|S_0|}}{2\beta_0} + \frac{1}{2}\sqrt{\left[\left(\frac{1}{\nu_0}+ \frac{1}{\nu_1}\right)^2 + 8\nu_1\log\left(\frac{\rho}{1-\rho}\right)n\beta_0 \right]\frac{|S_0|}{n\beta_0^2}}\\
    &\leq \frac{\sqrt{|S_0|}}{2n\beta_0}\left[\left(\frac{1}{\nu_0}+ \frac{1}{\nu_1}\right) + \sqrt{\left(\frac{1}{\nu_0}+ \frac{1}{\nu_1}\right)^2 + 8\nu_1\log\left(\frac{\rho}{1-\rho}\right)n\beta_0}\right]\\
    &\leq \frac{\sqrt{|S_0|}}{2n\beta_0}\left[2\left(\frac{1}{\nu_0}+ \frac{1}{\nu_1}\right) + 2\sqrt{2}\sqrt{\nu_1\log\left(\frac{\rho}{1-\rho}\right)n\beta_0}\right]\\
    &= \frac{\sqrt{|S_0|}}{n\beta_0}\left[\left(\frac{1}{\nu_0}+ \frac{1}{\nu_1}\right) + \sqrt{2\nu_1\log\left(\frac{\rho}{1-\rho}\right)n\beta_0}\right]\\
    &= \frac{\sqrt{|S_0|}}{\beta_0}\left(\frac{1}{n\nu_0}+ \frac{1}{n\nu_1}\right) + \sqrt{\frac{|S_0|}{\beta_0}}\left(\sqrt{\frac{2\nu_1}{n}\log\left(\frac{\rho}{1-\rho}\right)} \right)
\end{align*}

Using $\frac{1}{n\nu_1} = C_1\sqrt{\frac{\log(p+q)}{n}}; \quad \frac{1}{n\nu_0} = C_0\sqrt{\frac{\log(p+q)}{n}}$ 
\begin{align*}
    \|\Delta\|_F &\leq  \frac{\sqrt{|S_0|}}{\beta_0}\left(\frac{1}{n\nu_0}+ \frac{1}{n\nu_1}\right) + \sqrt{\frac{|S_0|}{\beta_0}}\left(\sqrt{\frac{2\nu_1}{n}\log\left(\frac{\rho}{1-\rho}\right)} \right)\\
    & = \underbrace{\frac{C_0 + C_1}{\beta_0}\sqrt{\frac{|S_0|\log(p+q)}{n}}}_{\text{Term 1}} + \underbrace{\frac{1}{\sqrt{\beta_0}}\sqrt{\frac{2\log(\rho/1-\rho)|S_0|}{n\sqrt{n}C_1\sqrt{\log(p+q)}}}}_{\text{Term 2}}\\
    &:= \varepsilon_n
\end{align*} 

Given $(p+|S_0|)\log(p+q) = O(n) \implies |S_0|\log(p+q) = O(n)$. Thus Term 1 $\rightarrow 0$. 

Term 2 $=\left(\sqrt{\frac{2}{C_1\beta_0}}\right)\left(\sqrt{\frac{\log(\rho/1-\rho)}{\sqrt{n}}}\right)\left(\sqrt{\frac{|S_0|\log(p+q)}{n}}\right)\left(\frac{1}{\log(p+q)^{3/4}}\right)\rightarrow 0$ since when $p$ is large enough i.e. $p \approx 1-1/n \implies \frac{\log(\rho/1-\rho)}{\sqrt{n}} \approx \frac{\log(n-1)}{\sqrt{n}} \rightarrow 0$ and $1/(p+q) \rightarrow 0$

Thus we have $\|\Delta\|_F \leq \varepsilon_n \rightarrow 0$

\noindent \textbf{Proof for $\|\Delta\|_F^2 \leq r_0$ always:}

let $\|\Delta\|_F^2 > r_0$, define $b_n = \frac{2\beta_0\lambda_{max}(\Lambda_0)}{\lambda_{max}(\Lambda)} \geq \frac{2\beta_0\lambda_{max}(\Lambda_0)}{R}$ Then, 

\begin{align*}
    \|\Delta\|_F & \leq \frac{\sqrt{|S_0|}}{b_n}\left(\frac{1}{n\nu_0}+ \frac{1}{n\nu_1}\right) + \sqrt{\frac{|S_0|}{b_n}}\left(\sqrt{\frac{2\nu_1}{n}\log\left(\frac{\rho}{1-\rho}\right)} \right)\\
    &\leq \frac{\sqrt{|S_0|}\lambda_{max}(\Lambda)}{2\beta_0\lambda_{max}(\Lambda_0)}\left(\frac{1}{n\nu_0}+ \frac{1}{n\nu_1}\right) + \sqrt{\frac{|S_0|\lambda_{max}(\Lambda)}{2\beta_0\lambda_{max}(\Lambda_0)}}\left(\sqrt{\frac{2\nu_1}{n}\log\left(\frac{\rho}{1-\rho}\right)} \right)\\
    &< \frac{R}{2\lambda_{min}(\Lambda_0)}\left(\frac{\sqrt{|S_0|}}{\beta_0}\left(\frac{1}{n\nu_0}+ \frac{1}{n\nu_1}\right) \right) + \sqrt{\frac{R}{2\lambda_{min}(\Lambda_0)}}\left(\sqrt{\frac{|S_0|}{\beta_0}\frac{2\nu_1}{n}\log\left(\frac{\rho}{1-\rho}\right)} \right)\\
    &< \frac{\sqrt{r_0}}{\varepsilon_n}\left(\frac{\sqrt{|S_0|}}{\beta_0}\left(\frac{1}{n\nu_0}+ \frac{1}{n\nu_1}\right) \right) + \sqrt{\frac{\sqrt{r_0}}{\varepsilon_n}}\left(\sqrt{\frac{|S_0|}{\beta_0}\frac{2\nu_1}{n}\log\left(\frac{\rho}{1-\rho}\right)} \right)\\
    & \text{ Since }  \varepsilon_n = \frac{\sqrt{|S_0|}}{\beta_0}\left(\frac{1}{n\nu_0}+ \frac{1}{n\nu_1}\right) + \sqrt{\frac{|S_0|}{\beta_0}}\left(\sqrt{\frac{2\nu_1}{n}\log\left(\frac{\rho}{1-\rho}\right)} \right)\\
    &< \sqrt{r_0} + r_0^{1/4}\varepsilon_n^{1/2}
\end{align*}

Thus for large n, $\varepsilon_n \rightarrow 0 \implies \|\Delta\|_F < \sqrt{r_0}$

But we had $\|\Delta\|_F^2 > r_0$ which leads to a contradiction. Thus $\|\Delta\|_F^2 \leq r_0$. 

\end{proof}
}

\section{Experimental Results}
In this Section, we first provide some simulation studies, followed by an analysis of a bike-share data set using our proposed method for Bayesian variable selection for graphical models (denoted by BVS.GM).  In all the experimental studies, we estimate $B$ using the plug-in estimator, as it demonstrates good convergence results even when $p$ is reasonably large.

\subsection{Simulation Studies}
We conduct a comparative analysis of various methods against BVS.GM in terms of estimation and structure recovery. The methods under comparison include Graphical LASSO (or GLASSO), which was introduced in \citet{friedman2008sparse}, CAPME, which was introduced in \citet{cai2013covariate}, and Bayes CRF (or BCRF), which was introduced in \citet{gan2022bayesian}. For our simulation setup, we generate the covariate matrix X from a zero-mean multivariate normal distribution. We construct X with a sparse tri-diagonal Toeplitz precision matrix $\Omega_{xx} := Toeplitz(0.3,1,0.3)$. This means that the diagonal of $\Omega_{xx}$ is set to 1, while the sub-diagonal and super-diagonal elements are set to 0.3. The true parameters of the conditional random field, denoted as $(\Theta^0,\Lambda^0)$, are generated with high sparsity as follows:

\textbf{$\Lambda^0$:} We create this matrix to contain a fixed number of non-zero elements ($s_{\Lambda^0}$) in the upper triangular off-diagonal section. The non-zero elements are generated from a uniform distribution. Since the matrix is symmetric, we generate values only for the upper-triangular off-diagonal part. To ensure that the matrix remains positive semi-definite, we set the diagonal elements to be 0.2 greater than the sum of the off-diagonal elements in their respective rows, i.e., $\Lambda^0_{ii} = \sum_{j\neq i}|\Lambda^0_{ij}| + 0.2$.

\textbf{$\Theta^0$:} We design this matrix to have a $70\%$ percent of the rows entirely filled with zeros.  The zero rows ensure that some covariates $X$ exist that do not impact the response. For the remaining rows, we generate the non-zero elements using two different methods. The first method involves sampling a fixed number of non-zero elements ($s_{\Theta^0}$) independently from a uniform distribution. Consequently, each non-zero element is treated as independent, with the maximum signal strength being the largest magnitude among the non-zero elements. In the second method, we use a randomized approach to determine the number of non-zero elements in each row, randomly sampled from $Unif(0.1*p,0.5*p)$. Then, these non-zero elements are generated from a uniform sphere with a constant norm. For example, consider a scenario with $p = 10$ where the number of non-zero elements in a row is randomly determined to be 3. In this case, the non-zero elements are generated from a 3-dimensional uniform sphere with a fixed norm. Here, the fixed norm determines the signal strength, and the elements within a row are no longer treated as independent.

Finally, we generate the response variable $Y$ given $X$ using the covariate-adjusted graphical model, specifically, $Y \sim N(B^0X,(\Lambda^0)^{-1})$, where $B^0 = -(\Lambda^0)^{-1}(\Theta^0)^T$ represents the true regression coefficient matrix. In the case of lower-dimensional simulation scenarios, we consider seven distinct values for the number of observations, $N = (20, 40, 100, 200, 500, 1000, 10000)$. For the higher-dimensional simulation scenarios, we explore four options for $N = (100, 500, 1000, 2000)$. The reported results are obtained as averages over $20$ replications for each specific value of N. The performance metrics we consider are: 
\begin{itemize}
    \item Estimation error: Using the Frobenius norm of the difference between the estimated parameter and the true parameter. 

    \item Structure recovery: Using MCC (Matthews Correlation Coefficient) which is defined as $$MCC = \frac{(TP*TN)-(FP*FN)}{\sqrt{(TP+FP)(TP+FN)(TN+FP)(TN+FN)}}$$ where TP denotes the true positives, TN denotes the true negatives, FP denotes the false positives and FN denotes the false negatives. 

    \item Column recovery for B: Using MCC for the columns where $0$ denotes a column which is fully 0, and $1$ when the column contains any non-zero element. 
\end{itemize}

We present the results for three simulation scenarios that we believe most effectively showcase the strengths of our method. Additionally, we have conducted numerous other simulations to assess the performance of our method in different setups, and these results are provided in the Supplementary Materials. 

\subsubsection{Set-up 1: Low signal strength, low dimensional setting with independent non-zero elements in Theta} In this setup, we consider the number of responses $p = 10$, the number of covariates $q = 50$, the number of non-zero elements in the non-zero rows of $\Theta^0$ ($s_{\Theta^0}$) = 10 and the number of non-zero elements in $\Lambda^0$ ($s_{\Lambda^0}$) = 5. This implies that $\Lambda^0$ has $10$ non-zero off-diagonal elements but since it is symmetric, only $5$ are uniquely estimated. $\Lambda^0$ also has p=10 non-zero diagonal elements. The non-zero elements for both $\Theta^0$ and $\Lambda^0$ are generated from $Unif\left([-0.2,-0.1]\cup [0.1,0.2]\right)$. Thus, the minimum signal strength is 0.2. The results for this setup are tabulated in tables \ref{tab:setup1err},\ref{tab:setup1mcc} and \ref{tab:setup1mccb}.

\subsubsection{Set-up 2: Low signal strength, low dimensional setting with non-
zero elements dependent in Theta} In this setup, we once again consider $p = 10$, $q = 50$ and $s_{\Lambda^0}$ = 5. The non-zero elements for $\Lambda^0$ are still generated from $Unif\left([-0.2,-0.1]\cup [0.1,0.2]\right)$. But instead of the non-zero elements in $\Theta^0$ being independent, they are generated from a uniform sphere with norm = 0.5 using the method described above. Thus, these non-zero elements are dependent and have a low signal strength since the norm of a row is fixed to be less than $0.5$. The results for this setup are tabulated in tables \ref{tab:setup2err},\ref{tab:setup2mcc} and \ref{tab:setup2mccb}.

\subsubsection{Set-up 3: Low signal strength, high dimensional setting with non-
zero elements dependent in Theta} This setup is a higher dimensional analogue of the previous setup. Here, we consider the number of responses $p = 50$, the number of covariates $q = 100$ and the number of non-zero elements in $\Lambda$ ($s_{\Lambda^0}$) = 100. The non-zero elements in $\Theta^0$ are generated from a uniform sphere with norm = 0.5. The results for this setup are tabulated in tables \ref{tab:setup3err},\ref{tab:setup3mcc} and \ref{tab:setup3mccb}.

\subsubsection{Discussion on the simulation results}
The tabulated results from our simulation experiments clearly demonstrate the superiority of BVS.GM across a majority of settings. Even in cases where BVS.GM does not emerge as the absolute best, its performance remains quite close to the top-performing method. Notably, when evaluating the structure recovery capabilities, BVS.GM excels in both element-wise and column-selection aspects, outperforming the competition by a significant margin in most settings. This showcases the effectiveness of our method in recovering relevant covariates and conducting precise variable selection.
It is worth noting that CAPME, which exclusively focuses on estimating the regression coefficient matrix $B$, exhibits nearly perfect performance when the number of observations is very large. However, its selection consistency drops significantly when the number of observations is less than or comparable to the number of covariates. In the high-dimensional setting, these results become even more pronounced.

\begin{table}[htp]
\centering
{\footnotesize	
\begin{tabular}{|c|l|lllllll|}
\hline
                            & \multicolumn{1}{c|}{}                         & \multicolumn{7}{c|}{Error (Frobenius Norm)}                                                                                                                                                                                                                                           \\ \cline{3-9} 
\multirow{-2}{*}{} & \multicolumn{1}{c|}{\multirow{-2}{*}{Method/N}} & 20                                & 40                                & 100                               & 200                               & 500                               & 1000                              & 10000                             \\ \hline
                            & BVS.GM                                        & {\color[HTML]{000000} \textbf{0.500}} & {\color[HTML]{000000} \textbf{0.500}} & {\color[HTML]{000000} \textbf{0.388}} & {\color[HTML]{000000} \textbf{0.327}} & {\color[HTML]{000000} \textbf{0.138}} & {\color[HTML]{000000} \textbf{0.076}} & {\color[HTML]{000000} \textbf{0.028}} \\ \cline{2-2}
                            & GLASSO                                        & 0.574                                 & 0.577                                 & 0.428                                 & 0.413                                 & 0.239                                 & 0.197                                 & 0.080                                 \\ \cline{2-2}
                            & CAPME                                         & 3.268                                 & 3.326                                 & 1.486                                 & 0.961                                 & 0.379                                 & 0.198                                 & 0.066                                 \\
\hline

\end{tabular}}
\caption{Simulation results for the estimation accuracy of the different methods under Setup 1.}
\label{tab:setup1err}
\vspace{5mm}

{\footnotesize
\begin{tabular}{|c|l|lllllll|}
\hline
                            & \multicolumn{1}{c|}{}                         & \multicolumn{7}{c|}{Selection Consistency (MCC)}                                                                                                                                                                                                                                      \\ \cline{3-9} 
\multirow{-2}{*}{} & \multicolumn{1}{c|}{\multirow{-2}{*}{Method/N}} & 20                                & 40                                & 100                               & 200                               & 500                               &1000                              &10000                             \\ \hline
                            & BVS.GM                                        & {\color[HTML]{000000} 0.060}          & {\color[HTML]{000000} 0.152}          & {\color[HTML]{000000} \textbf{0.433}} & {\color[HTML]{000000} \textbf{0.588}} & {\color[HTML]{000000} \textbf{0.650}} & {\color[HTML]{000000} \textbf{0.690}} & {\color[HTML]{000000} \textbf{0.624}} \\ \cline{2-2}
                            & GLASSO                                        & {\color[HTML]{000000} \textbf{0.124}} & 0.136                                 & 0.242                                 & 0.175                                 & 0.272                                 & 0.408                                 & 0.130                                 \\ \cline{2-2}
                            & CAPME                                         & 0.037                                 & 0.011                                 & 0.002                                 & 0.000                                 & 0.005                                 & 0.026                                 & 0.335                                 \\
\hline
\end{tabular}}
\caption{Simulation results for the element-wise structure recovery of the different methods under Setup 1.}
\label{tab:setup1mcc}

\vspace{5mm}

{\footnotesize
\begin{tabular}{|l|lllllll|}
\hline
\multicolumn{1}{|c|}{}                   & \multicolumn{7}{c|}{Column Selection (MCC)}                                                                                                                                                                                            \\ \cline{2-8} 
\multicolumn{1}{|c|}{\multirow{-2}{*}{Method/N}} & 20                                & {\color[HTML]{000000} 40}         & 100                               & 200                               & 500                               & 1000                              & 10000                             \\ \hline
Our                                      & {\color[HTML]{000000} \textbf{0.058}} & {\color[HTML]{000000} \textbf{0.187}} & {\color[HTML]{000000} \textbf{0.422}} & {\color[HTML]{000000} \textbf{0.542}} & {\color[HTML]{000000} \textbf{0.598}} & {\color[HTML]{000000} \textbf{0.624}} & 0.582                                 \\
GLASSO                                   & {\color[HTML]{330001} 0.161}          & {\color[HTML]{000000} 0.062}          & 0.159                                 & 0.000                                 & 0.163                                 & 0.313                                 & 0.000                                 \\
CAPME                                    & 0.053                                 & {\color[HTML]{000000} 0.000}          & 0.000                                 & 0.000                                 & 0.000                                 & 0.096                                 & {\color[HTML]{000000} \textbf{1.000}} \\
BCRF                                     & {\color[HTML]{000000} 0.134}          & {\color[HTML]{000000} 0.204}          & {\color[HTML]{000000} 0.285}          & {\color[HTML]{000000} 0.375}          & {\color[HTML]{000000} 0.465}          & 0.435                                 & 0.275                                 \\ \hline
\end{tabular}}
\caption{Simulation results for the column structure recovery for $B$ of the different methods under Setup 1.}
\label{tab:setup1mccb}
\end{table}

\begin{table}[htp]
\centering
{\footnotesize
\begin{tabular}{|c|l|lllllll|}
\hline
                            & \multicolumn{1}{c|}{}                         & \multicolumn{7}{c|}{Error (Frobenius Norm)}                                                                                                                                                                                                                                           \\ \cline{3-9} 
\multirow{-2}{*}{} & \multicolumn{1}{c|}{\multirow{-2}{*}{Method/N}} & 20                                & 40                                & 100                               & 200                               & 500                               & 1000                              &10000                             \\ \hline
                            & BVS.GM                                        & {\color[HTML]{000000} 1.406}          & {\color[HTML]{000000} 1.202}          & {\color[HTML]{000000} 0.943}          & {\color[HTML]{000000} \textbf{0.486}} & {\color[HTML]{000000} \textbf{0.271}} & {\color[HTML]{000000} \textbf{0.187}} & {\color[HTML]{000000} \textbf{0.058}} \\ \cline{2-2}
                            & GLASSO                                        & {\color[HTML]{000000} \textbf{1.345}} & 1.123                                 & 0.846                                 & 0.734                                 & 0.730                                 & 0.464                                 & 0.169                                 \\ \cline{2-2}
                            & CAPME                                         & 3.753                                 & 5.691                                 & 2.276                                 & 1.312                                 & 0.535                                 & 0.337                                 & 0.179                                 \\
\hline
\end{tabular}}
\caption{Simulation results for the estimation accuracy of the different methods under Setup 2.}
\label{tab:setup2err}
\vspace{5mm}
{\footnotesize
\begin{tabular}{|c|l|lllllll|}
\hline
                            & \multicolumn{1}{c|}{}                         & \multicolumn{7}{c|}{Selection Consistency (MCC)}                                                                                                                                                                                                                                      \\ \cline{3-9} 
\multirow{-2}{*}{} & \multicolumn{1}{c|}{\multirow{-2}{*}{Method/N}} & 20                                & 40                                & 100                               &200                               & 500                               & 1000                              & 10000                             \\ \hline
                            & BVS.GM                                        & {\color[HTML]{000000} \textbf{0.221}} & {\color[HTML]{000000} \textbf{0.374}} & {\color[HTML]{000000} \textbf{0.550}} & {\color[HTML]{000000} \textbf{0.696}} & {\color[HTML]{000000} \textbf{0.756}} & {\color[HTML]{000000} \textbf{0.808}} & {\color[HTML]{000000} \textbf{0.848}} \\ \cline{2-2}
                            & GLASSO                                        & {\color[HTML]{000000} 0.144}          & 0.191                                 & 0.190                                 & 0.231                                 & 0.346                                 & 0.281                                 & 0.233                                 \\ \cline{2-2}
                            & CAPME                                         & 0.101                                 & 0.011                                 & 0.013                                 & 0.000                                 & 0.004                                 & 0.053                                 & 0.438                                 \\
\hline
\end{tabular}}
\caption{Simulation results for the element-wise structure recovery of the different methods under Setup 2.}
\label{tab:setup2mcc}
\vspace{5mm}
{\footnotesize
\begin{tabular}{|l|lllllll|}
\hline

\multicolumn{1}{|c|}{\multirow{0}{*}{Method/N}}                   & \multicolumn{7}{c|}{Column Selection (MCC)}                                                                                                                                                                                            \\ \cline{2-8} 
\multicolumn{1}{|c|}{\multirow{-2}{*}{}} &20                                & {\color[HTML]{000000} 40}         & 100                               & 200                               & 500                               & 1000                              & 10000                             \\ \hline
Our                                      & {\color[HTML]{000000} \textbf{0.219}} & {\color[HTML]{000000} \textbf{0.420}} & {\color[HTML]{000000} \textbf{0.622}} & {\color[HTML]{000000} \textbf{0.708}} & {\color[HTML]{000000} \textbf{0.690}} & {\color[HTML]{000000} \textbf{0.778}} & 0.801                                 \\
GLASSO                                   & 0.112                                 & {\color[HTML]{000000} 0.136}          & 0.020                                 & 0.000                                 & 0.230                                 & 0.000                                 & 0.000                                 \\
CAPME                                    & 0.183                                 & {\color[HTML]{000000} 0.014}          & 0.000                                 & 0.000                                 & 0.000                                 & 0.000                                 & {\color[HTML]{000000} \textbf{1.000}} \\
BCRF                                     & {\color[HTML]{000000} 0.215}          & {\color[HTML]{000000} 0.388}          & {\color[HTML]{000000} 0.530}          & {\color[HTML]{000000} 0.668}          & {\color[HTML]{000000} 0.645}          & 0.624                                 & 0.405                                 \\ \hline
\end{tabular}}
\caption{Simulation results for the column structure recovery for $B$ of the different methods under Setup 2.}
\label{tab:setup2mccb}
\end{table}

\begin{table}[htp]
\centering{\footnotesize
\begin{tabular}{|c|l|llll|}
\hline
                            & \multicolumn{1}{c|}{}                         & \multicolumn{4}{c|}{{\color[HTML]{333333} Error   (Frobenius Norm)}}                                                                                                                                                         \\ \cline{3-6} 
\multirow{-2}{*}{} & \multicolumn{1}{c|}{\multirow{-2}{*}{Method/N}} & \multicolumn{1}{l|}{100}                               & \multicolumn{1}{l|}{500}                               & \multicolumn{1}{l|}{ 1000}                              & 2000                              \\ \hline
                            & BVS.GM                                        & \multicolumn{1}{l|}{{\color[HTML]{333333} 2.313}}          & \multicolumn{1}{l|}{{\color[HTML]{333333} 1.423}}          & \multicolumn{1}{l|}{{\color[HTML]{000000} \textbf{0.773}}} & {\color[HTML]{000000} \textbf{0.503}} \\
                            & {\color[HTML]{333333} GLASSO}                 & \multicolumn{1}{l|}{{\color[HTML]{000000} \textbf{2.269}}} & \multicolumn{1}{l|}{{\color[HTML]{000000} 1.516}}          & \multicolumn{1}{l|}{1.412}                                 & 0.932                                 \\
                            & {\color[HTML]{333333} CAPME}                  & \multicolumn{1}{l|}{{\color[HTML]{333333} 16.594}}         & \multicolumn{1}{l|}{{\color[HTML]{333333} 2.055}}          & \multicolumn{1}{l|}{1.123}                                 & 0.835                                 \\
\multirow{-4}{*}{Theta}     & BCRF                                          & \multicolumn{1}{l|}{{\color[HTML]{000000} 2.342}}          & \multicolumn{1}{l|}{{\color[HTML]{000000} \textbf{1.203}}} & \multicolumn{1}{l|}{0.874}                                 & 0.634                                 \\
\hline
\end{tabular}}
\caption{Simulation results for the estimation accuracy of the different methods under Setup 3.}
\label{tab:setup3err}
\vspace{5mm}
{\footnotesize
\begin{tabular}{|c|l|llll|}
\hline
                            & \multicolumn{1}{c|}{}                         & \multicolumn{4}{c|}{Error   (Frobenius Norm)}                                                                                                                                                                                \\ \cline{3-6} 
\multirow{-2}{*}{} & \multicolumn{1}{c|}{\multirow{-2}{*}{Method/N}} & \multicolumn{1}{l|}{100}                               & \multicolumn{1}{l|}{500}                               & \multicolumn{1}{l|}{1000}                              &2000                              \\ \hline
                            & BVS.GM                                        & \multicolumn{1}{l|}{{\color[HTML]{000000} \textbf{0.257}}} & \multicolumn{1}{l|}{{\color[HTML]{000000} \textbf{0.588}}} & \multicolumn{1}{l|}{{\color[HTML]{000000} \textbf{0.700}}} & {\color[HTML]{000000} \textbf{0.785}} \\
                            & GLASSO                                        & \multicolumn{1}{l|}{{\color[HTML]{000000} 0.214}}          & \multicolumn{1}{l|}{{\color[HTML]{000000} 0.330}}          & \multicolumn{1}{l|}{0.510}                                 & 0.149                                 \\
                            & CAPME                                         & \multicolumn{1}{l|}{{\color[HTML]{000000} 0.000}}          & \multicolumn{1}{l|}{{\color[HTML]{000000} 0.000}}          & \multicolumn{1}{l|}{0.000}                                 & 0.012                                 \\
\multirow{-4}{*}{Theta}     & BCRF                                          & \multicolumn{1}{l|}{{\color[HTML]{000000} 0.219}}          & \multicolumn{1}{l|}{{\color[HTML]{000000} 0.534}}          & \multicolumn{1}{l|}{0.632}                                 & 0.699                                 \\
\hline
\end{tabular}}
\caption{Simulation results for the element-wise structure recovery of the different methods under Setup 3.}
\label{tab:setup3mcc}
\vspace{5mm}
{\footnotesize
\begin{tabular}{|l|llll|}
\hline
                         & \multicolumn{4}{c|}{Column Selection MCC}                                                                                                                     \\ \cline{2-5} 
\multirow{-2}{*}{Method/N} & 100                               & {\color[HTML]{000000} 500}        & 1000                              & 2000                              \\ \hline
Our                      & {\color[HTML]{000000} \textbf{0.453}} & {\color[HTML]{000000} \textbf{0.617}} & {\color[HTML]{000000} \textbf{0.681}} & {\color[HTML]{000000} \textbf{0.695}} \\
GLASSO                   & 0.067                                 & {\color[HTML]{000000} 0.000}          & 0.066                                 & 0.000                                 \\
CAPME                    & 0.000                                 & {\color[HTML]{000000} 0.000}          & 0.000                                 & 0.000                                 \\
BCRF                     & 0.449                                 & 0.483                                 & 0.488                                 & 0.462                                 \\ \hline
\end{tabular}}
\caption{Simulation results for the column structure recovery for $B$ of the different methods under Setup 3.}
\label{tab:setup3mccb}
\end{table}
\subsection{Application to Capital Bikeshare Dataset}

\begin{figure}[h]
    \centering
    \includegraphics[scale = 0.3]{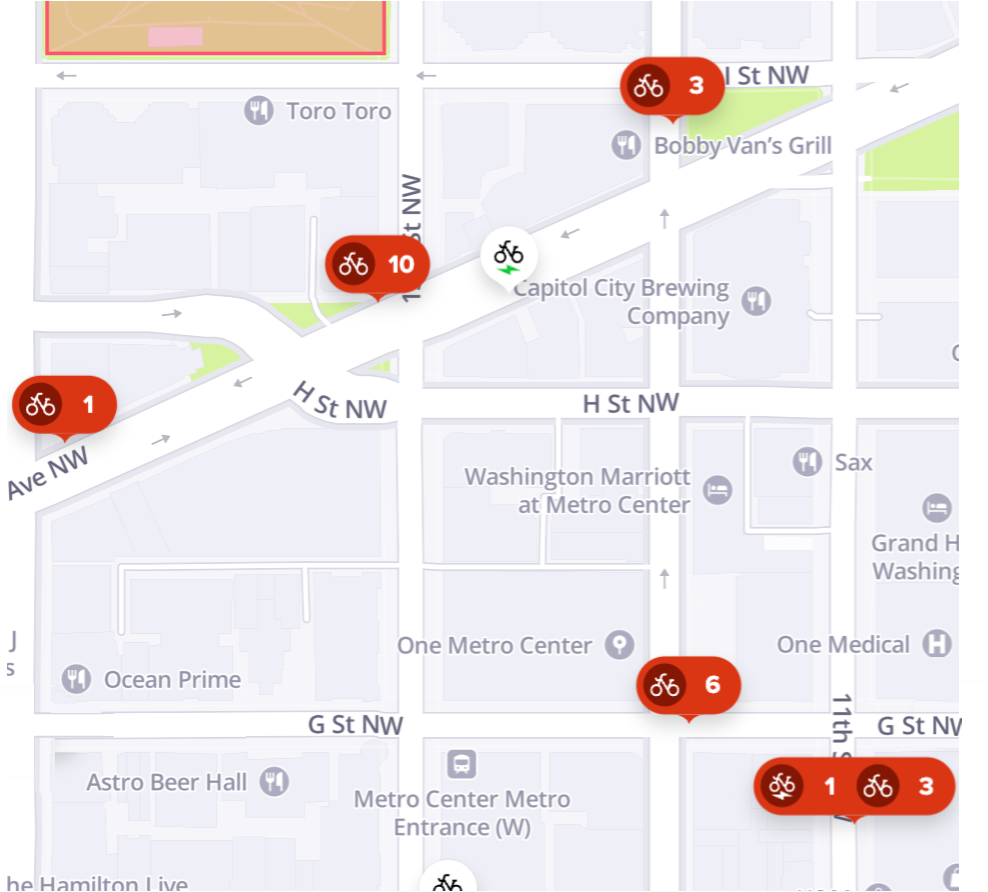}
     \caption{Locations of the Capital Bikeshare bike stations that are considered for our analysis.}
    \label{fig:loccapbike}
\end{figure}

We apply our approach to analyze the capital bike-share dataset and forecast daily bike demand across five nearby bike stations during three distinct time frames: morning, afternoon, and evening. Capital Bikeshare\footnote{Data available at \url{https://capitalbikeshare.com/system-data}.} is Washington DC’s bike-share system, with 700+ stations and 6,000 bikes across the metro area. Our analysis focuses on five specific bike stations: Metro Center / 12th \& G St NW, 14th St \& New York Ave NW, 14th \& G St NW, 13th St \& New York Ave NW, and 11th \& F St NW (see figure \ref{fig:loccapbike})\footnote{Image taken from \url{https://capitalbikeshare.com/\#homepage_map}.}. We predict bike demand on a particular day for each of these stations across the three time periods - morning (before 12 p.m.), afternoon (12 p.m. - 5 p.m.) and evening (after 5 p.m.). 

We are motivated to apply our method to this dataset because of the dynamic nature of bike-sharing systems. The system allows customers to rent bikes from any bike-sharing station using an app and then return the bike to any bike-sharing station at their convenience. This flexibility requires the company managing the bike-sharing system to relocate bikes between stations in response to demand fluctuations. Accurately predicting this demand can greatly enhance the efficiency of these systems. Additionally, customers seeking to rent bikes tend to go towards nearby stations with available bikes, when one station has limited availability. Consequently, the demand for bikes at one station is closely intertwined with the demand at nearby stations, creating a high degree of correlation. Our approach accounts for these interdependencies to provide more effective predictions for optimizing bike-sharing operations.

As mentioned earlier, our response variables correspond to the demand for bikes at the five different bike stations across three time periods on a given day $(T)$, resulting in a total of $p = 15$ responses. The covariates used in our experiment are the bike demand at these five stations during the same time periods over the last three days $(T-1, T-2, T-3),$ giving us $q = 45$ covariates. Our goal is to get accurate predictions by leveraging the information we get from the conditional dependency structure between the responses and between the responses and the past demand. 

\begin{figure}
    \centering
    \includegraphics[scale = 0.15]{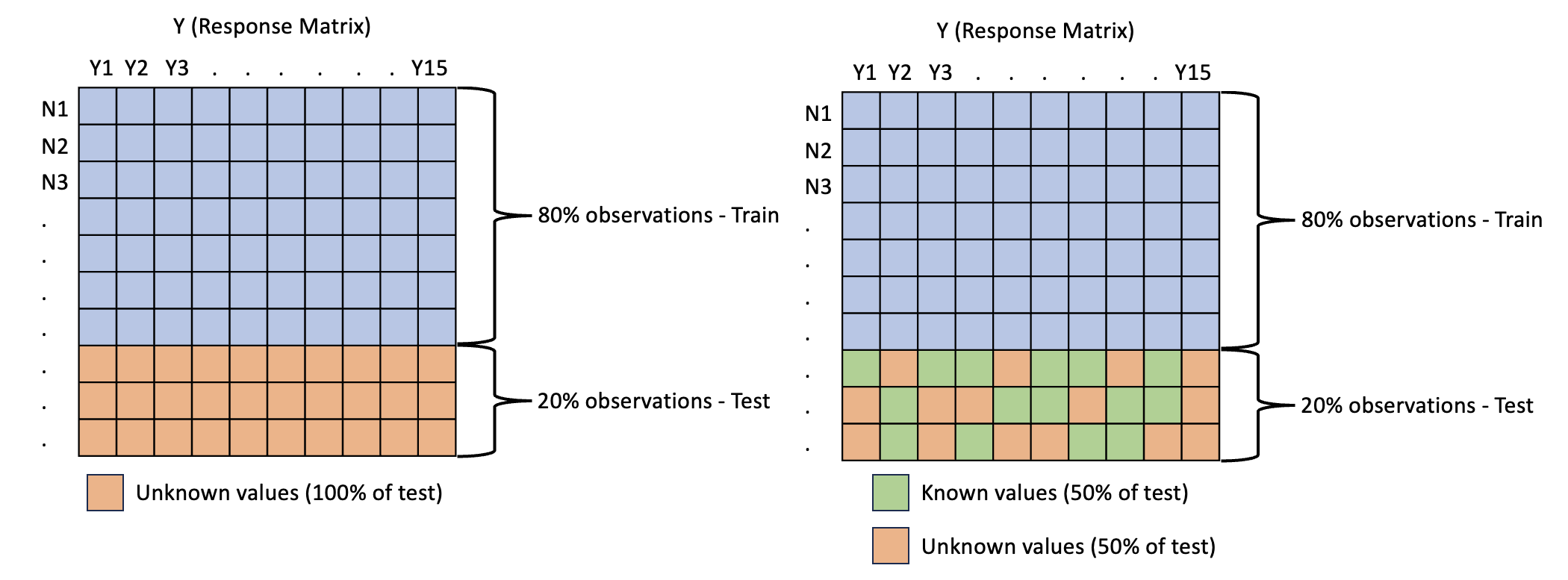}
    \caption{Two different training and testing setups. Left - Train and test with an 80 - 20 split. Right - Train and test with an 80 - 20 split, but half of the test responses are considered to be known.}
    \label{fig:traintestsplit}
\end{figure}

We consider two different extents for this data. In the initial configuration, we have data for January 2023 $(n = 31)$; in the second configuration, we have data from January - March 2023 $(n = 90)$. We also create the training and testing set in two different ways. In the first setup, we split the data into train and test with 80\% of the data belonging to the training set and 20\% belonging to the test. In the next set of experiments, we maintain the 80\% training set while distributing the remaining 20\% from the test set differently. Here, we assume that half of the values in the test set are known and the other half are unknown. We solely predict the unknown values.  These training and testing set-ups are demonstrated in figure \ref{fig:traintestsplit}. Comparing the two setups helps in evaluating the impact of incorporating the response dependencies (estimated via $\hat\Lambda$) into the predictions on the performance of the methods.

 We compare the prediction accuracy of our method against Graphical LASSO (or GLASSO), CAPME and Bayes CRF (or BCRF). The tuning parameters for each of these methods have been chosen from 5-fold cross-validation and the average prediction errors are evaluated by:

$$\bar{Err} = \frac 1n\sum_{d = 1}^{n}\|Y_d - \hat{Y_d}\|_2,$$
where $d$ denotes a particular day. 

 The average prediction errors for all the methods are provided in Table \ref{reallifetable}. From the results, we can see that our method outperforms all the other methods consistently. Additionally, it is interesting to note that when the sample size ($n = 31$) is less than the number of covariates ($p = 45$), our method exhibits significantly better performance. Conversely, when the sample size is larger, Graphical LASSO and Bayes CRF have comparable performance. This is because
our method uses the induced group sparsity to select only the relevant variables which is critical when the number of covariates is large. 
 
It is also worth highlighting that the CAPME method performs poorly when confronted with scenarios where all test values are unknown. This is due to the omission of the second stage of CAPME in this setup, wherein it predicts response dependencies. Consequently, the full potential of CAPME remains untapped in this context. In contrast, when half of the test values are unknown, it leverages the value of $\Lambda$ to predict these unknown responses, showcasing a more comprehensive utilization of its capabilities. This is reflected in the lower magnitude of errors for the other methods as well.

\begin{table}[]
\centering
\begin{tabular}{|c|cc|cc|}
\hline
                         & \multicolumn{2}{c|}{All test values unknown}                                                         & \multicolumn{2}{c|}{Half test values unknown}                                                        \\ \cline{2-5} 
\multirow{-2}{*}{Method} & \multicolumn{1}{c|}{ N = 31 }                                 &N = 90& \multicolumn{1}{c|}{ N = 31 }                                 &N = 90\\ \hline
BVS.GM                   & \multicolumn{1}{c|}{{\color[HTML]{000000} \textbf{18.734}}} & {\color[HTML]{000000} \textbf{22.222}} & \multicolumn{1}{c|}{{\color[HTML]{000000} \textbf{10.439}}} & {\color[HTML]{000000} \textbf{12.483}} \\
GLASSO                   & \multicolumn{1}{c|}{25.657}                                 & 22.687                                 & \multicolumn{1}{c|}{16.832}                                 & 16.907                                 \\
BCRF                     & \multicolumn{1}{c|}{26.638}                                 & 24.345                                 & \multicolumn{1}{c|}{14.681}                                 & 12.770                                 \\
CAPME                    & \multicolumn{1}{c|}{53.160}                                 & 108.424                                & \multicolumn{1}{c|}{39.895}                                 & 23.269                                 \\ \hline
\end{tabular}
\caption{Average prediction error for capital bike-share demand}
\label{reallifetable}
\end{table}

\section{Conclusion} In the space of estimating the relationship within high dimensional responses and covariates, we introduce a Bayesian model designed to simultaneously estimate three distinct sparsity structures:  the conditional dependency structure among the responses, the conditional dependency structure between the responses and the covariates, and the regression coefficient matrix. The proposed methodology bridges a significant gap in the literature on variable selection and sparse estimation for
high-dimensional graphical models which currently focus either on the sparse estimation of the precision matrix or the regression coefficient matrix. Our empirical results demonstrate the capability of our method to be particularly useful when the signal strength is low.  

{
\bibliography{references.bib}

\begin{thebibliography}{}

\bibitem[Bai et~al., 2020]{Bai2020SpikeandSlabML}
Bai, R., Ro{\v{c}}kov{\'a}, V., and George, E.~I. (2020).
\newblock {Spike-and-Slab Meets LASSO: A Review of the Spike-and-Slab LASSO}.
\newblock {\em Handbook of Bayesian Variable Selection}.

\bibitem[Cai et~al., 2013]{cai2013covariate}
Cai, T.~T., Li, H., Liu, W., and Xie, J. (2013).
\newblock {Covariate-adjusted precision matrix estimation with an application
  in genetical genomics}.
\newblock {\em Biometrika}, 100(1):139--156.

\bibitem[Candes and Tao, 2007]{candes2007dantzig}
Candes, E. and Tao, T. (2007).
\newblock {The Dantzig selector: Statistical estimation when $p$ is much larger
  than $n$}.
\newblock {\em The Annals of Statistics}, 35(6):2313--2351.

\bibitem[Castillo et~al., 2015]{castillo2015bayesian}
Castillo, I., Schmidt-Hieber, J., and van~der Vaart, A. (2015).
\newblock {Bayesian linear regression with sparse priors}.
\newblock {\em Annals of Statistics}, 43(5):1986--2018.

\bibitem[Consonni et~al., 2017]{consonni2017objective}
Consonni, G., La~Rocca, L., and Peluso, S. (2017).
\newblock {Objective Bayes covariate-adjusted sparse graphical model
  selection}.
\newblock {\em Scandinavian Journal of Statistics}, 44(3):741--764.

\bibitem[Deshpande et~al., 2019]{deshpande2019simultaneous}
Deshpande, S.~K., Ro{\v{c}}kov{\'a}, V., and George, E.~I. (2019).
\newblock {Simultaneous variable and covariance selection with the multivariate
  spike-and-slab LASSO}.
\newblock {\em Journal of Computational and Graphical Statistics},
  28(4):921--931.

\bibitem[Fan and Li, 2001]{SCAD}
Fan, J. and Li, R. (2001).
\newblock {Variable selection via nonconcave penalized likelihood and its
  oracle properties}.
\newblock {\em Journal of the American Statistical Association},
  96(456):1348--1360.

\bibitem[Friedman et~al., 2008]{friedman2008sparse}
Friedman, J., Hastie, T., and Tibshirani, R. (2008).
\newblock {Sparse inverse covariance estimation with the graphical LASSO}.
\newblock {\em Biostatistics}, 9(3):432--441.

\bibitem[Gan et~al., 2019]{gan2019bayesian}
Gan, L., Narisetty, N.~N., and Liang, F. (2019).
\newblock {Bayesian regularization for graphical models with unequal
  shrinkage}.
\newblock {\em Journal of the American Statistical Association},
  114(527):1218--1231.

\bibitem[Gan et~al., 2022]{gan2022bayesian}
Gan, L., Narisetty, N.~N., and Liang, F. (2022).
\newblock {Bayesian estimation of Gaussian conditional random fields}.
\newblock {\em Statistica Sinica}, 32:131--152.

\bibitem[Honorio et~al., 2012]{honorio2012variable}
Honorio, J., Samaras, D., Rish, I., and Cecchi, G. (2012).
\newblock Variable selection for gaussian graphical models.
\newblock In {\em Artificial Intelligence and Statistics}, pages 538--546.
  PMLR.

\bibitem[Ishwaran and Rao, 2005]{ishwaran2005spike}
Ishwaran, H. and Rao, J.~S. (2005).
\newblock {Spike and slab variable selection: frequentist and Bayesian
  strategies}.
\newblock {\em The Annals of Statistics}, 33(2):730--773.

\bibitem[Lafferty et~al., 2001]{lafferty2001conditional}
Lafferty, J., McCallum, A., and Pereira, F.~C. (2001).
\newblock {Conditional random fields: Probabilistic models for segmenting and
  labeling sequence data}.
\newblock {\em Machine Learning}, 46(1-3):283--334.

\bibitem[Li et~al., 2022]{li2022transfer}
Li, S., Cai, T.~T., and Li, H. (2022).
\newblock {Transfer learning in large-scale gaussian graphical models with
  false discovery rate control}.
\newblock {\em Journal of the American Statistical Association}, pages 1--13.

\bibitem[Li and McCormick, 2019]{li2019expectation}
Li, Z.~R. and McCormick, T.~H. (2019).
\newblock {An expectation conditional maximization approach for Gaussian
  graphical models}.
\newblock {\em Journal of Computational and Graphical Statistics},
  28(4):767--777.

\bibitem[Loh and Wainwright, 2013]{loh2013regularized}
Loh, P.-L. and Wainwright, M.~J. (2013).
\newblock {Regularized M-estimators with nonconvexity: Statistical and
  algorithmic theory for local optima}.
\newblock {\em Advances in Neural Information Processing Systems}, 26.

\bibitem[Mohammadi et~al., 2023]{mohammadi2023accelerating}
Mohammadi, R., Massam, H., and Letac, G. (2023).
\newblock {Accelerating Bayesian structure learning in sparse Gaussian
  graphical models}.
\newblock {\em Journal of the American Statistical Association},
  118(542):1345--1358.

\bibitem[Narisetty and He, 2014]{narisetty2014bayesian}
Narisetty, N.~N. and He, X. (2014).
\newblock {Bayesian variable selection with shrinking and diffusing priors}.
\newblock {\em The Annals of Statistics}, 42(2):789--817.

\bibitem[Osborne et~al., 2020]{Osborne2020LatentNE}
Osborne, N., Peterson, C.~B., and Vannucci, M. (2020).
\newblock {Latent Network Estimation and Variable Selection for Compositional
  Data Via Variational EM}.
\newblock {\em Journal of Computational and Graphical Statistics}, 31:163 --
  175.

\bibitem[Radosavljevic et~al., 2014]{radosavljevic2014neural}
Radosavljevic, V., Vucetic, S., and Obradovic, Z. (2014).
\newblock {Neural gaussian conditional random fields}.
\newblock In {\em Machine Learning and Knowledge Discovery in Databases:
  European Conference, ECML PKDD 2014, Nancy, France, September 15-19, 2014.
  Proceedings, Part II 14}, pages 614--629. Springer.

\bibitem[Ravikumar et~al., 2011]{ravikumar2011high}
Ravikumar, P., Wainwright, M.~J., Raskutti, G., and Yu, B. (2011).
\newblock {High-dimensional covariance estimation by minimizing
  $\ell_1$-penalized log-determinant divergence}.
\newblock {\em Electronic Journal of Statistics}, 5:935 – 980.

\bibitem[Rigollet, 2015]{Mit_regr}
Rigollet, P. (2015).
\newblock {High dimensional statistics lecture notes}.
\newblock
  \url{https://ocw.mit.edu/courses/18-s997-high-dimensional-statistics-spring-2015/resources/mit18_s997s15_chapter2/}.

\bibitem[Ro{\v{c}}kov{\'a}, 2018]{rovckova2018bayesian}
Ro{\v{c}}kov{\'a}, V. (2018).
\newblock {Bayesian estimation of sparse signals with a continuous
  spike-and-slab prior}.
\newblock {\em The Annals of Statistics}, page 401 – 437.

\bibitem[Ro{\v{c}}kov{\'a} and George, 2014]{rovckova2014emvs}
Ro{\v{c}}kov{\'a}, V. and George, E.~I. (2014).
\newblock {EMVS: The EM approach to Bayesian variable selection}.
\newblock {\em Journal of the American Statistical Association},
  109(506):828--846.

\bibitem[Ro{\v{c}}kov{\'a} and George, 2016]{rovckova2016fast}
Ro{\v{c}}kov{\'a}, V. and George, E.~I. (2016).
\newblock {Fast Bayesian factor analysis via automatic rotations to sparsity}.
\newblock {\em Journal of the American Statistical Association},
  111(516):1608--1622.

\bibitem[Ro{\v{c}}kov{\'a} and George, 2018]{rovckova2018spike}
Ro{\v{c}}kov{\'a}, V. and George, E.~I. (2018).
\newblock {The spike-and-slab LASSO}.
\newblock {\em Journal of the American Statistical Association},
  113(521):431--444.

\bibitem[Rothman et~al., 2010]{rothman2010sparse}
Rothman, A.~J., Levina, E., and Zhu, J. (2010).
\newblock {Sparse multivariate regression with covariance estimation}.
\newblock {\em Journal of Computational and Graphical Statistics},
  19(4):947--962.

\bibitem[Sohn and Kim, 2012]{sohn2012joint}
Sohn, K.-A. and Kim, S. (2012).
\newblock {Joint estimation of structured sparsity and output structure in
  multiple-output regression via inverse-covariance regularization}.
\newblock In {\em Artificial Intelligence and Statistics}, pages 1081--1089.
  PMLR.

\bibitem[Wytock and Kolter, 2013]{wytock2013sparse}
Wytock, M. and Kolter, Z. (2013).
\newblock {Sparse Gaussian conditional random fields: Algorithms, theory, and
  application to energy forecasting}.
\newblock In {\em International Conference on Machine Learning}, pages
  1265--1273. PMLR.

\bibitem[Yang et~al., 2021]{yang2021gembag}
Yang, X., Gan, L., Narisetty, N.~N., and Liang, F. (2021).
\newblock {GemBag: Group estimation of multiple Bayesian graphical models}.
\newblock {\em The Journal of Machine Learning Research}, 22(1):2450--2497.

\bibitem[Yin and Li, 2011]{yin2011sparse}
Yin, J. and Li, H. (2011).
\newblock {A sparse conditional Gaussian graphical model for analysis of
  genetical genomics data}.
\newblock {\em The Annals of Applied Statistics}, 5(4):2630.

\bibitem[Yuan and Zhang, 2014]{yuan2014partial}
Yuan, X.-T. and Zhang, T. (2014).
\newblock {Partial Gaussian graphical model estimation}.
\newblock {\em IEEE Transactions on Information Theory}, 60(3):1673--1687.

\bibitem[Zhang and Li, 2022]{zhang2022high}
Zhang, J. and Li, Y. (2022).
\newblock {High-dimensional gaussian graphical regression models with
  covariates}.
\newblock {\em Journal of the American Statistical Association}, pages 1--13.

\end{thebibliography}


\begin{thebibliography}{}

\bibitem[Gan et~al., 2022]{gan2022bayesian}
Gan, L., Narisetty, N.~N., and Liang, F. (2022).
\newblock {Bayesian estimation of Gaussian conditional random fields}.
\newblock {\em Statistica Sinica}, 32:131--152.

\bibitem[Rigollet, 2015]{Mit_regr}
Rigollet, P. (2015).
\newblock {High dimensional statistics lecture notes}.
\newblock
  \url{https://ocw.mit.edu/courses/18-s997-high-dimensional-statistics-spring-2015/resources/mit18_s997s15_chapter2/}.

\bibitem[Rinaldo, 2016]{CMU_regr}
Rinaldo, A. (2016).
\newblock {Advanced Statistical Theory lecture notes}.
\newblock
  \url{https://www.stat.cmu.edu/~arinaldo/Teaching/36755/F16/Scribed_Lectures/9_26_scribe_notes.pdf}.

\bibitem[Wytock and Kolter, 2013]{wytock2013sparse}
Wytock, M. and Kolter, Z. (2013).
\newblock {Sparse Gaussian conditional random fields: Algorithms, theory, and
  application to energy forecasting}.
\newblock In {\em International Conference on Machine Learning}, pages
  1265--1273. PMLR.

\bibitem[Yuan and Zhang, 2014]{yuan2014partial}
Yuan, X.-T. and Zhang, T. (2014).
\newblock {Partial Gaussian graphical model estimation}.
\newblock {\em IEEE Transactions on Information Theory}, 60(3):1673--1687.

\end{thebibliography}
\bibliographystyle{apalike}
}

\end{document}


%
%
%

\title{Bayesian Variable Selection and Sparse Estimation for
High-Dimensional Graphical Models}

\bigskip

%
%

\authors{Anwesha Chakravarti\inst{1}, Naveen N. Narisetty\inst{1}, Feng Liang\inst{1}
     }

\smallskip

%
%

 \university{
 \inst{1} University of Illinois at Urbana Champaign, Department of Statistics, United States, anwesha5@illinois.edu \\}

\begin{center}
    \textbf{\large{Supplementary Material}}
\end{center}
\tableofcontents
\section{Properties of the log-likelihood function}\label{liksec}
In this section, we provide the derivatives of the log-likelihood function along with useful results concerning it. The log-likelihood function is as follows: 
\begin{equation*}
    l(\Theta,\Lambda) = \frac{n}{2}\left(\log\det(\Lambda) - \Tr(S_{yy}\Lambda + 2S_{xy}^T\Theta + \Lambda^{-1}\Theta^TS_{xx}\Theta)\right).
\end{equation*}
The gradient of the likelihood function can then be computed to be as follows:
\begin{equation*}
    \nabla l(\Theta,\Lambda) = \begin{bmatrix} \nabla_
\Theta l(\Lambda,\Theta)\\
\nabla_
\Lambda l(\Lambda,\Theta)
\end{bmatrix} = \begin{bmatrix}
\frac{n}{2}\left(-2S_{xy}-2S_{xx}\Theta\Lambda^{-1}\right)\\
\frac{n}{2}\left(-S_{yy}+\Lambda^{-1}+\Lambda^{-1}\Theta^TS_{xx}\Theta\Lambda^{-1}\right)
\end{bmatrix},
\end{equation*}
and the Hessian of the likelihood function is as follows:
\begin{align*}
    \nabla^2 l(\Theta,\Lambda) &= \begin{bmatrix} \nabla^2_
{\Theta,\Theta} l(\Theta,\Lambda) & \nabla^2_
{\Theta,\Lambda} l(\Theta,\Lambda)\\
\nabla^2_
{\Theta,\Lambda} l(\Theta,\Lambda) & \nabla^2_
{\Lambda,\Lambda} l(\Theta,\Lambda)
\end{bmatrix} \\
&= \frac{n}{2} \begin{bmatrix}
-2\Lambda^{-1}\bigotimes S_{xx} & 2\Lambda^{-1}\bigotimes \Lambda^{-1}\Theta^TS_{xx}\\
2\Lambda^{-1}\bigotimes S_{xx}\Theta\Lambda^{-1} & -\Lambda^{-1}\bigotimes \big(\Lambda^{-1}+2\Lambda^{-1}\Theta^TS_{xx}\Theta\Lambda^{-1}\big)
\end{bmatrix}.
\end{align*}
Note that we can expand the likelihood function $l(\Theta,\Lambda)$ within a small neighbourhood of the actual parameter value $(\Theta^0,\Lambda^0)$ as follows:
$$\nabla l\big((\Theta^0,\Lambda^0) +\Delta\big) = \nabla l(\Theta^0,\Lambda^0) + Hvec(\Delta)+R(\Delta),$$
where $R(\Delta)$ is the residual. The following lemma, inspired by Proposition 4 in \citet{yuan2014partial} and by Lemma 3 in \citet{wytock2013sparse}, provides us with bounds for $\nabla l(\Theta,\Lambda)$ and $R(\Delta)$ which we later use to prove our results. 

\begin{Lemma}
\label{Lemma1}
Assume that the data is generated from a GCRF model with true parameter $(\Theta^0,\Lambda^0)$. Let $K^* = 8\max_i\Sigma^0_{ii} + 8max_i\left((\Lambda^0)^{-1}\Theta^T\Sigma^0_{xx}\Theta(\Lambda^0)^{-1}\right)_{ii}$, 

\begin{enumerate}
    \item We have $\|\nabla l(\Theta^0,\Lambda^0)\|_{\infty} \leq K_*\sqrt{n\log(10(p+q)^2/\epsilon_0)}$ with probability $1-\varepsilon_0$ given the sample size $n > \log(10(p+q)^2/\varepsilon_0)$; where $\varepsilon_0$ is any constant in $(0,1)$.
    \item If $\|\Delta\|_\infty \leq \frac{1}{d} \min\{\frac{1}{3c_{\Lambda^0}},\frac{c_{\Theta^0}}{2}\},$ then $\frac{2}{n}\|R(\Delta)\|_\infty \leq 1854d^2c_{\Theta^0}^2c_{\Lambda^0}^4\rho_2\|\Delta\|_\infty$, 
\end{enumerate} 
where $\rho_2=1.5\lambda_{max}(\Sigma_{xx}^0)$, $\Sigma^0$ denotes the covariance matrix of $(X,Y)$, $\Sigma_{xx}^0$ denotes the covariance matrix of X and $d$ denotes the maximum degree of the vertices corresponding to variables in the graphical model of the GCRF (i.e each row of $((\Theta^0)^T,\Lambda^0)$  has at most $d$ nonzero entries). 
\end{Lemma}
In the above lemma, the upper bound of $K^*$ is determined by the maximum variance of all the variables. If the covariates and response variables variances are capped, $K^*$ can be bounded above by a constant that is not a function of n. The proof for this Lemma is straightforward using the proofs of Proposition 4 in \citet{yuan2014partial} and Lemma 3 in \citet{wytock2013sparse}. 

 \section{Properties of the penalty functions}
In this Section, we prove the properties of the penalty function presented in Proposition 1 of the main paper, along with some other essential properties on the bounds of the derivatives of the penalty function. 

In our work, we use two types of penalty functions.
The first corresponds to a mixed spike and slab prior, denoted as $Pen_{\text{MSS}}$ and is defined on a vector parameter. This hierarchical prior first encourages the entire vector to be zero, and if not, it then promotes sparsity among the individual elements.  The second prior corresponds to the ordinary spike and slab prior, denoted as $Pen_{\text{SS}}$, is defined for a scalar parameter. The two penalty functions $Pen_{\text{MSS}}$ and $Pen_{\text{SS}}$ are defined as follows:
 \begin{align}\label{eq:penalties}
    Pen_{\text{MSS}}(\Phi_i)&:= \log\Bigg[\rho\prod_j\left(\frac{\eta}{2\vone}e^{-\frac{|\Phi_{ij}|}{\vone}}+\frac{1-\eta}{2\vzero}e^{-\frac{|\Phi_{ij}|}{\vzero}}\right)+\nonumber\\&\quad \quad \quad \quad \quad  (1-\rho)\prod_j\frac{1}{2\vzero}e^{-\frac{|\Phi_{ij}|}{\vzero}}\Bigg] \text{ and }\\
 Pen_{\text{SS}}(\Phi_{ij})&:=\log\left(\frac{\eta}{2\vone}\exp\left\{-\frac{|\Phi_{ij}|}{\vone}\right\} + \frac{(1-\eta)}{2\vzero}\exp\left\{-\frac{|\Phi_{ij}|}{\vzero}\right\}\right).
\end{align}
Thus, $\sum_iPen_{\text{MSS}}(\Theta_i) = -\log\Pi(\Theta)$ and $\sum_{i<j}Pen_{\text{SS}}(\Lambda_{ij})= -\log\Pi(\Lambda)$. Before we prove proposition 1, we state a proposition which gives the explicit forms for the derivatives of the penalty functions. 

Through the rest of the supplementary materials, we use the superscript $\Theta$ or $\Lambda$ for our parameters to denote that they correspond to the respective matrix.  For example, $\et$ corresponds to $\Theta$. We omit the superscript when the same condition applies to both sets of parameters. 

\begin{prop} \label{prop1} Let $Z(\Lambda_{ij})$ and $W(\Theta_i)$ be binary random variables defined as 
$$Z(\Lambda_{ij}) = \begin{cases}
    \frac{1}{\vlone} & \text{w.p.  } \eta_1(\Lambda_{ij})\\
    \frac{1}{\vlzero} & \text{w.p.  } 1-\eta_1(\Lambda_{ij})
\end{cases}  \text{ and } W(\Theta_i) = \begin{cases}
    \frac{1}{\vtone} & \text{w.p.  } \eta_1(\Theta_{ij})\eta_2(\Theta_i)\\
    \frac{1}{\vtzero} & \text{w.p.  } 1-\eta_1(\Theta_{ij})\eta_2(\Theta_i) 
\end{cases},$$
with $\eta_1(.)$ and $\eta_2(.)$ as defined in Section 3.2 of the main paper. Then, the following hold

\begin{itemize}
    \item[(i)] The first and second derivatives of $Pen_{SS}(\Lambda_{ij})$  are given by
    $$\frac{\partial}{\partial|\Lambda_{ij}|} Pen_{SS}(\Lambda_{ij}) =  EZ(\Lambda_{ij}) \text{ and } \frac{\partial^2}{\partial\Lambda_{ij}^2}Pen_{SS}(\Lambda_{ij}) = -Var(Z(\Lambda_{ij})).$$
\item[(ii)] The first and second derivatives of $Pen_{MSS}(\Theta_i)$  are given by
$$ \nabla_{|\Theta|}Pen_{MSS}(\Theta_i) =  EW(\Theta_{i}) \text{ and } \nabla^2_{\Theta}Pen_{MSS}(\Theta_i) = -Var(W(\Theta_{i})).$$
\end{itemize}
\end{prop}

\begin{proof} The proof for (i) can be found in the Supplementary materials for \citet{gan2022bayesian}. We only prove (ii). We have from (\ref{eq:penalties}), $Pen_{MSS}(\Theta) = -\log\bigg(\rho S_1(\Theta_i)+(1-\rho)S_2(\Theta_i)\bigg)$. Thus, the first derivative
\begin{equation}
    \frac{\partial Pen_{MSS}(\Theta_i)}{\partial |\Theta_{ij}|} = -\frac{\rho\frac{\partial}{\partial \Theta_{ij}}S_1(\Theta_{i})+(1-\rho)\frac{\partial}{\partial \Theta_{ij}}S_2(\Theta_{i})}{\rho S_1(\Theta_i)+(1-\rho)S_2(\Theta_i)}.
    \label{eq5}
\end{equation}
Here,
\begin{equation*}
    \frac{\partial}{\partial |\Theta_{ij}|}S_1(\Theta_{i}) 
    = -S_1(\Theta_i)\left(\frac{1}{\vtone}\eta_1(\Theta_{ij})+\frac{1}{\vtzero}(1-\eta_1(\Theta_{ij}))\right) \text{ and }
\end{equation*}
\begin{equation*}
    \frac{\partial}{\partial |\Theta_{ij}|}S_2(\Theta_{i}) 
    = -\frac{S_2(\Theta_{i})}{\vtzero}.
\end{equation*}
Putting the above values in (\ref{eq5}) we get
\begin{align*}
   \frac{\partial Pen_{MSS}(\Theta)}{\partial |\Theta_{ij}|} &=  \frac{\rho S_1(\Theta_i)\left(\frac{1}{\vtone}\eta_1(\Theta_{ij})+\frac{1}{\vtzero}(1-\eta_1(\Theta_{ij}))\right) +(1-\rho)\frac{S_2(\Theta_{i})}{\vtzero}}{\rho S_1(\Theta_i)+(1-\rho)S_2(\Theta_i)}\\
   &= \frac{1}{\vtone}\big(\eta_1(\Theta_{ij})\eta_2(\Theta_i)\big)+\frac{1}{\vtzero}\big(1-\eta_1(\Theta_{ij})\eta_2(\Theta_i)\big)\\
   &= \bigg(EW(\Theta_i)\bigg)_j.
\end{align*}
For the second derivative we first compute the derivatives of $\eta_1(\Theta_{ij})$ and $\eta_2(\Theta_i)$:
\begin{align*}
    \frac{\partial}{\partial |\Theta_{ij}|}\eta_1(\Theta_{ij}) 
    &= -\frac{1}{\vtone}\eta_1(\Theta_{ij})+\eta_1(\Theta_{ij})\left(\frac{1}{\vtone}\eta_1(\Theta_{ij})+\frac{1}{\vtzero}(1-\eta_1(\Theta_{ij}))\right)\\
    &= -\eta_1(\Theta_{ij})(1-\eta_1(\Theta_{ij}))\left(\frac{1}{\vtone}-\frac{1}{\vtzero}\right) \text{ and }
\end{align*}
\begin{align*}
    \frac{\partial}{\partial |\Theta_{ij}|}\eta_2(\Theta_{i}) 
    &= \frac{\rho(1-\rho)\big(\rho S_1'(\Theta_i)+(1-\rho)S_2'(\Theta_i)\big)}{\big(\rho S_1(\Theta_i)+(1-\rho)S_2(\Theta_i)\big)^2}\\
    &=-\frac{\rho(1-\rho)S_1(\Theta_i)S_2(\Theta_i)\left(\frac{1}{\vtone}\eta_1(\Theta_{ij}) +\frac{1}{\vtzero}(1-\eta_1(\Theta_{ij}))-\frac{1}{\vtzero}\right)}{\big(\rho S_1(\Theta_i)+(1-\rho)S_2(\Theta_i)\big)^2}\\
    &= -\eta_1(\Theta_{ij})\eta_2(\Theta_i)(1-\eta_2(\Theta_{i}))\left(\frac{1}{\vtone}-\frac{1}{\vtzero}\right).
\end{align*}Thus we have,
\begin{align*}
     \frac{\partial^2 Pen_{MSS}(\Theta)}{\partial \Theta_{ij}^2} &= \left(\frac{1}{\vtone}-\frac{1}{\vtzero}\right)\bigg(\eta_2(\Theta_i)\frac{\partial}{\partial |\Theta_{ij}|}\eta_1(\Theta_{ij})+\eta_1(\Theta_{ij})\frac{\partial}{\partial |\Theta_{ij}|}\eta_2(\Theta_{i})\bigg)\\
    &= -\left(\frac{1}{\vtone}-\frac{1}{\vtzero}\right)^2\bigg(\eta_1(\Theta_{ij})\eta_2(\Theta_i)(1-\eta_1(\Theta_{ij})\eta_2(\Theta_i))\bigg) \text{ and }
\end{align*}\begin{align*}
     \frac{\partial^2 Pen_{MSS}(\Theta)}{\partial \Theta_{ij}\Theta_{ij'}}
     &= \left(\frac{1}{\vtone}-\frac{1}{\vtzero}\right)\eta_1(\Theta_{ij})\bigg( \frac{\partial}{\partial |\Theta_{ij'}|}\eta_2(\Theta_{i})\bigg)\\
     &= -\left(\frac{1}{\vtone}-\frac{1}{\vtzero}\right)^2\bigg(\eta_1(\Theta_{ij}\eta_1(\Theta_{ij'}))\eta_2(\Theta_i)(1-\eta_2(\Theta_i))\bigg),
\end{align*}
which gives us $\nabla^2_{\Theta}Pen_{MSS}(\Theta_i) = -Var(W(\Theta_{i}))$. 
\end{proof}

\noindent \textbf{Proof for Proposition 1:}
 Note that the penalty function is concave since the second derivatives of the two penalty functions are negative semi-definite. The second part of the proposition follows directly from (i) in Proposition \ref{prop1}.
The following proposition gives some useful bounds for the derivatives of the mixed spike and slab prior $Pen_{\text{MSS}}(.)$ and the spike and slab prior $Pen_{\text{SS}}(.)$. 
\begin{prop}
\label{prop2}
    Let us assume the following conditions hold 
$$\left(\frac{\vtone}{\vtzero}\right)^{p+2}\frac{(1-\rho)}{\rho \cdot \eta^p}\leq \epsilon_1(p+q)^{\epsilon_2}, \quad 
\left(\frac{v_1}{v_0}\right)^3\frac{(1-\eta)}{\eta}
\leq \epsilon_1(p+q)^{\epsilon_2},$$
    $$\frac{1}{n\vone}<\frac{C_1}{1+2\epsilon_1}\sqrt{\frac{\log(p+q)}{n}},\quad\frac{1}{n\vzero}>C_4\sqrt{\frac{\log(p+q)}{n}},$$
    where $\epsilon_1,\epsilon_2,C_1,C_3$ and $C_4$ are positive constants with $C_3>C_1$, $C_4>C_1$ and $\epsilon_2 = (C_3 - C_1)c_H(C_4 - C_1)$.  
Then for all $ \Theta_{ij}$ such that 
$|\Theta_{ij}| \geq 2(C_3 - C_1)c_H\sqrt{\frac{\log(p+q)}{n}}$ and for all $ \Lambda_{ij}$ such that $|\Lambda_{ij}| \geq 2(C_3 - C_1)c_H\sqrt{\frac{\log(p+q)}{n}}$  we have,

\begin{itemize}
    \item For the mixed spike and slab prior $Pen_{MSS}(\Theta_i)$
 $$\frac{1}{n}\left|\frac{\partial}{\partial |\Theta_{ij}|}Pen_{MSS}(\Theta_i)\right| \leq C_1\sqrt{\frac{\log(p+q)}{n}},\quad \frac{1}{2n}\left|\frac{\partial^2}{\partial \Theta_{ij}^2}Pen_{MSS}(\Theta_i)\right| \leq C_1^2\frac{\log(p+q)}{4(p+q)^{\epsilon_2}}$$ and for  $j' \neq j$
$$\frac{1}{2n}\left|\frac{\partial}{\partial \Theta_{ij}\Theta_{ij'}}Pen_{MSS}(\Theta_i)\right| \leq C_1^2\frac{\log(p+q)}{8(p+q)^{\epsilon_2}}$$

\item For the spike and slab prior $Pen_{SS}(\Lambda_{ij})$
$$\frac{1}{n}\left|\frac{\partial}{\partial |\Lambda_{ij}|}Pen_{SS}(\Lambda_{ij})\right| \leq C_1\sqrt{\frac{\log(p+q)}{n}},\quad \frac{1}{2n}\left|\frac{\partial^2}{\partial \Lambda_{ij}^2}Pen_{SS}(\Lambda_{ij})\right| \leq C_1^2\frac{\log(p+q)}{8(p+q)^{\epsilon_2}}$$
\end{itemize}
\end{prop}

\begin{proof}
For the mixed spike and slab prior, we have shown in Proposition \ref{prop1} that $$\frac{\partial}{\partial|\Theta_{ij}|}Pen_{MSS}(\Theta_i) =   \frac{\eta_1(\Theta_{ij})\eta_2(\Theta_i)}{\vtone} + \frac{1-\eta_1(\Theta_{ij})\eta_2(\Theta_i)}{\vtzero},$$ and 
$$\frac{\partial}{\partial\Theta_{ij}^2}Pen_{MSS}(\Theta_i) = -\eta_1(\Theta_{ij})\eta_2(\Theta_i)\bigg(1- \eta_1(\Theta_{ij})\eta_2(\Theta_i)\bigg)\left(\frac{1}{\vtzero}-\frac{1}{\vtone}\right)^2.$$
Thus, to find the upper bounds for these derivatives, we first compute bounds for $\eta_1(.)$ and $\eta_2(.)$. The bound for $\eta_1(\Theta_{ij})$:
\begin{align*}
    \eta_1(\Theta_{ij}) &= \frac{\frac{\et}{2\vtone}e^{-\frac{|\Theta_{ij}|}{\vtone}}}{\frac{\et}{2\vtone}e^{-\frac{|\Theta_{ij}|}{\vtone}}+\frac{1-\et}{2\vtzero}e^{-\frac{|\Theta_{ij}|}{\vtzero}}}
    = \left(1+ \frac{1-\et}{\et}\frac{\vtone}{\vtzero}e^{-|\Theta_{ij}|\left(\frac{1}{\vtzero}-\frac{1}{\vtone}\right)}\right)^{-1}\\
    &\geq \left(1+\left(\frac{\vtzero}{\vtone}\right)^2\epsilon_1(p+q)^{-\epsilon_2}\right)^{-1}\\
    & \geq 1-\left(\frac{\vtzero}{\vtone}\right)^2\epsilon_1(p+q)^{-\epsilon_2},
\end{align*}
where we get the third inequality using the conditions $\left(\frac{\vtone}{\vtzero}\right)^3\frac{(1-\et)}{\et}\leq \epsilon_1(p+q)^{\epsilon_2}$, and noting that since $\frac{1}{n\vtone}<\frac{C_1}{1+2\epsilon_1}\sqrt{\frac{\log(p+q)}{n}}$,
    $\frac{1}{n\vtzero}>C_4\sqrt{\frac{\log(p+q)}{n}}$ and $|\Theta_{ij}| \geq 2(C_3 - C_1)c_H\sqrt{\frac{\log(p+q)}{n}}$,
\begin{align}
    \label{eq necc}
    |\Theta_{ij}|\left(\frac{1}{\vtzero}-\frac{1}{\vtone}\right) &\geq 2(C_3 - C_1)c_Hn\left(C_4-\frac{C_1}{1+2\epsilon_1}\right)\frac{\log(p+q)}{n} \nonumber\\
    &\geq \log(p+q)^{2\epsilon_2}.
\end{align}
The bound for $\eta_2(\Theta_{i})$:
\begin{align*}
    \eta_2(\Theta_{i}) &= \frac{\rho S_1(\Theta_i)}{\rho S_1(\Theta_i) + (1-\rho) S_2(\Theta_i)}\geq \left(1+ \left(\frac{1-\rho}{\rho}\right)\prod_j\frac{ \frac{1}{2\vtzero}e^{-\frac{|\Theta_{ij}|}{\vtzero}}}{\frac{\et}{2\vtone}e^{-\frac{|\Theta_{ij}|}{\vtone}}} \right)^{-1}\\
    &= \left(1+ \frac{1-\rho}{\rho (\et)^p}\left(\frac{\vtone}{\vtzero}\right)^p e^{-\sum_j|\Theta_{ij}|\left(\frac{1}{\vtzero}-\frac{1}{\vtone}\right)}\right)^{-1}\\
    &\geq \left(1+ \left(\frac{\vtzero}{\vtone
    }\right)^2\epsilon_1(p+q)^{-\epsilon_0}\right)^{-1}\\
    &\geq 1 - \left(\frac{\vtzero}{\vtone
    }\right)^2\epsilon_1(p+q)^{-\epsilon_2},
\end{align*}
where we get the fourth inequality by using the condition $\left(\frac{\vtone}{\vtzero}\right)^{p+2}\frac{(1-\rho)}{\rho(\et)^p}\leq \epsilon_1(p+q)^{\epsilon_2}$ and using (\ref{eq necc}) along with noting that $\log(p+q)^{2p\epsilon_2}>\log(p+q)^{2\epsilon_2}$. Thus we have, 
\begin{equation}
    \label{eq 8}
    \eta_1(\Theta_{ij})\eta_2(\Theta_{i}) \geq \left(1 - \left(\frac{\vtzero}{\vtone
    }\right)^2\epsilon_1(p+q)^{-\epsilon_2}\right)^2 \geq 1 - \frac{2\epsilon_1}{(p+q)^{\epsilon_2}}\left(\frac{\vtzero}{\vtone}\right)^2, 
\end{equation}
which implies that the first derivative
\begin{align*}
    \frac 1n \left|\frac{\partial}{\partial|\Theta_{ij}|}Pen_{MSS}(\Theta_i)\right| &=   \frac 1n \left(\frac{\eta_1(\Theta_{ij})\eta_2(\Theta_i)}{\vtone} + \frac{1-\eta_1(\Theta_{ij})\eta_2(\Theta_i)}{\vtzero}\right)\\
    & \leq \frac{1}{n\vtone}+ \frac{1}{n\vtzero}\frac{2\epsilon_1}{(p+q)^{\epsilon_2}}\left(\frac{\vtzero}{\vtone}\right)^2\\
    &\leq \frac{1}{n\vtone}(1+2\epsilon_1)\\
    &\leq C_1 \sqrt{\frac{\log(p+q)}{n}}.
\end{align*}
Here we get the second inequality using (\ref{eq 8}) and since $\eta_1(\Theta_{ij})\eta_2(\Theta_{i}) < 1$. We get the third inequality using $\vtzero<\vtone$ and we get the final inequality using the condition $\frac{1}{n\vtone}<\frac{C_1}{1+2\epsilon_1}\sqrt{\frac{\log(p+q)}{n}}$. For the second derivative, we have 
\begin{align*}
    \frac{1}{2n}\left|\frac{\partial}{\partial\Theta_{ij}^2}Pen_{MSS}(\Theta_i)\right| &= \frac{1}{2n}\eta_1(\Theta_{ij})\eta_2(\Theta_i)\bigg(1- \eta_1(\Theta_{ij})\eta_2(\Theta_i)\bigg)\left(\frac{1}{\vtzero}-\frac{1}{\vtone}\right)^2\\
    &\leq \frac{1}{2n} \frac{1}{(\vtone)^2}\left(\frac{\vtone}{\vtzero}\right)^2(1-\eta_1(\Theta_{ij})\eta_2(\Theta_{i}))\\
    &\leq \frac{C_1^2}{(1+2\epsilon_1)^2}\frac{\epsilon_1}{(p+q)^{\epsilon_2}}\log(p+q)\\
    & \leq C_1^2\frac{\log(p+q)}{4(p+q)^{\epsilon_2}}.
\end{align*}
We get the second inequality using $\eta_1(\Theta_{ij})\eta_2(\Theta_{i}) < 1$ and $\vtzero<\vtone$, the third one using (\ref{eq 8}) and the condition $\frac{1}{n\vtone}<\frac{C_1}{1+2\epsilon_1}\sqrt{\frac{\log(p+q)}{n}}$, and the final inequality using $(1+2\epsilon_1)^2 > 4\epsilon_1$. 
Finally, when $j' \neq j$, we get
\begin{align*}
    \frac{1}{2n}\left|\frac{\partial}{\partial\Theta_{ij}\Theta_{ij'}}Pen_{MSS}(\Theta_i)\right| &= \frac{1}{2n}\eta_1(\Theta_{ij})\eta_1(\Theta_{ij'})\eta_2(\Theta_i)\left(1- \eta_2(\Theta_i)\right)\left(\frac{1}{\vtzero}-\frac{1}{\vtone}\right)^2\\
    &\leq \frac{1}{2n}\left(\frac{1}{\vtzero}-\frac{1}{\vtone}\right)^2(1- \eta_2(\Theta_i))\\
    &\leq \frac{1}{2n} \frac{1}{(\vtone)^2}\frac{\epsilon_1}{(p+q)^{\epsilon_2}}\\
    & \leq C_1^2\frac{\log(p+q)}{8(p+q)^{\epsilon_2}},
\end{align*}
where we get the second inequality using $\eta_1(\Theta_{ij'})\eta_2(\Theta_i)<1$, the third inequality using the bound for $\eta_2(\Theta_i)$ proved above and the final inequality using $(1+2\epsilon)^2 > 4\epsilon_1$ and the condition $\frac{1}{n\vtone}<\frac{C_1}{1+2\epsilon_1}\sqrt{\frac{\log(p+q)}{n}}$. Similarly, using the bounds for $\eta_1(.)$, we derive the bounds for the derivatives of $Pen_{SS}$ as well.
\begin{align*}
    \frac 1n \left|\frac{\partial}{\partial|\Lambda_{ij}|}Pen_{SS}(\Lambda_{ij})\right| &=   \frac 1n \left(\frac{\eta_1(\Lambda_{ij})}{\vlone} + \frac{1-\eta_1(\Lambda_{ij})}{\vlzero}\right)\leq \frac 1n \left(\frac{1}{\vlone}+ \left(\frac{\vlzero}{\vlone}\right)^2\frac{\epsilon_1}{(p+q)^{\epsilon_2}\vlzero}\right)\\
    &\leq \frac{1}{n\vlone}\left(1+ \left(\frac{\vlzero}{\vlone}\right)\frac{\epsilon_1}{(p+q)^{\epsilon_2}}\right)\\
    &\leq C_1\sqrt{\frac{\log(p+q)}{n}},
\end{align*}
and the second derivative
\begin{align*}
    \frac{1}{2n}\left|\frac{\partial}{\partial\Lambda_{ij}^2}Pen_{SS}(\Lambda_{ij})\right| &= \frac{1}{2n}\eta_1(\Lambda_{ij})\bigg(1- \eta_1(\Lambda_{ij})\bigg)\left(\frac{1}{\vlzero}-\frac{1}{\vlone}\right)^2\\
    &\leq \frac{1}{2n} \frac{1}{(\vlone)^2}\left(\frac{\vlone}{\vlzero}\right)^2(1-\eta_1(\Lambda_{ij}))\\
    &\leq \frac{C_1^2}{2(1+2\epsilon_1)^2}\frac{\epsilon_1}{(p+q)^{\epsilon_2}}\log(p+q)\\
    & \leq C_1^2\frac{\log(p+q)}{8(p+q)^{\epsilon_2}}.
\end{align*}
\end{proof}

\section{Marginal Inclusion Probabilities for $(\Theta,\Lambda)$}
This Section describes how the marginal probabilities mentioned in Section 3.4 are computed.  

\noindent \textbf{Marginal Inclusion Probability for $\Theta$:}
\begin{align*}
    P(r_{ij}^\Theta=1|\Theta) &= P(r_{ij}^\Theta=1|\Theta_{i})= P(r_i^\Theta =1|\Theta_i)P(r_{ij}^\Theta=1|r_i^\Theta=1,\Theta_{ij}).
\end{align*}
The first term can be simplified as follows:
\begin{align*}
    P(r_i =1|\Theta_i) &= \frac{P(\Theta_i|r_i^\Theta =1)P(r_i^\Theta=1)}{P(\Theta_i)}\\
    &= \frac{\rho S_1(\Theta_i)}{\rho S_1(\Theta_i)+(1-\rho) S_2(\Theta_i)}\\
    &= \eta_2(\Theta_i).
\end{align*}
The second term can be simplified as follows:
\begin{align*}
    P(r_{ij}^\Theta=1|r_i^\Theta=1,\Theta_{ij}) &= \frac{P(\Theta_{ij}|r_{ij}^\Theta = 1)P(r_{ij}^\Theta =1|r_i^\Theta = 1)P(r_i^\Theta =1))}{P(\Theta_{ij}|r_i^\Theta =1)P(r_i^\Theta=1)}\\
    &= \frac{\el LP(\Theta_{ij},\vtone)}{\el LP(\Theta_{ij},\vtone)+(1-\el) LP(\Theta_{ij},\vtzero)}\\
    &= \eta_1(\Theta_{ij}).
\end{align*}
Thus $P(r_i^\Theta =1|\Theta_i)P(r_{ij}^\Theta=1|r_i^\Theta=1,\Theta{ij}) = \eta_2(\Theta_{i})\eta_1(\Theta_{ij})$.

\noindent \textbf{Marginal Inclusion Probability for $\Lambda$:} 
\begin{align*}
     P(r_{ij}^\Lambda=1|\Lambda) &= P(r_{ij}^\Lambda=1|\Lambda_{ij})= \frac{P(\Lambda_{ij}|r_{ij}^\Lambda = 1)P(r_{ij}^\Lambda = 1)}{P(\Lambda_{ij})}\\
    &= \frac{\el LP(\Lambda_{ij},\vlone)}{\el LP(\Lambda_{ij},\vlone)+(1-\el) LP(\Lambda_{ij},\vlzero)}\\
    &=  \eta_1(\Lambda_{ij}).
\end{align*}
    
\section{The EM Algorithm}
In this section, we elaborate on the details of the EM algorithm and describe the computations required to derive the E-Step and the M-Step. The full posterior distribution computed from the prior distribution described in Section 3.1 in terms of the binary indicator matrices $R^\Lambda = ((r_{ij}^\Lambda))$ and $R^\Theta = ((r_{ij}^\Theta))$ is as follows: 
\begin{align}
    \Pi(\Lambda,\Theta, R^\Theta,R^\Lambda) &\propto \CondProb{Y}{\Lambda,\Theta,X}\bigg(\prod_{i<j}\CondProbPi{\Lambda_{ij}}{r_{ij}^{\Lambda}}\Pi(r_{ij}^{\Lambda})\bigg) \nonumber\\
    &\times \Bigg(\prod_{i}\CondProbPi{r_i^\Theta}{\rho}\left(\prod_j\CondProbPi{\theta_{ij}}{r_{ij}^\Theta}\CondProbPi{r_{ij}^{\Theta}}{r_i^\Theta}\right)\Bigg) \nonumber\\
    &\times \indicator{\Lambda \succ 0, \|\Lambda\|_2 \leq R}.
\label{fullpost}
\end{align}
Note that the binary indicator matrices $R^\Lambda$ and $R^\Theta$ are latent variables which indicate whether the element in the corresponding matrix $(\Lambda,\Theta)$ is non-zero. 

\subsection{E-Step}

Let $p_{ij}^\Theta,p_{ij}^\Lambda$  be the estimates for the posterior inclusion probabilities $r_{ij}^\Theta,r_{ij}^\Lambda$ respectively. 

\noindent \textbf{E-Step for $\Lambda$:}
The posterior probability for $r_{ij}^\Lambda$ given the estimate of $\Lambda$ from the previous iteration is
\begin{align*}
    p_{ij}^{\Lambda^{(t)}} &= \Expect{r_{ij}^\Lambda}= \CondProb{r_{ij}^\Lambda}{\Lambda^{(t)}}= \eta_1(\Lambda_{ij}^{(t)}).
\end{align*}
To complete the E-step, we need to take the expectation of the objective function to get the $Q$ function, which we minimize in the M-step. 
\begin{align*}
    Q(\Lambda|\Lambda^{(t)}) &= -l_{\Theta}(\Lambda) +\sum_{i<j}\left(\frac{p_{ij}^{\Lambda(t)}}{\vlone}+\frac{1-p_{ij}^{\Lambda(t)}}{\vtzero} \right)|\Lambda_{ij}|+const\\
    &\equiv -l_{\Theta}(\Lambda) +\sum_{i<j}\left(\frac{\eta_1(\Lambda_{ij}^{(t)})}{\vlone}+\frac{1-\eta_1(\Lambda_{ij}^{(t)})}{\vlzero} \right)|\Lambda_{ij}|,
\end{align*}
where $l_{\Theta}(\Lambda)$ is the likelihood function $l(\Theta,\Lambda)$ in terms of $\Lambda$ keeping $\Theta$ constant.

\noindent\textbf{E-Step for $\Theta$:} The posterior probability for $r_{ij}^\Theta$ given the estimate of $\Theta$ from the previous iteration is
\begin{align*}
    p_{ij}^{\Theta^{(t)}} &= \Expect{r_{ij}^\Theta}
    = \CondProb{r_{ij}^\Theta}{\Theta^{(t)}}
    = \eta_1(\Theta_{ij}^{(t)})\eta_2(\Theta_{i}^{(t)}).
\end{align*}
Taking the expectation of the objective function to get the $Q-$ function, which we minimize in the M-step:
\begin{align*}
    Q(\Theta|\Theta^{(t)}) &= -l_{\Lambda}(\Theta) +\sum_{i<j}\left(\frac{p_{ij}^{\Theta(t)}}{\vtone}+\frac{1-p_{ij}^{\Theta(t)}}{\vtzero} \right)|\Theta_{ij}|+const\\
    &\equiv -l_{\Lambda}(\Theta) +\sum_{i<j}\left(\frac{\eta_1(\Theta_{ij}^{(t)})\eta_2(\Theta_{i}^{(t)})}{\vtone}+\frac{1-\eta_1(\Theta_{ij}^{(t)})\eta_2(\Theta_{i}^{(t)})}{\vtzero} \right)|\Theta_{ij}|,
\end{align*}
where $l_{\Lambda}(\Theta)$ is the likelihood function $l(\Theta,\Lambda)$ in terms of $\Theta$ keeping $\Lambda$ constant. 

\subsection{M-Step}

To do the M-step, we first make a second-order approximation of the likelihood function $l(\Theta + \Delta_{\Theta}, \Lambda + \Delta_\Lambda)$ around $(\Theta,\Lambda)$
$$g(\Delta_{\Theta},\Delta_\Lambda) = l(\Theta,\Lambda) + vec\big(\nabla l(\theta,\Lambda)\big)^Tvec(\Delta_\Theta,\Delta_\Lambda)+ \frac12vec\big(\nabla l(\theta,\Lambda)\big)^T\nabla^2 l(\Theta,\Lambda)vec(\Delta_\Theta,\Delta_\Lambda) $$
\noindent where $vec(\Delta_\Theta,\Delta_\Lambda) = \begin{bmatrix}
\Delta_\theta\\
\Delta_\Lambda
\end{bmatrix}$. Note that we have \begin{align*}
    vec\big(\nabla l(\theta,\Lambda)\big)^Tvec(\Delta_\Theta,\Delta_\Lambda) &= \Tr\left(\big(\nabla l(\theta,\Lambda)\big)^T\begin{bmatrix}
\Delta_\theta\\
\Delta_\Lambda\end{bmatrix}\right)= \Tr(\nabla_\Theta l(\theta,\Lambda)^T\Delta_\Theta+\nabla_\Lambda l(\theta,\Lambda)^T\Delta_\Lambda).
\end{align*} 
Using the gradient and Hessian of the likelihood function from Section \ref{liksec} along with the identity $\left(vec(B^T)\right)^T\left(A^T \bigotimes C\right)vec(D) = \Tr(ABCD)$ we get
\begin{align*}
    g(\Delta_{\Theta},\Delta_\Lambda) &= \frac n2 \log\det(\Lambda) - \frac n2 tr \left(S_{yy}\Lambda+2S_{xy}^T\Theta +\Lambda^{-1}\Theta^TS_{xx}\Theta\right) \\
    & \quad +tr\bigg[\frac n2 \left(\Lambda^{-1} - S_{yy} +\Lambda^{-1}\Theta^TS_{xx}\Theta\Lambda^{-1}\right)\Delta_\Lambda - \frac n2\left(2S_{xy}^T+2\Lambda^{-1}\Theta^TS_{xx}\right)\Delta_\Theta\bigg]\\
    & \quad +\frac n4 tr\bigg[-\Lambda^{-1}\Delta_\Lambda^T\left(\Lambda^{-1}+2\Lambda^{-1}\Theta^TS_{xx}\Theta\Lambda^{-1}\right)\Delta_\Lambda + 2(2\Lambda^{-1})\Delta_\Lambda^T\Lambda^{-1}\Theta^TS_{xx}\Delta_\Theta \\
    & \quad \quad \quad \quad \quad - 2\Lambda^{-1}\Delta_\Theta^TS_{xx}\Delta_\Theta\bigg]\\
    &= \frac n2\log\det(\Lambda) -\frac n2 \Tr(S_{yy}\Lambda) - n\Tr(S_{xy}^T\Theta) - \frac n2\Tr(\Lambda^{-1}\Theta^TS_{xx}\Theta)  +\frac n2 \Tr(\Lambda^{-1}\Delta_\Lambda) \\
    & \quad - \frac n2 \Tr(S_{yy}\Delta_\Lambda) + \frac n2 \Tr(\Lambda^{-1}\Theta^TS_{xx}\Theta\Lambda^{-1}\Delta_\Lambda) -\frac n4\Tr(\Lambda^{-1}\Delta_\Lambda\Lambda^{-1}\Delta_\Lambda) \\
    & \quad -\frac n2 \Tr(\Lambda^{-1}\Delta_\Lambda\Lambda^{-1}\Theta^TS_{xx}\Theta\Lambda^{-1}\Delta_\Lambda) + n\Tr(\Lambda^{-1}\Delta_\Lambda\Lambda^{-1}\Theta^TS_{xx}\Delta_\Theta) \\
    & \quad - n\Tr(S_{xy}^T\Delta_\Theta) -n\Tr(\Lambda^{-1}\Theta^TS_{xx}\Delta_\Theta) - \frac n2\Tr(\Lambda^{-1}\Delta_\Theta^TS_{xx}\Delta_\Theta).
\end{align*}
Given this second-order approximation, at each step, we estimate the direction of change $(\Delta_\Theta,\Delta_\Lambda)$ as 
\begin{align}
    \argmin_{\Delta_\Theta,\Delta_\Lambda} \bigg\{& -g(\Delta_{\Theta},\Delta_\Lambda) +\sum_{\substack{i \in \{1,2,\dots,p\}\\j \in \{1,2,\dots,p\}\\
i<j}}\bigg(\frac{(1-p_{ij}^{\Lambda(t)})}{\vlzero} + \frac{p_{ij}^{\Lambda(t)}}{\vlone}\bigg)|\Lambda_{ij}+\Delta_{\Lambda_{ij}}| \nonumber\\ &+ \sum_{\substack{i \in \{1,2,\dots,q\}\\j \in \{1,2,\dots,p\}}}\bigg(\frac{(1-p_{ij}^{\Theta(t)})}{\vlzero} + \frac{p_{ij}^{\Theta(t)}}{\vlone}\bigg)|\Theta_{ij}+\Delta_{\Theta_{ij}}| \bigg\}.
\label{eq 7}
\end{align}
We solve \ref{eq 7} by implementing a cyclic coordinate descent algorithm. Here, we update the value of $\Delta_{\Theta_{ij}} \rightarrow \Delta_{\Theta_{ij}} + ue_ie_j^T$ and of $\Delta_{\Lambda_{ij}} \rightarrow \Delta_{\Lambda_{ij}} + u(e_ie_j^T+e_je_i^T)$ (due to the symmetric nature of $\Lambda$). We iterate over all the coordinates to find the full updating direction $\Delta_\Theta,\Delta_\Lambda$. Due to the separable nature of the objective function, we can divide the optimization into three subproblems.
\begin{enumerate}
    \item Update entries of $\Delta_\Theta$,
    \item Update off-diagonal entries of $\Delta_\Lambda$,
    \item Update diagonal entries of $\Delta_\Lambda$.
\end{enumerate}

We will show below that each of these sub-problems simplifies to the simple lasso optimization function, i.e. take the form 
$$\argmin_d \left(\frac12 ad^2+bd+\tau|c+d|\right)$$
The closed-form solution to the above objective function is 
$$d = -c + S_{\tau/a}\left(c - \frac ba\right),$$ where $S_\tau(x) = sign(x)\max(|x|-\tau,0)$
\subsubsection{Updating entries of $\Delta_\Theta$ }

Consider $g\big(\Delta_\Theta+ue_ie_j^T,\Delta_\Lambda\big)$.
\begin{align*}
    g\big((\Delta_\Theta+ue_ie_j^T),\Delta_\Lambda\big) &= n\Tr\bigg[\left(\Lambda^{-1}\Delta_\Lambda\Lambda^{-1}\Theta^TS_{xx} - S_{xy}^T - \Lambda^{-1}\Theta^TS_{xx}\right)ue_ie_j^T\bigg]\\
    &\quad \quad -\frac n2 \Tr\bigg[\Lambda^{-1}\left(\Delta_\Theta+ue_ie_j^T\right)^TS_{xx}\left(\Delta_\Theta+ue_ie_j^T\right)\bigg]\\
    &= n\bigg[\left(\Lambda^{-1}\Delta_\Lambda\Lambda^{-1}\Theta^TS_{xx} - S_{xy}^T - \Lambda^{-1}\Theta^TS_{xx}-\Lambda^{-1}\Delta_\Theta^TS_{xx}\right)_{ji}u \\&  \quad \quad- \Lambda^{-1}_{jj}(S_{xx})_{ii}u^2\bigg].
\end{align*}
Thus, our optimization problem is as follows:
\begin{align*}
    \argmax_u &-n\bigg[\left(\Lambda^{-1}\Delta_\Lambda\Lambda^{-1}\Theta^TS_{xx} - S_{xy}^T - \Lambda^{-1}\Theta^TS_{xx}-\Lambda^{-1}\Delta_\Theta^TS_{xx}\right)_{ji}u - \Lambda^{-1}_{jj}(S_{xx})_{ii}u^2\bigg]\\ &+\left(\frac{p_{ij}^{\Theta(t)}}{\vtone}+\frac{1-p_{ij}^{\Theta(t)}}{\vtzero}\right)|\Theta_{ij}+ \Delta_{\Theta_{ij}}+u|.
\end{align*}
The above problem has the form of the simple lasso optimization problem, and thus, we can find a,b,c and $\tau$ as follows
$$\begin{cases}
& a = n\Lambda^{-1}_{jj}(S_{xx})_{ii}\\
&b = n \left(-\Lambda^{-1}\Delta_\Lambda\Lambda^{-1}\Theta^TS_{xx} + S_{xy}^T + \Lambda^{-1}\Theta^TS_{xx}+\Lambda^{-1}\Delta_\Theta^TS_{xx}\right)_{ji}\\
& c = \Theta_{ij}+ \Delta_{\Theta_{ij}}\\
& \tau = \left(\frac{p_{ij}^{\Theta(t)}}{\vtone}+\frac{1-p_{ij}^{\Theta(t)}}{\vtzero}\right).
\end{cases}$$
\subsubsection{Updating off-diagonal entries of $\Delta_\Lambda$}

Let $A = \Lambda^{-1}\Theta^TS_{xx}\Theta\Lambda^{-1}$. Consider $g\big(\Delta_\Theta,\Delta_\Lambda+u(e_ie_j^T+e_je_i^T)\big)$.

\begin{align*}
    g\big(\Delta_\Theta,\Delta_\Lambda+u(e_ie_j^T+e_je_i^T)\big) &= \frac n2tr\bigg[\big(\Lambda^{-1}-S_{yy}+A+2\Lambda^{-1}\Delta_\Theta^TS_{xx}\Theta\Lambda^{-1} - \Lambda^{-1}\Delta_\Lambda\Lambda^{-1}\\
    &\quad \quad -2\Lambda^{-1}\Delta_\Lambda\Lambda^{-1}\Theta^TS_{xx}\Theta\Lambda^{-1}\big)u(e_ie_j^T+e_je_i^T)\\
    & \quad \quad - \frac 12 \Lambda^{-1}\left(\Delta_\Lambda + u(e_ie_j^T+e_je_i^T)\right)\Lambda^{-1}\left(\Delta_\Lambda+u(e_ie_j^T+e_je_i^T)\right)\bigg]\\
    &= n\bigg[\Lambda^{-1}_{ij} - (S_{yy})_{ij}+A_{ij} + (\Lambda^{-1}\Delta_\Theta^TS_{xx}\Theta\Lambda^{-1})_{ij+ji}\\
    &\quad \quad   - (\Lambda^{-1}\Delta_\Lambda\Lambda^{-1})_{ij} - (\Lambda^{-1}\Delta_\Lambda A)_{ij+ji}\bigg]u \\&  \quad \quad -\frac n2 \bigg[\Lambda^{-2}_{ij}+ \Lambda^{-1}_{jj}\Lambda^{-1}_{ii} + 2\Lambda^{-1}_{ij}A_{ij} + \Lambda^{-1}_{ii}A_{jj} + \Lambda^{-1}_{jj}A_{ii}\bigg]u^2.
\end{align*}
Thus, our optimization problem is as follows:
\begin{align*}
    \argmax_u & -n\bigg[\Lambda^{-1}_{ij} - (S_{yy})_{ij}+A_{ij} + (\Lambda^{-1}\Delta_\Theta^TS_{xx}\Theta\Lambda^{-1})_{ij+ji} - (\Lambda^{-1}\Delta_\Lambda\Lambda^{-1})_{ij}  \\&- (\Lambda^{-1}\Delta_\Lambda A)_{ij+ji}\bigg]u    +\frac n2 \bigg[\Lambda^{-2}_{ij}+ \Lambda^{-1}_{jj}\Lambda^{-1}_{ii} + 2\Lambda^{-1}_{ij}A_{ij} + \Lambda^{-1}_{ii}A_{jj} + \Lambda^{-1}_{jj}A_{ii}\bigg]u^2\\&+\left(\frac{p_{ij}^{\Lambda(t)}}{\vlone}+\frac{1-p_{ij}^{\Lambda(t)}}{\vlzero}\right)|\Lambda_{ij}+ \Delta_{\Lambda_{ij}}+u|.
\end{align*}
Once again, the above problem has the form of the simple lasso optimization problem, and thus we can find a,b,c and $\tau$ as follows
$$\begin{cases}
& a = \frac n2 \bigg(\Lambda^{-2}_{ij}+ \Lambda^{-1}_{jj}\Lambda^{-1}_{ii} + 2\Lambda^{-1}_{ij}A_{ij} + \Lambda^{-1}_{ii}A_{jj} + \Lambda^{-1}_{jj}A_{ii}\bigg)\\
&b = -n \bigg(\Lambda^{-1}_{ij} - (S_{yy})_{ij}+A_{ij} + (\Lambda^{-1}\Delta_\Theta^TS_{xx}\Theta\Lambda^{-1})_{ij+ji} - (\Lambda^{-1}\Delta_\Lambda\Lambda^{-1})_{ij}  - (\Lambda^{-1}\Delta_\Lambda A)_{ij+ji}\bigg)\\
& c = \Lambda_{ij}+ \Delta_{\Lambda_{ij}}\\
& \tau = \left(\frac{p_{ij}^{\Lambda(t)}}{\vlone}+\frac{1-p_{ij}^{\Lambda(t)}}{\vlzero}\right).
\end{cases}$$
\subsubsection{Updating diagonal entries of $\Delta_\Lambda$}
Consider $g\big(\Delta_\Theta,\Delta_\Lambda+ue_ie_iT\big)$.
\begin{align*}
    g\big(\Delta_\Theta,\Delta_\Lambda+ue_ie_i^T\big) &= \frac n2tr\bigg[\big(\Lambda^{-1}-S_{yy}+A+2\Lambda^{-1}\Delta_\Theta^TS_{xx}\Theta\Lambda^{-1} - \Lambda^{-1}\Delta_\Lambda\Lambda^{-1}\\&  \quad \quad 
    -2\Lambda^{-1}\Delta_\Lambda\Lambda^{-1}\Theta^TS_{xx}\Theta\Lambda^{-1}\big)ue_ie_i^T\\
    & \quad \quad - \frac 12 \Lambda^{-1}\left(\Delta_\Lambda + ue_ie_i^T\right)\Lambda^{-1}\left(\Delta_\Lambda+ue_ie_i^T\right)\bigg]\\
    &= n\bigg[\Lambda^{-1}_{ii} - (S_{yy})_{ii}+A_{ii} + 2(\Lambda^{-1}\Delta_\Theta^TS_{xx}\Theta\Lambda^{-1})_{ii} - (\Lambda^{-1}\Delta_\Lambda\Lambda^{-1})_{ii}\\
    &\quad \quad  - 2(\Lambda^{-1}\Delta_\Lambda A)_{ii}\bigg]u  -n\bigg[\Lambda^{-2}_{ii} + 2\Lambda^{-1}_{ii}A_{ii} \bigg]u^2.
\end{align*}
Thus, our optimization problem is as follows:
\begin{align*}
    \argmax_u & -n\bigg[\Lambda^{-1}_{ii} - (S_{yy})_{ii}+A_{ii} + 2(\Lambda^{-1}\Delta_\Theta^TS_{xx}\Theta\Lambda^{-1})_{ii} - (\Lambda^{-1}\Delta_\Lambda\Lambda^{-1})_{ii}  - 2(\Lambda^{-1}\Delta_\Lambda A)_{ii}\bigg]u \\&   +n \bigg[\Lambda^{-2}_{ii} + 2\Lambda^{-1}_{ii}A_{ii}\bigg]u^2.
\end{align*}
The above problem is the simple lasso optimization problem, and thus, we can find a,b,c and $\tau$ as follows:

$$\begin{cases}
& a =  n \bigg(\Lambda^{-2}_{ii} + 2\Lambda^{-1}_{ii}A_{ii}\bigg)\\
&b = -n \bigg(\Lambda^{-1}_{ii} - (S_{yy})_{ii}+A_{ii} + 2(\Lambda^{-1}\Delta_\Theta^TS_{xx}\Theta\Lambda^{-1})_{ii} - (\Lambda^{-1}\Delta_\Lambda\Lambda^{-1})_{ii}  - 2(\Lambda^{-1}\Delta_\Lambda A)_{ii}\bigg)\\
& c = \Lambda_{ii}+ \Delta_{\Lambda_{ii}}\\
& \tau = 0.
\end{cases}$$

\subsubsection{Updating the estimates of $\Theta$ and $\Lambda$}
Given the estimates of $\Delta_\Theta$ and $\Delta_\Lambda$ that we get by solving \ref{eq 7}, we compute the estimates of $\Theta$ and $\Lambda$ by
$$\Theta \rightarrow \Theta + \alpha\Delta_\Theta$$
$$\Lambda \rightarrow \Lambda +\alpha\Delta_\Lambda,$$
where $\alpha$ is the step size decided according to three conditions: \begin{enumerate}
    \item Armijo's Rule
    \item Positive Definiteness of $\Lambda$ and
    \item Boundedness of the $\Lambda$
\end{enumerate}

\subsection{Techniques for enhancing computational efficiency}
\subsubsection{Updating $(\Delta_\Theta\Lambda^{-1},\Delta_\Lambda\Lambda^{-1})$ instead of $(\Delta_\Theta,\Delta_\Lambda)$}
Note that all the optimization problems in the M-Step above, which are functions of $(\Delta_\Theta,\Delta_\Lambda)$, can be reparametrized as functions of $(\Delta_\Theta\Lambda^{-1},\Delta_\Lambda\Lambda^{-1})$. This, combined with the fact that updating an element in $\Delta_\Theta$ (or $\Delta_\Lambda$) is the same as updating the corresponding row in $\Delta_\Theta\Lambda^{-1}$ (or $\Delta_\Lambda\Lambda^{-1}$) helps us increase the computational efficiency of our algorithm. We do this by computing the value of $(\Delta_\Theta\Lambda^{-1},\Delta_\Lambda\Lambda^{-1})$ at the start of the M-Step and then updating their appropriate row in each iteration of the coordinate decent, instead of naively updating $\Delta_\Theta$ and $\Delta_\Lambda$ and then computing $(\Delta_\Theta\Lambda^{-1},\Delta_\Lambda\Lambda^{-1})$ in each iteration. 

\subsubsection{Updating only on an Active Set}
Since the matrices we want to estimate are sparse, we leverage this information to make our algorithm faster by only updating the elements in the active set during our coordinate descent instead of updating all the elements. We define the active set as follows:

\begin{itemize}
    \item For $\Theta$:
    \begin{equation}\left|\big(\nabla_\Theta l(\Theta,\Lambda)\big)_{ij}\right|> \left|\big(\nabla_\Theta Pen(\Theta)\big)_{ij}\right| \quad \text{ or } \quad \Theta_{ij} \neq 0; \label{thetaactiveset} \end{equation}
    \item For $\Lambda$:
    \begin{equation}\left|\big(\nabla_\Lambda l(\Theta,\Lambda)\big)_{ij}\right|> \left|\big(\nabla_\Lambda Pen(\Lambda)\big)_{ij}\right| \quad \text{ or } \quad \Lambda_{ij} \neq 0, \label{lambdaactiveset}\end{equation}
\end{itemize}
where $Pen(\Theta) = \sum_i Pen_{MSS}(\Theta_i)$ and $Pen(\Lambda) = \sum_{i<j} Pen_{SS}(\Lambda_{ij})$. The rationale is that if an element was estimated as 0 in the previous iteration and the penalty's derivative at the coordinate is greater than the derivative of the likelihood function, the element will not shift from 0 during that coordinate descent iteration anyway. This trick leads to a substantial speedup in our algorithm, particularly when the matrices we estimate are highly sparse. 

\section{Proof of technical lemmas}
In Section 4.1 of our main paper, we state the following property of our likelihood function, which we now present as a lemma.
\begin{Lemma}
If $1/\max\{\vtone,\vlone\}>2\|l(\Theta^0,\Lambda^0)\|_{\infty}$, then for any $(\Theta,\Lambda) = (\Theta^0,\Lambda^0) + \Delta$ such that $(\Theta^0,\Lambda^0) \in HPD$, we have 
$$\|\Delta_{(S^0_{row})^C}\|_1\leq  \alpha \|\Delta_{S^0_{row}}\|_1,$$ where $\alpha = 1 + 2\max\{\vtone,\vlone\}/\min\{\vtzero,\vlzero\}.$
\label{Lemma3}
\end{Lemma}
\begin{proof}
Since $(\Theta,\Lambda) \in HPD$ we have, $L(\Theta,\Lambda) < L(\Theta^0,\Lambda^0)$. Thus, 
\begin{align*}
    -l(\Theta^0,\Lambda^0) &+ \sum_iPen_{MSS}(\Theta_i^0) +\sum_{i<j}Pen_{SS}(\Lambda_{ij}^0)\\ &\geq -l(\Theta,\Lambda) + \sum_iPen_{MSS}(\Theta_i) +\sum_{i<j}Pen_{SS}(\Lambda_{ij}).
\end{align*}
\begin{align} \label{eq9}
        \sum_i\bigg(Pen_{MSS}(\Theta_i^0)&-Pen_{MSS}(\Theta_i)\bigg) + \sum_{i<j}\left(Pen_{SS}(\Lambda_{ij}^0)-Pen_{SS}(\Lambda_{ij})\right)\nonumber\\& \geq l(\Theta^0,\Lambda^0)-l(\Theta,\Lambda)\geq -\inprod{\nabla l(\Theta,\Lambda)}{\Delta}\nonumber\\
        & \geq -\|l(\Theta^0,\Lambda^0)\|_{\infty}\|\Delta\|_1\nonumber\\
        & \geq -\|l(\Theta^0,\Lambda^0)\|_{\infty}\left(\|\Delta_{S^0_{row}}\|_1+\|\Delta_{(S^0_{row})^C}\|_1\right),
    \end{align}
where we get the second inequality due to the convexity of the negative likelihood function $-l(\Theta,\Lambda)$ \citep{yuan2014partial} and the third using Holder's inequality. Next, we look at each of the terms on the LHS. Let us partition $\Delta$ into $(\Delta^\Theta,\Delta^\Lambda)$. Then,
\begin{align*}
    \sum_i\bigg(Pen_{MSS}(\Theta_i^0)-&Pen_{MSS}(\Theta_i)\bigg)\\ &=  \sum_{i:(i,j)\in S_{row}^{\Theta}}\left(Pen_{MSS}(\Theta_i^0)-Pen_{MSS}(\Theta_i)\right)\\ &\quad \quad \quad \quad+ \sum_{i:(i,j)\in (S_{row}^{\Theta})^C}\left(Pen_{MSS}(0)-Pen_{MSS}(\Theta_i)\right)\\
    &\leq  \sum_{i:(i,j)\in S_{row}^{\Theta}}\left(Pen_{MSS}(\Theta_i^0)-Pen_{MSS}(\Theta_i^0 +\Delta^\Theta_i)\right)\\ &\quad \quad \quad \quad- \sum_{i:(i,j)\in (S_{row}^{\Theta})^C}Pen_{MSS}(\Delta^\Theta_i)\\
    &\leq -\frac{1}{\vtone}\|\Delta^\Theta_{(S_{row}^{\Theta})^C}\|_1 + \frac{1}{\vtzero}\|\Delta^\Theta_{S_{row}^{\Theta}}\|_1.
\end{align*}
Note that in the first equality, we split the summation depending on whether the rows are \textbf{0}. Thus for $i:(i,j)\in (S_{row}^{\Theta})^C$, $\Theta_i = \textbf{0}$. This, along with the fact that $Pen_{MSS}(0)<0$, leads to the second inequality. We get the final inequality due to the following properties of the mixed spike and slab prior: For any vectors $A,B\in \mathbbm{R}^p$, $Pen_{MSS}(A) \geq \sum_{k = 1}^p|A_k|/\vtone$ and $Pen_{MSS}(B) - Pen_{MSS}(A) \leq \sum_{k = 1}^p|B_k-A_k|/\vtzero$. These properties follow from the concavity of the mixed spike and slab penalty function along with the form of its first derivative (Proposition \ref{prop1}). Similarly
\begin{align*}
    \sum_{i<j}\bigg(Pen_{SS}(\Lambda_{ij}^0)-&Pen_{SS}(\Lambda_{ij})\bigg) \\&=  \sum_{(i,j)\in S^{\Lambda}}\left(Pen_{SS}(\Lambda_{ij}^0)-Pen_{SS}(\Lambda_{ij})\right)\\ &\quad \quad \quad \quad+ \sum_{(i,j)\in (S^{\Lambda})^C}\left(Pen_{SS}(0)-Pen_{SS}(\Lambda_{ij})\right)\\
    &\leq  \sum_{(i,j)\in S^{\Lambda}}\left(Pen_{SS}(\Lambda_{ij}^0)-Pen_{SS}(\Lambda_{ij}^0 +\Delta^\Lambda)\right)\\ &\quad \quad \quad \quad- \sum_{(i,j)\in (S^{\Lambda})^C}Pen_{SS}(\Delta^\Lambda)\\
    &\leq -\frac{1}{\vlone}\|\Delta^\Lambda_{(S^{\Lambda})^C}\|_1 + \frac{1}{\vlzero}\|\Delta^\Lambda_{S^{\Lambda}}\|_1,
\end{align*}
where we get the second inequality because $Pen_{SS}(0)<0$ and the final inequality due to the following properties of the spike and slab prior: $Pen_{SS}(a) \geq |a|/\vtone$ and $Pen_{SS}(b) - Pen_{MSS}(a) \leq |b-a|/\vtzero$. Similar to before, these properties arise from the concavity of the spike and slab penalty function, coupled with the structure of its first derivative (Proposition \ref{prop1}). Thus we can bound the LHS of (\ref{eq9}) as follows:
\begin{align}
\label{eq10}
    \sum_i\bigg(Pen&_{MSS}(\Theta_i^0)-Pen_{MSS}(\Theta_i)\bigg) + \sum_{i<j}\left(Pen_{SS}(\Lambda_{ij}^0)-Pen_{SS}(\Lambda_{ij})\right)\nonumber \\
    &\leq  -\frac{1}{\vtone}\|\Delta^\Theta_{(S_{row}^{\Theta})^C}\|_1 + \frac{1}{\vtzero}\|\Delta^\Theta_{S_{row}^{\Theta}}\|_1 -\frac{1}{\vlone}\|\Delta^\Lambda_{(S^{\Lambda})^C}\|_1 + \frac{1}{\vlzero}\|\Delta^\Lambda_{S^{\Lambda}}\|_1\nonumber\\
    &\leq -\frac{1}{\max\{\vtone,\vlone\}}\|\Delta_{(S^0_{row})^C}\|_1+\frac{1}{\min\{\vtzero,\vlzero\}}\|\Delta_{S^0_{row}}\|_1.
\end{align} Using (\ref{eq9}) and (\ref{eq10}) we get \begin{align*}-\|l(\Theta^0,\Lambda^0)\|_{\infty}&\left(\|\Delta_{S^0_{row}}\|_1+\|\Delta_{(S^0_{row})^C}\|_1\right) \\&\leq -\frac{1}{\max\{\vtone,\vlone\}}\|\Delta_{(S^0_{row})^C}\|_1+\frac{1}{\min\{\vtzero,\vlzero\}}\|\Delta_{S^0_{row}}\|_1,
\end{align*}
\begin{align*}
    \implies \|\Delta_{(S^0_{row})^C}\|_1&\leq \|\Delta_{S^0_{row}}\|_1\frac{\left(\frac{1}{\min\{\vtzero,\vlzero\}}+\|l(\Theta^0,\Lambda^0)\|_{\infty}\right)}{\left(\frac{1}{\max\{\vtone,\vlone\}}-\|l(\Theta^0,\Lambda^0)\|_{\infty}\right)}\\
    &\leq \alpha \|\Delta_{S^0_{row}}\|_1,
\end{align*}
where we get the last inequality using the condition $1/\max\{\vtone,\vlone\}>2\|l(\Theta^0,\Lambda^0)\|_{\infty}$
\end{proof}

\section{Proof of theorems}
\subsection{Proof for Theorem 1}
\begin{proof}
    Consider a $(\Theta,\Lambda) \in HPD$ and let $\Delta := (\Delta_\Theta,\Delta_\Lambda) = (\Theta,\Lambda)-(\Theta^0,\Lambda^0)$. Let us also assume that $\|\Delta\|_F^2 < r_0$. We will show later that this always holds. Since $(\Theta,\Lambda) \in HPD$ we have, 
    $L(\Theta,\Lambda)\leq L(\Theta^0,\Lambda^0),$
    \begin{align*}
        \implies 0 \geq l(\Theta^0,\Lambda^0) -l(\Theta,\Lambda)& + \sum_i\left(Pen_{MSS}(\Theta_i)-Pen_{MSS}(\Theta_i^0)\right) \\ &\quad  + \sum_{i<j}\left(Pen_{SS}(\Lambda_{ij})-Pen_{SS}(\Lambda_{ij}^0)\right).
    \end{align*}
Since $\|\Delta\|_F^2 < r_0$ and $(\Theta,\Lambda) \in HPD$, the result for the LRSC in Section 4.1 of our main paper can be used to show that $l(\Theta^0,\Lambda^0) -l\big((\Theta^0,\Lambda^0)+\Delta\big)+ \inprod{\nabla l(\Theta^0,\Lambda^0)}{\Delta}\geq n\beta_0\|\Delta\|_F^2$. This along with the inequality (\ref{eq10}) from the proof of Lemma \ref{Lemma3}, gives us
\begin{align*}
    0&\geq n\beta_0\|\Delta\|_F^2 -\inprod{\nabla l(\Theta^0,\Lambda^0)}{\Delta} +\frac{1}{\max\{\vtone,\vlone\}}\|\Delta_{(S^0_{row})^C}\|_1\\&\quad \quad \quad-\frac{1}{\min\{\vtzero,\vlzero\}}\|\Delta_{S^0_{row}}\|_1\\
    & \geq n\beta_0\|\Delta\|_F^2+\left(\frac{1}{\max\{\vtone,\vlone\}}- \|\nabla l(\Theta^0,\Lambda^0)\|_{\infty}\right)\|\Delta_{(S^0_{row})^C}\|_1\\& \quad \quad \quad -\left(\frac{1}{\min\{\vtzero,\vlzero\}}+ \|\nabla l(\Theta^0,\Lambda^0)\|_{\infty}\right)\|\Delta_{S^0_{row}}\|_1\\
    & \geq n\beta_0\|\Delta\|_F^2-\left(\frac{1}{\min\{\vtzero,\vlzero\}}+ \|\nabla l(\Theta^0,\Lambda^0)\|_{\infty}\right)\|\Delta_{S^0_{row}}\|_1\\
    &\geq n\beta_0\|\Delta\|_F^2 - \left(\frac{1}{\min\{\vtzero,\vlzero\}}+ \|\nabla l(\Theta^0,\Lambda^0)\|_{\infty}\right)\|\Delta\|_F\sqrt{|S^0_{row}|},
\end{align*}
where we get the second inequality using Holders' and Triangle inequalities, the third inequality using the condition $1/\max\{\vtone,\vlone\}>2\|\nabla l(\Theta^0,\Lambda^0)\|_{\infty}$ which causes the second term to be greater than 0. We get the final inequality by once again using the Holders' inequality and noting that $\|\Delta_{S^0_{row}}\|_F \leq\|\Delta\|_F$.  Hence,
    \begin{align*}
        \|\Delta\|_F &\leq \left(\frac{1}{\min\{\vtzero,\vlzero\}}+ \|\nabla l(\Theta^0,\Lambda^0)\|_{\infty}\right)\frac{\sqrt{|S^0_{row}|}}{n\beta_0}\\
        &\leq \left(\frac{1}{n\min\{\vtzero,\vlzero\}}+\frac{1}{n\max\{\vtone,\vlone\}}\right)\frac{\sqrt{|S^0_{row}|}}{\beta_0}\\
        &\leq \frac{C_1+C_2}{\beta_0}\sqrt{\frac{|S^0_{row}|log(p+q)}{n}},
    \end{align*}
where we use the first condition, which gives the following relations for our hyperparameters: $$\frac{1}{n\min\{\vtzero,\vlzero\}}<\frac{C_1}{1+2\epsilon_1}\sqrt{\frac{\log(p+q)}{n}},$$
    $$\frac{1}{n\max\{\vtone,\vlone\}}<C_2\sqrt{\frac{\log(p+q)}{n}},$$ 
    for some positive constants $C_1$ and $C_2$, 
to get the last two inequalities.

Finally, all we need to show is that $\|\Delta\|_F^2 < r_0$. We will do so through proof by contradiction. Say $\|\Delta\|_F^2 \geq r_0$, then following the proof of Proposition 3 in \cite{yuan2014partial} we can show that 
    
    \begin{equation}
    \label{eq 12}
        \|\Delta\|_F \leq  \frac{C_1+C_2}{b_n}\sqrt{\frac{|S^0_{row}|log(p+q)}{n}},
    \end{equation}$\text{where }b_n = 2\beta_0 \frac{\lambda_{max}(\Lambda^0)}{\lambda_{max}(\Lambda)}\geq 2\beta_0 \frac{\lambda_{max}(\Lambda^0)}{R}$ due to the definition of R as the matrix norm bound of $\Lambda$. Using (\ref{eq 12}) and the second condition, we then get,
    $$\|\Delta\|_F \leq \frac{R}{2\lambda_{max}(\Lambda^0)}\epsilon_n< \sqrt{r_0},$$ which leads to a contradiction. Let us now consider $B = -\Lambda^{-1}\Theta^T$ where $(\Theta,\Lambda)\in HPD$. Then, for such a $B$ we get,
    \begin{align*}
        \|B - B^0\|_F &= \|\Lambda^{-1}\Theta^T - (\Lambda^0)^{-1}(\Theta^0)^T\|_F\\
        &= \|\Lambda^{-1}\Theta^T- (\Lambda^0)^{-1}\Theta^T+ (\Lambda^0)^{-1}\Theta^T  -(\Lambda^0)^{-1}(\Theta^0)^T\|_F\\
        &\leq \|\Lambda^{-1}- (\Lambda^0)^{-1}\|_F\|\Theta^T\|_F + \|(\Lambda^0)^{-1}\|_F\|\Theta^T - (\Theta^0)^T\|_F\\
        &\leq \left(\frac{F_\Lambda^2\epsilon_n}{1 - F_\Lambda\epsilon_n}\right)\left(F_\Theta + \epsilon_n\right) + F_\Lambda\epsilon_n\\
        & = \frac{F_\Lambda\epsilon_n}{1- F_\Lambda\epsilon_n}(1+F_\Lambda F_\Theta)
    \end{align*}
We use $\|A^{-1} - B^{-1}\| \leq \frac{\|A^{-1}\|^2\|A - B\|}{1 - \|A^{-1}\|\|A - B\|}$ for any sub-multiplicative norm and $\|A\| \leq \|A- B\| + \|B\|$ to get the fourth inequality above. 
\end{proof}
\subsection{Proof for Theorem 2 and uniqueness}
\textbf{Proof of Theorem 2:}
\begin{proof}
    We build a two-step construction-based proof for this theorem. In the first step, we consider a $(\Theta,\Lambda)$ that minimizes the negative log posterior function constrained on them taking 0 values in the inactive set and show that it satisfies $\|\Theta - \Theta^0\|_\infty \leq r_n$ and $\|\Lambda - \Lambda^0\|_\infty \leq r_n$. In the second step, we show that the $(\Theta,\Lambda)$ considered above is a local minimizer of the negative log posterior function $L(\Theta,\Lambda)$ in the HPD region, thus proving that a local minimizer of our objective function has the desired properties. We now elaborate on each of these steps. 

  \textbf{Step 1:} Consider the restricted optimization problem
    $$\min L(\theta,\Lambda) \text{ s.t. } \Theta_{(S^\Theta)^C}=0 \text{ and } \Lambda_{(S^\Lambda)^C}=0.$$
    Let $\mathcal{A}$ be the solution set for the above restricted optimization problem, i.e 
    $$\mathcal{A} = \{(\Theta,\Lambda): \nabla L(\Theta,\Lambda)\big|_{S^0:= S^\Theta\bigcup S^\Lambda} = 0, \Theta_{(S^\Theta)^C}=0 \text{ and } \Lambda_{(S^\Lambda)^C}=0\}.$$ 
We want to prove that $\exists (\hat\Theta,\hat\Lambda)\in \mathcal{A}$ such that  $\|\hat\Theta - \Theta^0\|_\infty \leq r_n$ and $\|\hat\Lambda - \Lambda^0\|_\infty \leq r_n$. Note that since $\hat\Theta_{(S^\Theta)^C}= \Theta^0_{(S^\Theta)^C}=0$ and $\hat\Lambda_{(S^\Lambda)^C} =\Lambda^0_{(S^\Lambda)^C}=0$, it suffices to prove that $\exists (\hat\Theta,\hat\Lambda)\in \mathcal{A}$ such that  $\|\left(\hat\Theta - \Theta^0\right)_{S^\Theta}\|_\infty \leq r_n$ and $\|\left(\hat\Lambda - \Lambda^0\right)_{S^\Lambda}\|_\infty \leq r_n$.
Recall, a $(\hat\Theta,\hat\Lambda) \in \mathcal{A}$ must satisfy the zero sub-gradient conditions: 
    \begin{equation}
    \label{eq13}
        -\bigg(\nabla_\Theta l(\hat\Theta,\hat\Lambda)\bigg)_{S^\Theta} + \bigg(U(\hat\Theta)\bigg)_{S^\Theta} = 0;
    \end{equation}\begin{equation}
    \label{eq14}
        -\bigg(\nabla_\Lambda l(\hat\Theta,\hat\Lambda)\bigg)_{S^\Lambda} + \bigg(V(\hat\Lambda)\bigg)_{S^\Lambda} = 0,
    \end{equation} where $U(\Theta)_{ij} =  \frac{\partial}{\partial|\Theta_{ij}|}Pen_{MSS}(\Theta_i)sgn(\Theta_{ij})$, $V(\Lambda)_{ij} =  \frac{\partial}{\partial|\Lambda_{ij}|}Pen_{SS}(\Lambda_{ij})sgn(\Lambda_{ij})$ and $sgn(x)$ is the sub-gradient of $|x|$ which takes the value $1$ if $x>0$, $-1$ if $x<0$ and lies in the interval $[-1,1]$ when $x = 0$. 
    
    The first order Taylor expansion of $\bigg(\nabla l(\hat\Theta,\hat\Lambda)\bigg) = \bigg(\nabla l(\Theta^0+\hat\Delta_\Theta,\Lambda^0+\hat\Delta_\Lambda)\bigg)$ at $(\Theta^0,\Lambda^0)$ is $$vec\left(\nabla l(\hat\Theta,\hat\Lambda)\right) = vec\left(\nabla l(\Theta^0,\Lambda^0)\right) + Hvec\left(\hat\Delta\right)+ R\left(\hat \Delta\right),$$ where $\hat\Delta := (\hat\Delta_\Theta,\hat\Delta_\Lambda)$, $H$ is the Hessian and $R\left(\hat \Delta\right)$ is the residual. Let $l_\Lambda(\Theta)$ be the likelihood function $l(\Theta, \Lambda)$ in terms of $\Theta$ keeping $\Lambda$ constant and $l_\Theta(\Lambda)$ be the likelihood function $l(\Theta, \Lambda)$ in terms of $\Lambda$ keeping $\Theta$ constant. Thus (\ref{eq13}) and (\ref{eq14}) can be expressed as 
    \begin{equation}
    \label{eq15}
        vec\left(\nabla l_{\Lambda^0}(\Theta^0)\right)_{S^\Theta} + H_{S^\Theta S^\Theta}vec\left(\hat\Delta_\Theta\right)_{S^\Theta}+ \left(R(\hat \Delta_\Theta)\right)_{S^\Theta} - vec\bigg(U(\hat\Theta)\bigg)_{S^\Theta} = 0;
    \end{equation}
    \begin{equation}
    \label{eq16}
        vec\left(\nabla l_{\Theta^0}(\Lambda^0)\right)_{S^\Lambda} + H_{S^\Lambda S^\Lambda}vec\left(\hat\Delta_\Lambda\right)_{S^\Lambda}+ \left(R(\hat \Delta_\Lambda)\right)_{S^\Lambda} - vec\bigg(V(\hat\Lambda)\bigg)_{S^\Lambda} = 0.
    \end{equation}
We first explore (\ref{eq15}) and obtain the results for $\Theta$. From (\ref{eq15}) we get
    $$ vec\left(\hat\Delta_\Theta\right)_{S^\Theta} = -H_{S^\Theta S^\Theta}^{-1}\left[vec\left(\nabla l_{\Lambda^0}(\Theta^0)\right)_{S^\Theta} + \left(R(\hat \Delta_\Theta)\right)_{S^\Theta} - vec\bigg(U(\hat\Theta)\bigg)_{S^\Theta}\right].$$
Let $F: \mathbbm{R}^{|S^\Theta|}\rightarrow\mathbbm{R}^{|S^\Theta|}$ be a mapping defined as
 $$F\left(vec\left(\hat\Delta_\Theta\right)_{S^\Theta}\right) =-H_{S^\Theta S^\Theta}^{-1}\left[vec\left(\nabla l_{\Lambda^0}(\Theta^0)\right)_{S^\Theta} + \left(R(\hat \Delta_\Theta)\right)_{S^\Theta} - vec\bigg(U(\hat\Theta)\bigg)_{S^\Theta}\right],$$ where $|S^\Theta|= card\{S^\Theta\}$. Additionally, we consider a $\left(\hat\Delta_\Theta\right)_{S^\Theta}$ such that $vec\left(\hat\Delta_\Theta\right)_{S^\Theta} \in \mathbbm{B}(r) \subset \mathbbm{B}(r_n)$ where $r = 4C_1c_H\sqrt{log(p+q)/n}$ and $\mathbbm{B}(r)$ denotes the vector $l_\infty$ norm ball with radius $r$  ($r<r_n$ since $C_1<C_3$).
    Then, we can bound $F\left(vec\left(\hat\Delta_\Theta\right)_{S^\Theta}\right)$ from above as follows:
    \begin{align*}
        \left\|F\left(vec\left(\hat\Delta_\Theta\right)_{S^\Theta}\right)\right\|_\infty &\leq \vertiii{H_{S^\Theta S^\Theta}^{-1}}_\infty\bigg[\left\|vec\left(\nabla l_{\Lambda^0}(\Theta^0)\right)_{S^\Theta}\right\|_\infty + \left\|\left(R(\hat \Delta_\Theta)\right)_{S^\Theta}\right\|_\infty \\
        & \quad \quad \quad \quad+ \left\|vec\bigg(U(\hat\Theta)\bigg)_{S^\Theta}\bigg\|_\infty\right]\\
        &= \frac{2}{n}c_H\bigg[\left\|vec\left(\nabla l_{\Lambda^0}(\Theta^0)\right)_{S^\Theta}\right\|_\infty + \left\|\left(R(\hat \Delta_\Theta)\right)_{S^\Theta}\right\|_\infty\\ &\quad \quad \quad \quad+ \left\|vec\bigg(U(\hat\Theta)\bigg)_{S^\Theta}\right\|_\infty\bigg].
    \end{align*}
Looking at each of the terms individually:

Term 1: $\frac2n c_H \left\|vec\left(\nabla l_{\Lambda^0}(\Theta^0)\right)_{S^\Theta}\right\|_\infty$
\begin{align*}
\frac2n c_H \left\|vec\left(\nabla l_{\Lambda^0}(\Theta^0)\right)_{S^\Theta}\right\|_\infty &\leq \frac2n c_H \left\|\nabla l(\Lambda^0,\Theta^0)\right\|_\infty\\
    & \leq c_H\left(\frac{C_1}{1+2\epsilon_1}\right)\sqrt{\frac{log(p+q)}{n}}\\
    &< C_3c_H\sqrt{\frac{log(p+q)}{n}}\\
    &= \frac{r_n}{4},
\end{align*}
where we get the second inequality using the first condition and the third inequality because $C_1<C_3$.

Term 2: $\frac 2n c_H \left\|\left(R(\hat \Delta_\Theta)\right)_{S^\Theta}\right\|_\infty$
\begin{align*}
    \frac 2n c_H \left\|\left(R(\hat \Delta_\Theta)\right)_{S^\Theta}\right\|_\infty &\leq  \frac{1}{4r_n}\left\|\left(\hat\Delta_\Theta\right)_{S^\Theta}\right\|_\infty^2\leq \frac{r_n}{4},
\end{align*}
where we get the first inequality using Lemma 2 and the second because $\left(\hat\Delta_\Theta\right)_{S^\Theta} \in \mathbbm{B}(r_n)$ which implies $\left\|\left(\hat\Delta_\Theta\right)_{S^\Theta}\right\|_\infty \leq r_n$.

Term 3: $\frac2n c_H\left\|vec\bigg(U(\hat\Theta)\bigg)_{S^\Theta}\right\|_\infty$
\begin{align*}
    \frac2n c_H\left\|vec\bigg(U(\hat\Theta)\bigg)_{S^\Theta}\right\|_\infty &\leq 2c_HC_1\sqrt{\frac{log(p+q)}{n}} < 2c_HC_3\sqrt{\frac{log(p+q)}{n}}= \frac{r_n}{2},
\end{align*}
where we get the first inequality using the upper bound of the first derivative of the penalty derived in Proposition \ref{prop2} which can be used since $|\Theta_{ij}|_{S^\Theta} \geq |\Theta^0_{ij}|_{S^\Theta}- |\hat\Delta_{ij}|_{S^\Theta} \geq r_n- r \geq 2(C_3 - C_1)\sqrt{\log(p+q)/n}$. Putting the results for each of the terms together, we get,
\begin{align*}
\left\|F\left(vec\left(\hat\Delta_\Theta\right)_{S^\Theta}\right)\right\|_\infty &\leq \frac{2}{n}c_H\bigg[\left\|vec\left(\nabla l_{\Lambda^0}(\Theta^0)\right)_{S^\Theta}\right\|_\infty + \left\|\left(R(\hat \Delta_\Theta)\right)_{S^\Theta}\right\|_\infty \\&\quad\quad\quad\quad+ \left\|vec\bigg(U(\hat\Theta)\bigg)_{S^\Theta}\right\|_\infty\bigg]\\
        &< \frac{r_n}{4}+\frac{r_n}{4}+\frac{r_n}{2}\\
        &= r_n.
    \end{align*}
This implies that for any $\left(\hat\Delta_\Theta\right)_{S^\Theta} \in \mathbbm{B}(r_n)$ we have $F\left(vec\left(\hat\Delta_\Theta\right)_{S^\Theta}\right) \in \mathbbm{B}(r_n)$. Thus, by Brouwers' fixed point theorem, there exists a fixed point $\left(\hat\Delta_\Theta\right)_{S^\Theta} \in \mathbbm{B}(r_n)$ such that $F\left(vec\left(\hat\Delta_\Theta\right)_{S^\Theta}\right) = vec\left(\hat\Delta_\Theta\right)_{S^\Theta}$ i.e $\left(\hat\Delta_\Theta\right)_{S^\Theta}$ follows the sub-gradient condition (\ref{eq13}). In other words, we have shown that there exists a $\hat\Theta \in \mathcal{A}$ such that $\|\left(\hat\Theta - \Theta^0\right)_{S^\Theta}\|_\infty \leq r_n$. 

Let us next explore  (\ref{eq16}) and obtain the results for $\Lambda$. From (\ref{eq16}) we get
    $$ vec\left(\hat\Delta_\Lambda\right)_{S^\Lambda} = -H_{S^\Lambda S^\Lambda}^{-1}\left[vec\left(\nabla l_{\Theta^0}(\Lambda^0)\right)_{S^\Lambda} + \left(R(\hat \Delta_\Lambda)\right)_{S^\Lambda} - vec\bigg(V(\hat\Lambda)\bigg)_{S^\Lambda}\right].$$
Similar to our computation above, we define a mapping, $G: \mathbbm{R}^{|S^\Lambda|}\rightarrow\mathbbm{R}^{|S^\Lambda|}$ such that,
    $$G\left(vec\left(\hat\Delta_\Lambda\right)_{S^\Lambda}\right) =-H_{S^\Lambda S^\Lambda}^{-1}\left[vec\left(\nabla l_{\Theta^0}(\Lambda^0)\right)_{S^\Lambda} + \left(R(\hat \Delta_\Lambda)\right)_{S^\Lambda} - vec\bigg(V(\hat\Lambda)\bigg)_{S^\Lambda}\right],$$ where $|S^\Lambda|= card\{S^\Lambda\}$ and consider a $\left(\hat\Delta_\Lambda\right)_{S^\Lambda}$ such that $vec\left(\hat\Delta_\Lambda\right)_{S^\Lambda} \in \mathbbm{B}(r) \subset \mathbbm{B}(r_n)$. 
    Following a similar proof to the one done above for $\Theta$ and using the upper bound of the first derivative of the penalty $Pen_{SS}$ derived in Proposition \ref{prop2} (which can be used since $|\Lambda_{ij}|_{S^\Lambda} \geq |\Lambda^0_{ij}|_{S^\Lambda}- |\hat\Delta_{ij}|_{S^\Lambda} \geq r_n- r \geq 2(C_3 - C_1)\sqrt{\log(p+q)/n}$), we can bound $G\left(vec\left(\hat\Delta_\Lambda\right)_{S^\Lambda}\right)$ from above as follows:
    \begin{align*}
        \left\|G\left(vec\left(\hat\Delta_\Lambda\right)_{S^\Lambda}\right)\right\|_\infty &\leq \vertiii{H_{S^\Lambda S^\Lambda}^{-1}}_\infty\bigg[\left\|vec\left(\nabla l_{\Theta^0}(\Lambda^0)\right)_{S^\Lambda}\right\|_\infty + \left\|\left(R(\hat \Delta_\Lambda)\right)_{S^\Lambda}\right\|_\infty \\
        & \quad \quad \quad \quad + \left\|vec\bigg(V(\hat\Lambda)\bigg)_{S^\Theta}\bigg\|_\infty\right]\\
        &= \frac{2}{n}c_H\bigg[\left\|vec\left(\nabla l_{\Theta^0}(\Lambda^0)\right)_{S^\Lambda}\right\|_\infty + \left\|\left(R(\hat \Delta_\Lambda)\right)_{S^\Lambda}\right\|_\infty \\
        & \quad \quad \quad \quad+ \left\|vec\bigg(V(\hat\Lambda)\bigg)_{S^\Lambda}\right\|_\infty\bigg]\\
        &< \frac{r_n}{4}+\frac{r_n}{4}+\frac{r_n}{2}\\
        &= r_n.
    \end{align*}
Thus, by Brouwers' fixed point theorem, there exists a $\hat\Lambda \in \mathcal{A}$ such that $\|\left(\hat\Lambda - \Lambda^0\right)_{S^\Lambda}\|_\infty \leq r_n$. 

\textbf{Step 2:}  We now show that the above solution to the restricted optimization problem is a local minimizer of our original optimization function $L(\Theta,\Lambda)$ in the HPD region. Note that since $(\Theta^0,\Lambda^0)$ belongs to the constrained set of the restricted optimization problem $(\hat\Theta,\hat\Lambda)$ is a minimizer of, $L(\hat\Theta,\hat\Lambda) \leq L(\Theta^0,\Lambda^0)$ and thus $(\hat\Theta,\hat\Lambda) \in HPD$. Thus, to prove the above, it suffices to show that there exists a small positive number $\epsilon$ with $L\big((\hat\Theta,\hat\Lambda)+\Delta\big)\geq L(\hat\Theta,\hat\Lambda)$ for any $\Delta$ such that $\|\Delta\|_\infty < \epsilon$ and $\big((\hat\Theta,\hat\Lambda)+\Delta\big) \in HPD$. We define $D(\Delta) = L\big((\hat\Theta,\hat\Lambda)+\Delta\big)- L(\hat\Theta,\hat\Lambda)$, then our problem boils down to showing $D(\Delta)\geq0$ for any $\Delta$ such that $\|\Delta\|_\infty < \epsilon$ and $\big((\hat\Theta,\hat\Lambda)+\Delta\big) \in HPD$. 
Taking $\Delta := (\Delta_\Theta, \Delta_\Lambda)$ and  considering $\|\Delta\|_\infty \leq r_0$ to be small enough, 
\begin{align}
\label{eq17}
    D(\Delta) & = L\big((\hat\Theta,\hat\Lambda)+\Delta\big)- L(\hat\Theta,\hat\Lambda)\nonumber\\
    &= -l\big((\hat\Theta,\hat\Lambda)+\Delta\big)+ l(\hat\Theta,\hat\Lambda) + \sum_iPen_{MSS}\left(\hat \Theta_i + (\Delta_\Theta)_i\right)\nonumber\\
    &\quad \quad \quad - \sum_iPen_{MSS}(\hat \Theta_i)  + \sum_{i<j}Pen_{SS}\left(\hat \Lambda_{ij} + (\Delta_\Lambda)_{ij}\right) \nonumber\\
    &\quad \quad \quad- \sum_iPen_{MSS}(\hat \Lambda_{ij}).
\end{align}
Since $(\hat\Theta,\hat\Lambda)$ and $\big((\hat\Theta,\hat\Lambda)+\Delta\big)$ both like in the HPD, using Lemma 1 and Lemma 2, we have 
\begin{align}
\label{eq18}
    l(\hat\Theta,\hat\Lambda)-l\big((\hat\Theta,\hat\Lambda)+\Delta\big) &\geq n\beta_0\|\Delta\|_F^2 - vec\left(\nabla l(\hat\Theta,\hat\Lambda)\right)^Tvec(\Delta)\nonumber\\
    & = n\beta_0\|\Delta\|_F^2 - \sum_{(i,j)\in S^0}\left(\nabla l(\hat\Theta,\hat\Lambda)\right)_{ij}\Delta_{ij}\nonumber\\
    &\quad \quad \quad- \sum_{(i,j)\in (S^0)^C}\left(\nabla l(\hat\Theta,\hat\Lambda)\right)_{ij}\Delta_{ij}\nonumber\nonumber\\
    &= n\beta_0\|\Delta\|_F^2 - \sum_{(i,j)\in S^\Theta}\left(\nabla_\Theta l(\hat\Theta,\hat\Lambda)\right)_{ij}\left(\Delta_\Theta\right)_{ij}\nonumber\\
    &\quad \quad \quad- \sum_{(i,j)\in (S^\Theta)^C}\left(\nabla_\Theta l(\hat\Theta,\hat\Lambda)\right)_{ij}\left(\Delta_\Theta\right)_{ij}\nonumber\\
    &\quad \quad \quad - \sum_{(i,j)\in S^\Lambda}\left(\nabla_\Lambda l(\hat\Theta,\hat\Lambda)\right)_{ij}\left(\Delta_\Lambda\right)_{ij}\nonumber\\
    &\quad \quad \quad- \sum_{(i,j)\in (S^\Lambda)^C}\left(\nabla_\Lambda l(\hat\Theta,\hat\Lambda)\right)_{ij}\left(\Delta_\Lambda\right)_{ij}.
\end{align}
Also, since the penalty functions $Pen_{MSS}$ is symmetric, we use a second-order Taylor expansion to get, 
\begin{align}
\label{eq19}
    \sum_i&Pen_{MSS}\left(\hat \Theta_i + (\Delta_\Theta)_i\right) - \sum_iPen_{MSS}(\hat \Theta_i)\nonumber\\ &= \sum_iPen_{MSS}\left(|\hat \Theta_i + (\Delta_\Theta)_i|\right) - \sum_iPen_{MSS}(|\hat \Theta_i|)\nonumber\\
    &= \sum_i U\left(\hat \Theta_i\right)^Tvec\left(|\hat \Theta_i + (\Delta_\Theta)_i| - |\hat \Theta_i|\right)\nonumber\\
    & \quad \quad\quad \quad +\frac12\sum_{i}vec\left(|\hat \Theta_i + (\Delta_\Theta)_i| - |\hat \Theta_i|\right)^T\nabla^2Pen_{MSS}\left(\hat \Theta_i + u\Delta_i\right)vec\left(|\hat \Theta_i + (\Delta_\Theta)_i| - |\hat \Theta_i|\right)\nonumber\\
    & \geq \sum_{(i,j)\in S^\Theta}\left(U(\hat \Theta_i)\right)_{ij}\left(|\hat \Theta_{ij} + (\Delta_\Theta)_{ij}| - |\hat \Theta_{ij}|\right)
      + \sum_{(i,j)\in (S^\Theta)^C}\left(U(\hat \Theta_i)\right)_{ij}\left(|\hat \Theta_{ij} + (\Delta_\Theta)_{ij}| - |\hat \Theta_{ij}|\right)\nonumber\\
    & \quad \quad \quad \quad -\frac12 \max_i\left\{\left\|\nabla^2Pen_{MSS}\left(\hat \Theta_i + u\Delta_i\right)\right\|_\infty\right\}\sum_i\sum_{j}\left(|\Theta_{ij}+(\Delta_\Theta)_{ij}|-|\Theta_{ij}|\right)^2\nonumber\\
    & \geq \sum_{(i,j)\in S^\Theta}\left(U(\hat \Theta_i)\right)_{ij}\left(|\hat \Theta_{ij} + (\Delta_\Theta)_{ij}| - |\hat \Theta_{ij}|\right)  + \sum_{(i,j)\in (S^\Theta)^C}\left(U(\hat \Theta_i)\right)_{ij}\left(|\hat \Theta_{ij} + (\Delta_\Theta)_{ij}| - |\hat \Theta_{ij}|\right)\nonumber \\&\quad \quad \quad \quad -\frac{nC_1^2\log(p+q)}{4(p+q)^{\epsilon_0}}\|\Delta_\Theta\|_F^2.
\end{align}
We use $x^TAx \geq -\|A\|_\infty \left(x^Tx\right)$ when $A$ is a negative semi-definite matrix and x is a vector to get the third inequality. The fourth inequality is due to the triangle inequality and the bound of the second derivative of $Pen_{MSS}$ derived in Proposition \ref{prop2}. Similarly since $Pen_{SS}$ is also symmetric we have, 
\begin{align}
\label{eq20}
    \sum_i\sum_jPen&_{SS}\left(\hat \Lambda_{ij} + (\Delta_\Lambda)_{ij}\right) - \sum_i\sum_jPen_{SS}(\hat \Lambda_{ij})\nonumber\\ &= \sum_i\sum_jPen_{SS}\left(|\hat \Lambda_{ij} + (\Delta_\Lambda)_{ij}|\right) - \sum_i\sum_jPen_{SS}(|\hat \Lambda_{ij}|)\nonumber\\& \geq \sum_{(i,j)\in S^\Lambda}\left(V(\hat \Lambda_{ij})\right)_{ij}\left(|\hat \Lambda_{ij} + (\Delta_\Lambda)_{ij}| - |\hat \Lambda_{ij}|\right)\nonumber\\
    &\quad \quad  + \sum_{(i,j)\in (S^\Lambda)^C}\left(V(\hat \Lambda_{ij})\right)_{ij}\left(|\hat \Lambda_{ij} + (\Delta_\Lambda)_{ij}| - |\hat \Lambda_{ij}|\right) -\frac{nC_1^2\log(p+q)}{8(p+q)^{\epsilon_0}}\|\Delta_\Lambda\|_F^2,
\end{align}
where we use the bound of the second derivative of $Pen_{SS}$ derived in Proposition \ref{prop2} to get the last inequality. 
Using the results from (\ref{eq18}),(\ref{eq19}) and (\ref{eq20}), and using the triangle inequality, we can bound $D(\Delta)$ from below as follows:

\begin{align*}
    D(\Delta)&\geq n\left(\beta_0-\frac{C_1^2\log(p+q)}{4(p+q)^{\epsilon_0}} \right)\|\Delta\|_F^2
    \\& \quad \quad+ \sum_{(i,j)\in S^\Theta}\left(U(\hat \Theta_{i})-\nabla_\Theta l(\hat\Theta,\hat\Lambda)\right)_{ij}\left(|\hat \Theta_{ij} + (\Delta_\Theta)_{ij}| - |\hat \Theta_{ij}|\right)\nonumber\\
    &\quad \quad  + \sum_{(i,j)\in (S^\Theta)^C}\left(U(\hat \Theta_{i})-\nabla_\Theta l(\hat\Theta,\hat\Lambda)\right)_{ij}\left(|\hat \Theta_{ij} + (\Delta_\Theta)_{ij}| - |\hat \Theta_{ij}|\right) \\
    & \quad \quad+\sum_{(i,j)\in S^\Lambda}\left(V(\hat \Lambda_{ij})-\nabla_\Lambda l(\hat\Theta,\hat\Lambda)\right)_{ij}\left(|\hat \Lambda_{ij} + (\Delta_\Lambda)_{ij}| - |\hat \Lambda_{ij}|\right)\nonumber\\
    &\quad \quad  + \sum_{(i,j)\in (S^\Lambda)^C}\left(V(\hat \Lambda_{ij})-\nabla_\Lambda l(\hat\Theta,\hat\Lambda)\right)_{ij}\left(|\hat \Lambda_{ij} + (\Delta_\Lambda)_{ij}| - |\hat \Lambda_{ij}|\right).
\end{align*}
Since $(\hat\Theta,\hat\Lambda) \in \mathcal{A}$,  we have 
$\sum_{(i,j)\in S^\Theta}\left(U(\hat \Theta_{i})-\nabla_\Theta l(\hat\Theta,\hat\Lambda)\right)_{ij} = 0$ and
$\sum_{(i,j)\in S^\Lambda}\bigg(V(\hat \Lambda_{ij})$\\ $-\nabla_\Lambda l(\hat\Theta,\hat\Lambda)\bigg)_{ij} = 0$ due to the zero sub-gradient conditions (\ref{eq13}) and (\ref{eq14}). Thus,
\begin{align*}
    D(\Delta)&\geq n\left(\beta_0-\frac{C_1^2\log(p+q)}{4(p+q)^{\epsilon_0}} \right)\|\Delta\|_F^2
    \\
    &\quad \quad+\sum_{(i,j)\in (S^\Theta)^C}\left(U(\hat \Theta_{i})-\nabla_\Theta l(\hat\Theta,\hat\Lambda)\right)_{ij}\left(|\hat \Theta_{ij} + (\Delta_\Theta)_{ij}| - |\hat \Theta_{ij}|\right) \\
    &\quad \quad  + \sum_{(i,j)\in (S^\Lambda)^C}\left(V(\hat \Lambda_{ij})-\nabla_\Lambda l(\hat\Theta,\hat\Lambda)\right)_{ij}\left(|\hat \Lambda_{ij} + (\Delta_\Lambda)_{ij}| - |\hat \Lambda_{ij}|\right).
\end{align*}
Note that for $(i,j)\in (S^\Theta)^C,\Theta_{ij} = 0$ and for $(i,j)\in (S^\Lambda)^C, \Lambda_{ij} = 0$ which implies,
\begin{align}
\label{eq21}
    D(\Delta)&\geq n\left(\beta_0-\frac{C_1^2\log(p+q)}{4(p+q)^{\epsilon_0}} \right)\|\Delta\|_F^2 - \sum_{(i,j)\in(S^\Theta)^C}\nabla l(\hat\Theta,\hat\Lambda)|\Delta_{ij}| - \sum_{(i,j)\in(S^\Lambda)^C}\nabla l(\hat\Theta,\hat\Lambda)|\Delta_{ij}|\nonumber\\& \quad\quad \quad \quad+ U(0)\sum_{(i,j)\in (S^\Theta)^C}|\Delta_{ij}|+V(0)\sum_{(i,j)\in (S^\Lambda)^C}|\Delta_{ij}|.
\end{align}
Looking at each of the terms individually, we first note that the derivatives of our penalty functions at 0 can be bounded from below as follows:
\begin{align*}
    U(0) &= \underbrace{\frac{1}{\vtone}\eta_2(0)\eta_1(0)}_{\geq0}+\frac{1}{\vtzero}\big(1-\underbrace{\eta_2(0)}_{<1}\eta_1(0)\big) \geq n\left(\frac{1}{n\vtzero}\big(1-\eta_1(0)\big)\right)\\
    & \geq nC_4\sqrt{\frac{\log(p+q)}{n}}\left(1+ \left(\frac{\et}{1-\et}\right)\left(\frac{\vtone}{\vtzero}\right)\right)^{-1} \text{ and}
\end{align*}
\begin{align*}
    V(0) &= \underbrace{\frac{1}{\vlone}\eta_1(0)}_{\geq0}+\frac{1}{\vlzero}\big(1-\eta_1(0)\big) \geq n\left(\frac{1}{n\vlzero}\big(1-\eta_1(0)\big)\right)\\
    & \geq nC_4\sqrt{\frac{\log(p+q)}{n}}\left(1+ \left(\frac{\el}{1-\el}\right)\left(\frac{\vlone}{\vlzero}\right)\right)^{-1},
\end{align*}
and the likelihood terms can be bounded from below as follows:
\begin{align*}
    - \sum_{(i,j)\in(S^\Theta)^C}\nabla l(\hat\Theta,\hat\Lambda)|\Delta_{ij}|&  \geq - \|\nabla l(\hat\Theta,\hat\Lambda)\|_\infty \sum_{(i,j)\in(S^\Theta)^C}|\Delta_{ij}|\\
    &= -\left(\|\nabla l(\Theta^0,\Lambda^0)\|_\infty+ c_H\frac{n}{2}\|\hat\Delta\|_\infty + \|R(\hat \Delta)\|_\infty \right)\sum_{(i,j)\in(S^\Theta)^C}|\Delta_{ij}|\\
    &\geq -\frac{nr_n}{2}\left(\frac{1}{2c_H}+ c_H\right)\sum_{(i,j)\in(S^\Theta)^C}|\Delta_{ij}|,
\end{align*}
where we get the second inequality due to a first-order Taylor expansion and the third inequality using the bounds derived in Step 1. Similarly, 
\begin{align*}
    - \sum_{(i,j)\in(S^\Lambda)^C}\nabla l(\hat\Theta,\hat\Lambda)|\Delta_{ij}|
    \geq -\frac{nr_n}{2}\left(\frac{1}{2c_H}+ c_H\right)\sum_{(i,j)\in(S^\Lambda)^C}|\Delta_{ij}|.
\end{align*}
Substituting the above results in (\ref{eq21}) we get, 

\begin{align*}
    D(\Delta) &\geq n\left(\beta_0-\frac{C_1^2\log(p+q)}{4(p+q)^{\epsilon_0}} \right)\|\Delta\|_F^2 \\&\quad  + n\left(C_4\sqrt{\frac{\log(p+q)}{n}}\left(1+ \left(\frac{\et}{1-\et}\right)\left(\frac{\vtone}{\vtzero}\right)\right)^{-1} - \frac{r_n}{2}\left(\frac{1}{2c_H}+ c_H\right)\right)\sum_{(i,j)\in(S^\Lambda)^C}|\Delta_{ij}|       \\&\quad  +n\left(C_4\sqrt{\frac{\log(p+q)}{n}}\left(1+ \left(\frac{\el}{1-\el}\right)\left(\frac{\vlone}{\vlzero}\right)\right)^{-1} - \frac{r_n}{2}\left(\frac{1}{2c_H}+ c_H\right)\right)\sum_{(i,j)\in(S^\Lambda)^C}|\Delta_{ij}|.
\end{align*}
Each term on the right-hand side of the above inequality is non-negative due to the conditions, and thus we have 
$$D(\Delta)\geq 0.$$
\end{proof}

\noindent \textbf{Proof for Uniqueness:}

We present the following lemma, which shows that our negative log posterior function is strongly convex. Once the lemma is applied, the proof for the uniqueness of our estimators is straightforward. 

\begin{Lemma}
\label{lemma4}
    The negative log posterior function $L(\Theta,\Lambda)$ is strongly convex with constant $m$, in the HPD region, if $\beta_0 \geq \left(\frac 14 + \frac p8\right)\frac{C_1^2\log(p+q)}{(p+q)^{\epsilon_0}}+ \frac{m}{2n}$ where $\epsilon_0 = (C_3-C_1)c_H(C_4 - C_1)$ with $C_3,C_1,C_4$ as defined in Proposition \ref{prop2} and $\beta_0$ as defined in Lemma \ref{Lemma1}.  
\end{Lemma}
\begin{proof} We have $L(\Theta,\Lambda) = -l(\Theta,\Lambda) + \sum_iPen_{MSS}(\Theta_i)+\sum_{i<j}Pen_{SS}(\Lambda_{ij})$. Let $Pen(\Theta) := \sum_iPen_{MSS}(\Theta_i)$ and $Pen(\Lambda):=\sum_{i<j}Pen_{SS}(\Lambda_{ij})$. Using the properties of the likelihood function from Section 4.1 of the main paper, we have for all points in the HPD, $$-l\left((\Theta^0,\Lambda^0)+\Delta)\right) \geq - l\left(\Theta^0,\Lambda^0\right) - \inprod{\nabla l\left(\Theta^0,\Lambda^0\right) }{\Delta} + n\beta_0\|\Delta\|_F^2.$$ Thus, $-l\left(\Theta,\Lambda\right)$ is strongly convex with constant $2n\beta_0$ in the HPD region.
    $$\implies -\nabla^2l\left(\Theta,\Lambda\right) \succeq 2n\beta_0I$$
Moreover, using the bounds for the penalty function $Pen_{MSS}$ from Proposition \ref{prop2} we have,  
$$\nabla^2Pen(\Theta) \succeq \begin{bmatrix}
        A& 0\\0&0
    \end{bmatrix},$$ where A is a $qp \times qp$ block diagonal matrix with the form $\begin{bmatrix}
        B & 0 & \dots & 0\\
        0 & B & \dots & 0\\
        \vdots &  & \ddots & \vdots\\
        0& 0 & \dots & B
    \end{bmatrix}$ and each block $B$ is a $p \times p$ matrix with $B_{ii} = -\frac{2nC_1^2\log(p+q)}{4(p+q)^{\epsilon_0}}$ and $B_{ij} = -\frac{2nC_1^2\log(p+q)}{8(p+q)^{\epsilon_0}}$ for $i = \{1,2,\dots p\}$, $j = \{1,2,\dots p\}$. Similarly, using the bounds for the penalty function $Pen_{SS}$ from Proposition \ref{prop2} we have, $$\nabla^2Pen(\Lambda) \succeq \begin{bmatrix}
        0& 0\\0&C
    \end{bmatrix},$$  where C is a $p^2 \times p^2$ diagonal matrix with $C_{ii} = -\frac{2nC_1^2\log(p+q)}{8(p+q)^{\epsilon_0}}$. 
Thus in the HPD region, for a fixed m, $$\nabla^2L(\Theta,\Lambda) - mI \succeq \begin{bmatrix}
        B + (2n\beta_0-m)I_{p\times p} & 0  && \ldots\\
       0& \ddots && 0\\
        \ldots & 0 & B + (2n\beta_0-m)I_{p\times p}&0\\
        \ldots & 0 & 0&C + (2n\beta_0-m)I_{p^2\times p^2}\\
    \end{bmatrix}.$$
Note that a block diagonal matrix is positive semi-definite if each block is positive semi-definite. Let us look at each of the blocks individually. 
$$C + (2n\beta_0-m)I_{p^2\times p^2} = diag\left(2n\beta_0 - m -\frac{2nC_1^2\log(p+q)}{8(p+q)^{\epsilon_0}} \right)\succeq 0,$$ since we have $0 \leq m \leq 2n\beta_0 - \frac{2nC_1^2\log(p+q)}{8(p+q)^{\epsilon_0}}$ from the condition. Next, we show $B + (2n\beta_0-m)I_{p\times p}$ is positive semi-definite by using the result: a symmetric diagonally dominant matrix with non-negative diagonal entries is positive-semi definite. The diagonal entries of $B + (2n\beta_0-m)I_{p\times p}$,
$$diag\left(B + (2n\beta_0-m)I_{p\times p}\right) = 2n\beta_0 - m -\frac{2nC_1^2\log(p+q)}{4(p+q)^{\epsilon_0}} \geq 0,$$ 
since we have $0 \leq m \leq 2n\beta_0 - \frac{2nC_1^2\log(p+q)}{4(p+q)^{\epsilon_0}}$ from the condition.  Thus $B + (2n\beta_0-m)I_{p\times p}$ has non-negative diagonal entries. 
Additionally, using the condition we have 
$$2n\beta_0 - m - \frac{2nC_1^2\log(p+q)}{4(p+q)^{\epsilon_0}}\geq \frac{2npC_1^2\log(p+q)}{8(p+q)^{\epsilon_0}},$$ which implies that the sum of the off-diagonal elements in a row is smaller than the diagonal element of that row in the matrix $B + (2n\beta_0-m)I_{p\times p}$, thus proving that $B + (2n\beta_0-m)I_{p\times p}$ is diagonally dominant. Therefore $B + (2n\beta_0-m)I_{p\times p}$ is positive semi-definite. This, combined with $C + (2n\beta_0-m)I_{p^2\times p^2}$ being positive semi-definite, gives us $$\nabla^2L(\Theta,\Lambda) - mI \succeq 0,$$ implying that the negative log posterior function is strongly convex with constant $m$ in the HPD region. 
\end{proof}

The proof for the uniqueness of the estimators is trivial since the negative log posterior $L(\Theta,\Lambda)$ is strongly convex in the HPD region. Thus, the stationary point obtained in Theorem 2 is the unique minimizer. 

\subsection{Proof for Corollary 2 and Corollary 3}
\textbf{Proof for Corollary 2:}
\begin{proof}
The upper bound for the Frobenius norm of the difference in the plug-in estimate and the true value of B:

    \begin{align*}
        \|\hat B - B^0\|_F &= \|\Lambda^{-1}\Theta^T - (\Lambda^0)^{-1}(\Theta^0)^T\|_F\\
        &= \|\Lambda^{-1}\Theta^T- (\Lambda^0)^{-1}\Theta^T+ (\Lambda^0)^{-1}\Theta^T  -(\Lambda^0)^{-1}(\Theta^0)^T\|_F\\
        &\leq \|\Lambda^{-1}- (\Lambda^0)^{-1}\|_F\|\Theta^T\|_F + \|(\Lambda^0)^{-1}\|_F\|\Theta^T - (\Theta^0)^T\|_F\\
        &\leq \left(\frac{F_\Lambda^2\epsilon_n}{1 - F_\Lambda\epsilon_n}\right)\left(F_\Theta + \epsilon_n\right) + F_\Lambda\epsilon_n\\
        & = \frac{F_\Lambda\epsilon_n}{1- F_\Lambda\epsilon_n}(1+F_\Lambda F_\Theta).
    \end{align*}
The fourth inequality uses the following results: $\|A^{-1} - B^{-1}\| \leq \frac{\|A^{-1}\|^2\|A - B\|}{1 - \|A^{-1}\|\|A - B\|}$ for any sub-multiplicative norm and $\|A\| \leq \|A- B\| + \|B\|$. 
\end{proof}
\noindent\textbf{Proof for Corollary 3:}

 We have $\Tilde{X}_{n\times \Tilde{q}} = (X_1,X_2,\dots,X_{\Tilde{q}})$ the design matrix comprising the variables selected. We estimate $\Tilde{B}$ using $p$ linear regression models, where each model $M_j$ is defined as: 
$$M_j: Y_{.j} = \Tilde{X}\Tilde{B}_{j.}^T+\varepsilon_j; \quad j = \{1,2,\dots, p\}$$
Every row of $\Tilde{B}$ (denoted as $\Tilde{B}_{j.}$) is estimated using the ordinary least squares estimator for model $M_j$. Let us denote this least squares estimator as $\hat{\Tilde{B}}_{j.}$. It is a well-known result \citep{Mit_regr,CMU_regr} that the least squares estimator satisfies with probability greater than $1-\mathcal{O}\{(p+q)^{-1}\}$

$$ \|\Tilde{X}\left(\hat{\Tilde{B}}_{j.}^T - \Tilde{B}_{j.}^T\right)\|_F \leq C\sqrt{\frac{\Tilde{q}+log(p+q)}{n}}$$
Using the result $\|\hat{\Tilde{B}}_{j.}^T - \Tilde{B}_{j.}^T\|_F^2\lambda_{min}^2(\Tilde{X}) \leq  \|\Tilde{X}\left(\hat{\Tilde{B}}_{j.}^T - \Tilde{B}_{j.}^T\right)\|_F^2$ and aggregating over the $p$ rows we get the final result. 

\section{More experimental results}
\subsection{Simulations}
\textbf{Set-up S1: High signal strength, low dimensional setting with independent non-zero elements in Theta} 

In this setup, we consider the number of responses $p = 10$, the number of covariates $q = 50$, the number of non-zero elements in the non-zero rows of $\Theta^0$ ($s_{\Theta^0}$) = 10 and the number of non-zero elements in $\Lambda^0$ ($s_{\Lambda^0}$) = 5. This implies that $\Lambda^0$ has 10 non-zero off-diagonal elements, but only 5 are uniquely estimated since it is symmetric. $\Lambda^0$ also has p=10 non-zero diagonal elements. The non-zero elements for both $\Theta^0$ and $\Lambda^0$ are generated from $Unif\left([-6,-4]\cup [4,6]\right)$. Thus, the minimum signal strength is 4. The results for this setup are tabulated in tables \ref{tab:setup1err},\ref{tab:setup1mcc} and \ref{tab:setup1mccb}.

\noindent \textbf{Set-up S2: High signal strength, low dimensional setting with non-
zero elements dependent in Theta} 

In this setup, we once again consider $p = 10$, $q = 50$ and $s_{\Lambda^0}$ = 5. The non-zero elements for $\Lambda^0$ are still generated from $Unif\left([-6,-4]\cup [4,6]\right)$. But instead of the non-zero elements in $\Theta^0$ being independent, they are generated from a uniform sphere with norm = 4 using the method described above. Thus, these non-zero elements are dependent and have a low signal strength since the row norm is fixed to be less than $4$. The results for this setup are tabulated in tables \ref{tab:setup2err},\ref{tab:setup2mcc} and \ref{tab:setup2mccb}.

\begin{table}[htp]
\centering
{\footnotesize	
\begin{tabular}{|c|l|lllllll|}
\hline
                            & \multicolumn{1}{c|}{}                         & \multicolumn{7}{c|}{Error (Frobenius Norm)}                                                                                                                                                                                                                                           \\ \cline{3-9} 
\multirow{-2}{*}{} & \multicolumn{1}{c|}{\multirow{-2}{*}{Method/N}} & 20                                & 40                                & 100                               & 200                               & 500                               & 1000                              & 10000                             \\ \hline
                            & BVS.GM                                        &  7.842& {\color[HTML]{000000} \textbf{5.869}} & {\color[HTML]{000000} \textbf{3.247}} & {\color[HTML]{000000} \textbf{1.551}} & {\color[HTML]{000000} \textbf{1.062}} & {\color[HTML]{000000} \textbf{1.111}} & {\color[HTML]{000000} \textbf{0.354}} \\ \cline{2-2}
                            & GLASSO                                        & 7.599                               & 7.441                                 & 5.986                                & 6.030                                 & 2.251                                 &2.261                                & 0.688                                 \\ \cline{2-2}
                            & CAPME                                         & 15.761                                & 14.625                                & 7.249                                & 3.118                                & 1.951                               & 1.486                                 & 1.017                                 \\
\hline

\end{tabular}}
\caption{Simulation results for the estimation accuracy of the different methods under Setup S1.}
\label{tab:setup1err}
\vspace{5mm}

{\footnotesize
\begin{tabular}{|c|l|lllllll|}
\hline
                            & \multicolumn{1}{c|}{}                         & \multicolumn{7}{c|}{Selection Consistency (MCC)}                                                                                                                                                                                                                                      \\ \cline{3-9} 
\multirow{-2}{*}{} & \multicolumn{1}{c|}{\multirow{-2}{*}{Method/N}} & 20                                & 40                                & 100                               & 200                               & 500                               &1000                              &10000                             \\ \hline
                            & BVS.GM                                        & \textbf{0.590}         & \textbf{0.857}         & 0.952 & {\color[HTML]{000000} \textbf{0.971}} & {\color[HTML]{000000} \textbf{0.932}} & {\color[HTML]{000000} \textbf{0.873}} & 0.817 \\ \cline{2-2}
                            & GLASSO                                        & 0.343 & 0.431                                & 0.275                                & 0.358                                & 0.108                                 & 0.130                                 & 0.106                                 \\ \cline{2-2}
                            & CAPME                                         & 0.065                                 & 0.052                                 & 0.035                                 & 0.061                                 & 0.171                                 & 0.281                                 & 0.352                                 \\
\hline
\end{tabular}}
\caption{Simulation results for the element-wise structure recovery of the different methods under Setup S1.}
\label{tab:setup1mcc}

\vspace{5mm}

{\footnotesize
\begin{tabular}{|l|lllllll|}
\hline
\multicolumn{1}{|c|}{}                   & \multicolumn{7}{c|}{Column Selection (MCC)}                                                                                                                                                                                            \\ \cline{2-8} 
\multicolumn{1}{|c|}{\multirow{-2}{*}{Method/N}} & 20                                & {\color[HTML]{000000} 40}         & 100                               & 200                               & 500                               & 1000                              & 10000                             \\ \hline
BVS.GM                                      & {\color[HTML]{000000} \textbf{0.593}} & 0.863& {\color[HTML]{000000} \textbf{0.995}} & {\color[HTML]{000000} \textbf{1.000}} & {\color[HTML]{000000} \textbf{1.000}} & {\color[HTML]{000000} \textbf{1.000}} & 0.962                                 \\
GLASSO                                   & {\color[HTML]{330001} 0.176}          & {\color[HTML]{000000} 0.469}          & 0.176                                & 0.156                                 & 0.000                                 & 0.000                                 & 0.000                                 \\
CAPME                                    & 0.196                                 & {\color[HTML]{000000} 0.245}          & 0.149                                 & 0.190                                & 0.515                                & 0.882                              & {\color[HTML]{000000} \textbf{1.000}} \\
BCRF                                     & {\color[HTML]{000000} 0.570}          & {\color[HTML]{000000} \textbf{0.891}}          & {\color[HTML]{000000} 0.992}          & {\color[HTML]{000000} 0.993}          & {\color[HTML]{000000} 0.964}          & 0.986                                 & 0.935                                 \\ \hline
\end{tabular}}
\caption{Simulation results for the column structure recovery for $B$ of the different methods under Setup S1.}
\label{tab:setup1mccb}
\end{table}

\begin{table}[htp]
\centering
{\footnotesize	
\begin{tabular}{|c|l|lllllll|}
\hline
                            & \multicolumn{1}{c|}{}                         & \multicolumn{7}{c|}{Error (Frobenius Norm)}                                                                                                                                                                                                                                           \\ \cline{3-9} 
\multirow{-2}{*}{} & \multicolumn{1}{c|}{\multirow{-2}{*}{Method/N}} & 20                                & 40                                & 100                               & 200                               & 500                               & 1000                              & 10000                             \\ \hline
                            & BVS.GM                                        &  10.206& 10.591 & {\color[HTML]{000000} \textbf{7.771}} & {\color[HTML]{000000} \textbf{3.455}} & {\color[HTML]{000000} \textbf{2.391
}} & {\color[HTML]{000000} \textbf{1.843}} & {\color[HTML]{000000} \textbf{1.552}} \\ \cline{2-2}
                            & GLASSO                                        & 10.211                          & 
9.833
                                     & 9.516
                              & 9.645
                            & 3.529
                                      &4.661                               & 
0.752                                \\ \cline{2-2}
                            & CAPME                                         & 14.378
&16.583
&8.512
&5.083
&2.694
&2.308
&2.031                                                                \\ \cline{2-2}
\multirow{-4}{*}{Theta}     & BCRF                                          &\textbf{9.564}
&\textbf{8.986}
&7.822
&6.222
&4.221
&3.926
&1.797                                 \\ \hline
                            & BVS.GM                                        & 12.671
&12.535
&10.180
&\textbf{5.349}
&3.604
&3.434
&2.115 \\
\hline

\end{tabular}}
\caption{Simulation results for the estimation accuracy of the different methods under Setup S2.}
\label{tab:setup2err}
\vspace{5mm}

{\footnotesize
\begin{tabular}{|c|l|lllllll|}
\hline
                            & \multicolumn{1}{c|}{}                         & \multicolumn{7}{c|}{Selection Consistency (MCC)}                                                                                                                                                                                                                                      \\ \cline{3-9} 
\multirow{-2}{*}{} & \multicolumn{1}{c|}{\multirow{-2}{*}{Method/N}} & 20                                & 40                                & 100                               & 200                               & 500                               &1000                              &10000                             \\ \hline
                            & BVS.GM                                        & \textbf{0.329}
&\textbf{0.565}
&\textbf{0.767}
&0.785
&0.827
&\textbf{0.851}
&0.792 \\ \cline{2-2}
                            & GLASSO                                        & 0.247
&0.364
&0.314
&0.435
&0.183
&0.228
&0.137                                 \\ \cline{2-2}
                            & CAPME                                         & 0.110
&0.093
&0.053
&0.116
&0.263
&0.375
&0.436                                 \\ \cline{2-2}
\multirow{-4}{*}{Theta}     & BCRF                                          & 0.322
&0.561
&0.720
&\textbf{0.809}
&\textbf{0.839}
&0.838
&0.851                                \\ \hline
                            & BVS.GM                                        & \textbf{0.545}
&\textbf{  0.587}
&0.634
&0.765
&\textbf{0.825}
&\textbf{0.856}
&0.839 \\ \cline{2-2}
                            & GLASSO                                        & 0.449
&0.404
&0.532
&0.587
&0.448
&0.384
&0.287                                 \\
\hline
\end{tabular}}
\caption{Simulation results for the element-wise structure recovery of the different methods under Setup S2.}
\label{tab:setup2mcc}

\vspace{5mm}

{\footnotesize
\begin{tabular}{|l|lllllll|}
\hline
\multicolumn{1}{|c|}{}                   & \multicolumn{7}{c|}{Column Selection (MCC)}                                                                                                                                                                                            \\ \cline{2-8} 
\multicolumn{1}{|c|}{\multirow{-2}{*}{Method/N}} & 20                                & {\color[HTML]{000000} 40}         & 100                               & 200                               & 500                               & 1000                              & 10000                             \\ \hline
BVS.GM                                      & 0.428
&0.704
&0.862
&\textbf{0.956}
&\textbf{0.963}
&\textbf{0.972}
&0.982 \\
GLASSO                                   & 0.158
&0.217
&0.000
&0.343
&0.000
&0.000
&0.000                                 \\
CAPME                                    & 0.282
&0.094
&0.121
&0.182
&0.584
&0.826
&1.000 \\
BCRF                                     & \textbf{0.456}
&\textbf{0.772}
&\textbf{0.902}
&0.953
&0.961
&0.956
&0.949                                 \\ \hline
\end{tabular}}
\caption{Simulation results for the column structure recovery for $B$ of the different methods under Setup 2.}
\label{tab:setup2mccb}
\end{table}

\subsection{Application to Bike-share data}
In the main paper, we consider two different extents of the data: data for January 2023 (n = 31) and data from three months, January - March 2023 (n = 90). In Table \ref{reallifetable} we present results for the data of the whole year (n = 365).

\begin{table}[ht]
\centering
\begin{tabular}{|l|l|l|}
\hline
Method                                 & All test values unknown                & Half test values unknown                 \\ \hline
BVS.GM & {\textbf{23.034}} & {\textbf{13.22729}} \\
GLASSO                                 & 23.958                                 & 13.55859                                \\
CAPME                                  & 80.556                                 & 45.13003                                 \\
BCRF                                   & 23.422                                 & 16.41848                                 \\ \hline
\end{tabular}
\caption{Average prediction error for capital bike-share demand (n = 365)}
\label{reallifetable}
\end{table}

{
\bibliography{references.bib}
\bibliographystyle{apalike}
}